\documentclass[aps,pra,reprint,twocolumn,superscriptaddress,floatfix,longbibliography]{revtex4-1}
\usepackage{graphicx,amsmath,amsfonts,amssymb,amsthm,xr}
\usepackage{epsfig,amsmath,amssymb,color,dsfont,upgreek,physics}
\usepackage{mathrsfs}
\usepackage{mathtools}
\usepackage{bbold}
\usepackage{comment}
\usepackage{braket}
\usepackage{float}

\usepackage[bookmarks=true,colorlinks,linkcolor=OrangeRed,urlcolor=NavyBlue,citecolor=RoyalBlue]{hyperref}
\usepackage[dvipsnames]{xcolor}
\usepackage{orcidlink}

\def\d{\mathrm d}
\definecolor{mygold}{rgb}{0.93,0.69,0.13}

\definecolor{mypurple}{rgb}{0.49,0.18,0.56}

\definecolor{mygreen}{rgb}{0,0.5,0}
\definecolor{myred}{rgb}{0.7,0,0}
\definecolor{myblue}{rgb}{0,0,1}
\newcommand{\X}{\hat{X}}
\newcommand{\Y}{\hat{Y}}
\newcommand{\Z}{\hat{Z}}

\newcommand{\n}{\hat{n}}

\newcommand{\Pd}{\hat{P}}
\newcommand{\Pu}{\hat{Q}}
\newcommand{\Heff}{\hat{H}_{\rm eff}}
\renewcommand{\c}{\hat{c}^{\phantom\dagger}}
\newcommand{\cd}{\hat{c}^{\dagger}}

\newcommand{\Hn}{\hat{H}_{0}}
\renewcommand{\d}{\downarrow}
\renewcommand{\u}{\uparrow}
\newcommand{\rd}{{\color{red}\downarrow}}
\newcommand{\ru}{{\color{red}\uparrow}}
\newcommand{\kb}{\ket{\beta}}

\begin{document}
\title{Mass-Assisted Local Deconfinement in a Confined \texorpdfstring{$\mathbb{Z}_2$}{Z2} Lattice Gauge Theory}

\author{Jean-Yves Desaules${}^{\orcidlink{0000-0002-3749-6375}}$}
\email{jean-yves.desaules@ist.ac.at}
\affiliation{Institute of Science and Technology Austria (ISTA), Am Campus 1, 3400 Klosterneuburg, Austria}
\author{Thomas Iadecola${}^{\orcidlink{0000-0002-5145-6441}}$}
\email{iadecola@iastate.edu}
\affiliation{Department of Physics and Astronomy, Iowa State University, Ames, IA 50011, USA}
\affiliation{Ames National Laboratory, Ames, IA 50011, USA}
\author{Jad C.~Halimeh${}^{\orcidlink{0000-0002-0659-7990}}$}
\email{jad.halimeh@physik.lmu.de}
\affiliation{Department of Physics and Arnold Sommerfeld Center for Theoretical Physics (ASC), Ludwig-Maximilians-Universit\"at M\"unchen, Theresienstra\ss e 37, D-80333 M\"unchen, Germany}
\affiliation{Munich Center for Quantum Science and Technology (MCQST), Schellingstra\ss e 4, D-80799 M\"unchen, Germany}
\affiliation{Dahlem Center for Complex Quantum Systems, Freie Universit\"at Berlin, 14195 Berlin, Germany}

\begin{abstract}
Confinement is a prominent phenomenon in condensed matter and high-energy physics that has recently become the focus of quantum-simulation experiments of lattice gauge theories (LGTs). As such, a theoretical understanding of the effect of confinement on LGT dynamics is not only of fundamental importance, but can lend itself to upcoming experiments. Here, we show how confinement in a $\mathbb{Z}_2$ LGT can be \textit{locally} avoided by proximity to a resonance between the fermion mass and the electric field strength. Furthermore, we show that this local deconfinement can become global for certain initial conditions, where information transport occurs over the entire chain. In addition, we show how this can lead to strong quantum many-body scarring starting in different initial states. Our findings provide deeper insights into the nature of confinement in $\mathbb{Z}_2$ LGTs and can be tested on current and near-term quantum devices.
\end{abstract}

\date{\today} 
\maketitle

\section{Introduction and model}
Lattice gauge theories (LGTs) were originally devised to probe the physics of quark confinement \cite{Wilson1974confinement}, but have since proven to be a useful tool in probing various phenomena in high-energy physics (HEP) \cite{Rothe_book}. With the advent of quantum simulators of HEP \cite{Dalmonte_review,Pasquans_review,Zohar_review,Alexeev_review,aidelsburger2021cold,zohar2021quantum,klco2021standard,Bauer_review,dimeglio2023quantum,halimeh2023coldatom,cheng2024emergent}, it is now possible to directly observe a plethora of HEP features in the laboratory \cite{Martinez2016,Klco2018,Goerg2019,Schweizer2019,Mil2020,Yang2020,Wang2021,Zhou2022,Mildenberger2022,Zhang2023observation,farrell2023scalable,angelides2023firstorder,charles2023simulating}. Confinement in LGTs has recently spurred great interest in the quantum-simulation community, with theoretical proposals \cite{Halimeh2022tuning,Cheng2022tunable} that have led to direct experimental realization \cite{Zhang2023observation}.

Another prominent phenomenon seemingly intimately connected to LGTs is quantum many-body scarring (QMBS) \cite{Bernien2017,Turner2018,Moudgalya2018,Zhao2020,Jepsen2021,Serbyn2020,Moudgalya_review,Chandran_review}. Of great interest to studies of ergodicity breaking in interacting models, QMBS relies on the existence of special nonthermal eigenstates equally spaced in energy across the entire spectrum, which is otherwise ergodic~\cite{BernevigEnt,Turner2018,Schecter2019}. These \textit{scarred} eigenstates have anomalously low bipartite entanglement entropy~\cite{BernevigEnt,lin2018exact}. Initializing the system in a state having large overlap with these scarred eigenstates leads to long-lived oscillations in the dynamics of local observables, a slow growth in the entanglement entropy, and persistent revivals in the wave function fidelity \cite{Bernien2017,Serbyn2020}. QMBS has been the subject of various recent experiments \cite{Bernien2017,Bluvstein2021,Bluvstein2022quantum,Su2022,Zhang2023Many-body,Dong2023Disorder}, and its connection to LGTs has been established in various models~\cite{Surace2020,Iadecola2020quantum,Banerjee2021,Halimeh2022robust,Hudomal2022,Desaules2022weak,Desaules2022prominent,aramthottil2022scar,biswas2022scars,Daniel2023,ebner2024entanglement,Sau2024,osborne2024quantum,budde2024quantum}.

$\mathbb{Z}_2$ LGTs have been the subject of substantial interest when it comes to confinement \cite{Borla2019,kebric2021confinement,Kebric2023njp}, including in higher spatial dimensions \cite{Homeier2023realistic,linsel2024percolation} and at finite temperature \cite{Kebric2023confinement,Fromm2023simulating}, and also when it comes to QMBS \cite{Iadecola2020quantum,aramthottil2022scar,Gustafson2023}. 

The model we consider in this work is the $1+1$D $\mathbb{Z}_2$ LGT with Hamiltonian \cite{Borla2019}
\begin{equation}
\label{eq:H-LGT}
\hat{H}{=}\sum_j\left[J\big(\cd_j\c_{j+1}{+}\text{H.c.}\big)\hat{\tau}^z_{j,j+1}
{+}h\hat{\tau}^x_{j,j+1}+\mu(-1)^j\n_j\right],     
\end{equation}
where $\cd_j/\c_j$ creates/annihilates a fermion on site $j$, $\n_j=\cd_j\c_j$ is the fermion density, $J$ is the fermionic hopping strength, $\mu$ is the fermionic mass, and $h$ is the electric field strength. 
The Pauli operators $\hat{\tau}^z_{j,j+1}$ and $\hat{\tau}^x_{j,j+1}$ represent the local gauge and electric fields at the link between the two neighboring sites $j$ and $j+1$. The generator of the $\mathbb{Z}_2$ gauge symmetry is
\begin{align}
    \hat{G}_j=-(-1)^{\n_j}\hat{\tau}^x_{j-1,j}\hat{\tau}^x_{j,j+1},
\end{align}
which has eigenvalues $g_j{=}{\pm}1$ called \textit{background charges}.
Note that the fermion mass term takes the form of a staggered chemical potential with opposite signs on even and odd sites.
This is the standard fermion mass term in the Kogut-Susskind formulation of LGT~\cite{Kogut1975}, where the two sublattices with positive/negative chemical potential map onto the particle-/hole-like degrees of freedom of a conventional Dirac spinor in the continuum limit.

In 1D and within the gauge-invariant sector $\{\ket{\Psi}\}$ satisfying $\hat{G}_j\ket{\Psi}=\ket{\Psi},\,\forall j$,  Eq.~\eqref{eq:H-LGT} can be expressed as a spin-1/2 Ising Hamiltonian~\cite{Borla2019}. The $\mathbb{Z}_2$ gauge variables are $\hat{\tau}^z_{j,j+1}\equiv \X_j$ and $\hat{\tau}^x_{j,j+1}\equiv \Z_j$, while the gauge-invariant fermion density is $\n_j\equiv (1-\Z_j\Z_{j+1})/2$, where $\X$, $\Y$ and $\Z$ denote the Pauli operators. Thus, fermions are not explicitly represented but correspond to domain walls between up and down spins, and fermion number conservation is manifested as conservation of the domain wall number $\hat{N}_\mathrm{DW}{=}\sum_j\n_j$. For clarity, we will use $\d$ and $\u$ when describing the state of the Ising spins, and $\circ$ (empty) and $\bullet$ (filled) when describing the state of the fermions of the original LGT. 

Using this mapping, we get the spin-$1/2$ Hamiltonian
\begin{equation}\label{eq:H-ZXZ}
\hspace*{-0.19cm}\hat{H}{=}\hspace{-0.08cm}\sum_{j=1}^L\hspace{-0.08cm}\left[\frac{J}{2}\Big(\X_j{-}\Z_{j{-}1}\X_{j}\Z_{j{+}1}\Big){+}h \Z_j{-}\frac{\mu}{2}({-}1)^j\Z_j\Z_{j{+}1}\right].\hspace{-0.12cm}
\end{equation}
For simplicity, we will focus on this formulation and use periodic boundary conditions (PBC) in the rest of this work. Without loss of generality, we also set $J{=}1$ and express $\mu$ and $h$ in units of $J$. In Appendix~\ref{app:syms}, we review the symmetries and properties of this model. We note that the dynamical term $\X_j{-}\Z_{j-1}\X_{j}\Z_{j+1}$ can also be written as $2\Pu_{j-1} \X_j \Pd_{j+1}{+}2\Pd_{j-1} \X_j \Pu_{j+1}$ with $\Pd_j{=}(1{-}\Z_j)/2{=}1{-}\Pu_j$. This means that a spin can be flipped only if its two nearest neighbors are in different states. Such flips cannot create new Ising domain walls, only move them. This makes the conservation of the number of domain walls $\hat{N}_\mathrm{DW}$ more apparent.   

The massless case $\mu=0$ of Eqs.~\eqref{eq:H-LGT}, \eqref{eq:H-ZXZ} was considered in Refs.~\cite{Borla2019,Yang2020fragmentation}, where it was shown that, in the limit $h{\gg}J$, the system becomes strongly confined such that the only allowed motion is the hopping of mesons (particle-antiparticle pairs). 
In fact, this process only happens at second order and so is suppressed by a factor of $J^2/h$; as such, in the limit $h\to \infty$ the system is completely frozen.
The addition of a fermion mass $\mu$, while physically well motivated, may appear inconsequential at first, and indeed the physics above persists in the limit $h{\gg}\mu{,}J$.

However, in this work, we show that adding a large mass $\mu\approx h$ counterintuitively leads to deconfinement. While this deconfinement is generally only local, from certain initial states information can propagate throughout the chain, exhibiting \textit{global} deconfinement. We also show that the effective dynamics at $\mu=h$ leads to perfect revivals from several states. Importantly, due to the special interplay of higher-order perturbative corrections, this leads to QMBS in the full model already for $\mu=h\approx J$.

\section{Mass Resonance and deconfinement}
To understand the effective dynamics at the resonance, we set $h{=}\mu$ in Eq.~\eqref{eq:H-ZXZ} and split it as $\hat H{=}\mu(\Hn {+} \lambda \hat V)$ with 
\begin{align}
    \Hn &=\sum_j \left(\Z_j-\frac{(-1)^j}{2}\Z_j\Z_{j+1}\right), \\ 
    \hat V &=\frac{1}{2}\sum_j\left(\X_j-\Z_{j-1}\X_j\Z_{j+1}\right),
\end{align}
 and $\lambda = J/\mu$. Taking the limit $\mu\to\infty$, we perform a Schrieffer-Wolff (SW)~\cite{Bravyi2011} transformation to obtain an effective Hamiltonian with an emergent conservation of $\Hn$. The leading-order contribution is $\mathcal{O}(\lambda)$ and obtained by projecting $\hat V$ into a degenerate eigenspace of $\hat H_0$:
\begin{equation}\label{eq:Heff}
        \Heff =J\sum_{j\text{ odd}}\Pu_{j-1}\X_j\Pd_{j+1}+J\sum_{j\text{ even}}\Pd_{j-1}\X_j\Pu_{j+1}.
\end{equation}
This Hamiltonian can be understood as only keeping processes where the change of energy due to $h$ is exactly compensated by that due to the mass term $\mu$.
It is worth noting that $\Heff$ can be turned into a translation-invariant Hamiltonian using the unitary transformation $\prod_{j \ {\rm 
 even}} \hat{X}_j$. We then have simply $ \Heff =\sum_{j}\Pd_{j-1}\X_j\Pu_{j+1}$. This is reminiscent of the rule-156 classical cellular automaton~\cite{Inokuchi156}. This kinetically constrained model was recently also considered in Refs.~\cite{Valencia2024Rydberg,Maity2024kinetically}, where a few of its properties were discussed. We will address these in the language of our spin model and in the context of LGTs, as well as show additional ones.

By construction, $\Heff$ conserves $\Hn$ and the number of domain walls. However, it exhibits further Hilbert space fragmentation~\cite{Sala20,Khemani20} within each of these symmetry sectors. This can be understood by the emergence of an extensive number of local conserved quantities $\hat{O}_j$ defined as \begin{equation}\label{eq:Oj}
    \hat{O}_j=\begin{cases}
      \Pu_j\Pu_{j+1}, & \text{for}\ j \ \text{odd} \\
      \Pd_j\Pd_{j+1}, & \text{for}\ j \ \text{even}.
    \end{cases}
\end{equation}
To prove that these operators commute with $\Heff$, let us us first focus on an odd site $j$ for which $\hat{O}_j=\Pu_j\Pu_{j+1}$. Then the two operators in $\Heff$ that can change the value of spins $j$ or $j+1$ are $\Pu_{j-1}\hat{X}_j\Pd_{j+1}$ and $\Pd_{j}\hat{X}_{j+1}\Pu_{j+2}$. When computing the product of the former  with $\hat{O}_j$, we find that it contains the term $\Pd_{j+1}\Pu_{j+1}=0$. The same annihilation happens for the product of $\hat{O}_j$ and $\Pd_{j}\hat{X}_{j+1}\Pu_{j+2}$ which contains $\Pu_{j}\Pd_{j}=0$. An analogous calculation can be done with even sites for which $\hat{O}_j=\Pd_j\Pd_{j+1}$.
As a result, $\Heff \hat{O}_j=0$ for all $j$. The Hermiticity of both operators implies $ \hat{O}_j\Heff=0$ and thus $\left[\Heff,\hat{O}_j\right]=0$.

As each $\hat{O}_j$ is a product of disjoint -- and so commuting -- single-site projectors, these operators are themselves projectors and have eigenvalues 0 and 1. The $\hat{O}_j$ are also diagonal in the computational basis, and for each basis state $\ket{n}$ we can define
$O_j=\langle n | \hat{O}_j|n \rangle \in \{0,1\}$. Almost all disconnected sectors can be uniquely identified by grouping together basis states with the same values of $O_j$ for all $j$. The only exception are the two N\'eel states $\ket{\cdots \uparrow\downarrow \uparrow\downarrow \cdots}$ which are both frozen -- and so in their own sector -- despite having the same values $O_j=0$ for all $j$. 

This classification according to the $O_j$ also allows us to compute the number of fragments in the Hilbert space. It is straightforward to see that if $O_j$ is $1$ then sites $j$ and $j+1$ must be up (for $j$ odd) and so both $O_{j-1}$ and $O_{j+1}$ must be $0$ as they contain $\Pd_j$ and $\Pd_{j+1}$ respectively.
Thus the number of possible configurations of the $O_j$ is equivalent to the number of states in the PXP model~\cite{Lesanovsky2012} with $L$ sites. In order to translate that to the number of fragments one needs to add one to this quantity to account for the two N\'eel states being disconnected despite having the same $O_j$. Using the analytical formula from Ref.~\cite{Turner2018} we arrive at a total of $f_{L-1}+f_{L+1}+1$ sectors, where $f_{n}$ is the $n$-th Fibonacci number.

\begin{figure}[t]
	\centering
	\includegraphics[width=0.9\linewidth]{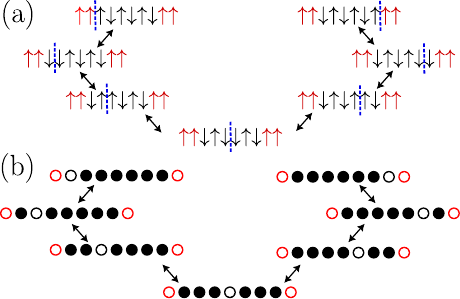}
\caption{(a) Example of the possible configurations in a part of the chain between two frozen domains. Frozen sites are highlighted in red, while the blue dashed line indicates the boundary between two inequivalent N\'eel domains. This boundary can move, and the effective dynamics in that part of the chain is that of a tight-binding chain with a single particle. (b) Matter configurations corresponding to the gauge configurations in (a), making apparent the nature of the particle hopping in the tight-binding chain. It corresponds to a hole on top of a fully packed background of matter.}
	\label{fig:sector_states}
\end{figure}

Beyond this fragmentation, one can prove that $\Heff$ is in fact integrable. Dynamics only arise when a few consecutive $O_j=0$. This is because $O_j=1$ (for $j$ odd) implies $\u\u$ while $O_j=0$ allows for some freedom as it is satisfied by $\u\d$, $\d\u$, or $\d \d$ (and similarly for $j$ even). If we consider $O_k=O_{k+r}=1$ but $O_j=0$ for $k<j<k+r$, then we have $r-2$ sites that can change. These states are not free, as they must respect $O_j=0$. So between the two frozen blocks, the state must locally resemble a N\'eel state with no $\u\u$ or $\d \d$. However, as we can still have $\d \d$ on odd-even sites and $\u \u$ on even-odd sites, we can have two different N\'eel domains and move the boundary between them. As such, we have an effective tight-binding chain with hopping rate $J$ and with $r-1$ sites, which can be mapped to free fermions, making it integrable.
As the full dynamics is simply composed of all disconnected parts evolving on their own, this means that $\Heff$ itself is integrable. 
This picture also allows us to compute the number of states in each $O_j$ sector by simply multiplying the number of states in each tight-binding chain.

We now look at this tight-binding chain from the vantage point of the LGT. Between two frozen blocks, there is a single mobile quasiparticle corresponding to the absence of a domain wall. In LGT language, this quasiparticle is simply a hole in a fully packed matter background. While this is not very intuitive, we can provide an illustrative example. Let us consider the state $\ket{\uparrow \uparrow \uparrow \downarrow\uparrow \downarrow\uparrow\downarrow\uparrow\uparrow \cdots}$. We can check that $O_1=O_9=1$ and that all the $O_j$ in between are equal to 0. All possible configurations respecting these constraints are shown in Fig.~\ref{fig:sector_states}(a), and it is straightforward to count that there are $7=8-1$ of them and that they form an effective tight-binding chain. In Fig.~\ref{fig:sector_states}(b), we also show the matter configurations corresponding to the same states. It immediately becomes apparent that the effective particle hopping in the tight-binding chain is a hole on top of a fully packed matter background.

\begin{figure}[t]
	\centering
	\includegraphics[width=\linewidth]{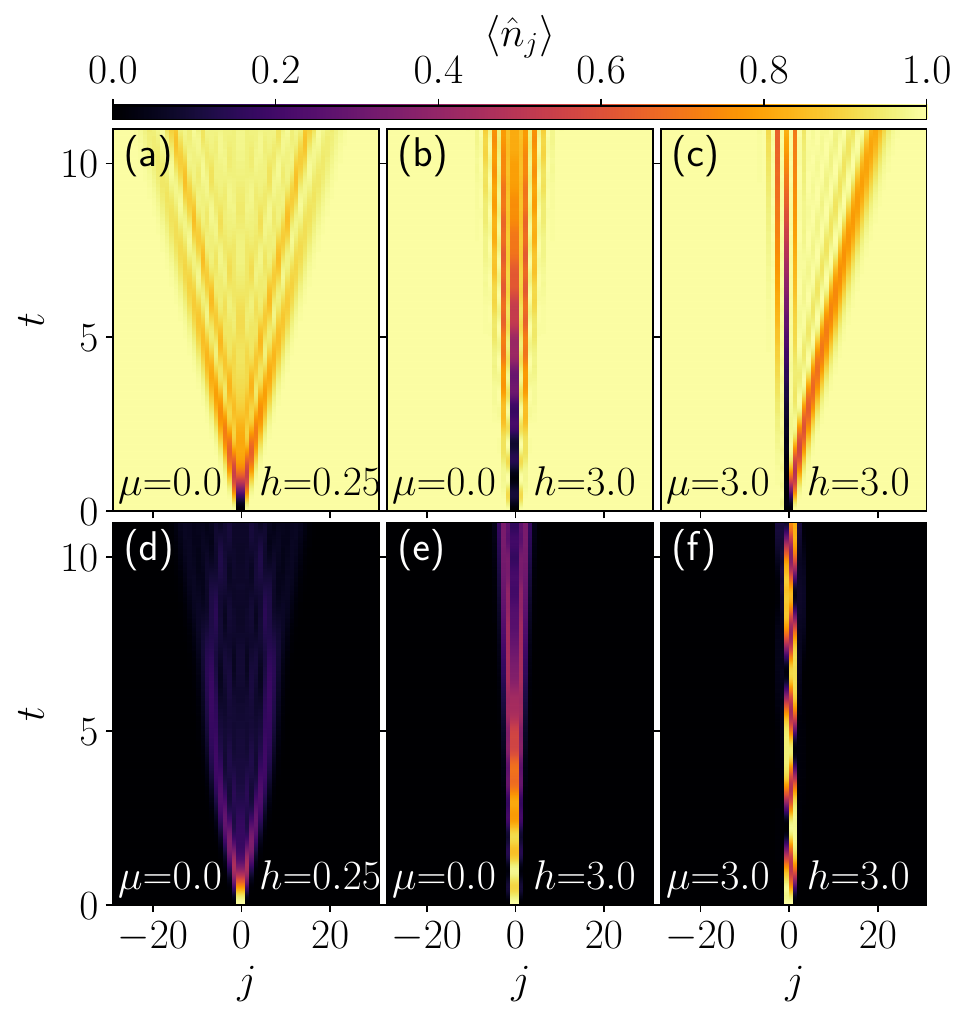}
    \caption{Dynamics of (a)-(c) holes and (d)-(f) particles in various parameter regimes for $L=60$. Panels (a) and (d) are for $h=0.25$ and $\mu=0$, in the ergodic regime. Panels (b) and (e) are for $h=3$ and $\mu=0$, where both holes and particles are confined as expected in the large $h$ limit. Panels (c) and (f) are at the resonance $h=\mu=3$. While particles are still confined, one of the two holes propagates freely.
    The data in this figure and all subsequent ones is obtained using exact diagonalization (ED) with all symmetries explicitly resolved.
	}\label{fig:matter_deconf}
\end{figure}

If we focus on the sector with only $O_1=1$, the hole can travel along the \textit{entirety} of the chain, even for $h=\mu \to \infty$. This is surprising, as in the limit $h\to \infty$ the system is completely frozen due to confinement \cite{Borla2019}. Thus, adding a very large mass actually favors the mobility of holes, despite intuition suggesting the opposite. We show the dynamics of holes and fermions in Fig.~\ref{fig:matter_deconf} for different parameter regimes, illustrating the deconfining effect that mass has on holes. We briefly note that in panel (c), the left hole remains frozen due to being part of an  $O_j=1$ and prevents the right hole from propagating to the left. This is not indicative of chirality of hole propagation in that regime.  

The difference between the mobility of holes and fermions upon the addition of resonant mass can be understood in the following way. When $\mu=0$, hopping a hole in a fully packed matter background changes the energy by an amount $\propto h$. However, for a sequence of hops, the sign of the energy change alternates with each hop. Suppose for example that hopping a hole from an even to an odd fermion site entails flipping a spin from up to down. The next hop will then require a spin flip from down to up, producing an energy shift of the opposite sign. Adding a mass term with $\mu=h$ exactly compensates for this alternating energy change, since the energy cost of having a fermion on an even vs.~an odd site also has an alternating sign. Meanwhile, hopping a fermion in a background of holes always requires the same type of spin flip, i.e.~always from up to down or always from down to up. So the addition of resonant mass can allow at most one consecutive move and only leads to local deconfinement of fermions.

We note that the emergent integrability is relatively robust as it is not destroyed by detuning between $\mu$ and $h$, as long as the detuning is small compared to these two parameters. Indeed, the effect of the detuning on the mobile holes is that of a staggered Z-field, see Appendix~\ref{app:det} for the derivation. This means that it does not kill the mapping to free fermions, and thus preserves the integrabiltiy of the effective model.

We emphasize that this mass-assisted deconfinement is not a generic feature. For example, in the spin-$1/2$ $\mathrm{U}(1)$ quantum link model \cite{Chandrasekharan1997,Wiese_review}, a similar resonance between mass and confining potential arises. While it also locally unfreezes the system, the resulting dynamics is strictly local and information cannot propagate~\cite{Desaules2024ergodicitybreaking}.

\section{QMBS away from integrability}
Thanks to the solvability of the model, it is straightforward to engineer states that exhibit periodic dynamics in the limit of $h{=}\mu {\to}\infty$. For that, we restrict to sectors where all eigenvalues are equally spaced. This implies that they need to be equally spaced in each disconnected part of the chain. However, for a tight-binding chain with equal couplings the eigenvalues are only equidistant for $1$, $2$, or $3$ sites~\cite{Kay2010perfect}. For a chain with $1$ site, everything is trivial as there is no dynamics and we avoid such cases. As the hopping strength is $J$, in the two-site case the eigenvalues are $\pm J$ while in the three-site case they are $\pm \sqrt{2}J$ and $0$. The energy spacings are not commensurate between these two cases, so we expect periodic dynamics in the full system only if we have a single type of chains. In that case, we can map every frozen domain with two or three states to a spin-$1/2$ and a spin-$1$, respectively. We can then identify three simple initial states that correspond to charge density waves (CDWs) for the matter sites. These states are $\ket{\Psi_2}{=}\ket{\rd\! \u\u\!\rd\rd\!\u\u\!\rd\!\cdots}{=}\ket{\bullet \circ\bullet\circ\cdots }$, $\ket{\Psi_3}{=}\ket{\rd\! \d\! \ru \ru \!\u\! \rd \rd\! \d\! \ru \ru\! \u\! \rd \cdots}{=}\ket{ \circ \bullet \circ \circ \bullet \circ  \cdots }$, and  $\ket{\Psi_4}{=}\ket{\rd\!\d\u\!\rd\rd\!\d\u\!\rd\!\cdots}{=}\ket{\circ \bullet \bullet \circ \circ \bullet \bullet \circ \cdots}$, where red denotes that a site is frozen. The state $\ket{\Psi_3}$ lives in a sector with $N/3$ effective spins-$1/2$, as each unfrozen site can freely flip between $0$ and $1$. Meanwhile, both $\ket{\Psi_2}$ and $\ket{\Psi_4}$ live in the same sector with $N/4$ effective spins-$1$. In each unfrozen part of the chain, only the three states $\ket{\u \d}{\equiv} \ket{-1}$, $\ket{\u \u}{\equiv}\ket{0}$ and $\ket{\d \u}{\equiv}\ket{+1}$ are allowed. Thus $\ket{\Psi_2}$ is akin to $\ket{0,0,0,\ldots}$ in the spin-$1$ language while $\ket{\Psi_4}$ is akin to $\ket{+1,+1,+1,\ldots}$. This means that $\ket{\Psi_2}$ will lead to state transfer to the entangled state where each spin-$1$ is in a superposition of $\ket{+1}$ and $\ket{-1}$, meaning that each cell is $(\ket{\rd\!\u\d\!\rd}{+}\ket{\rd\!\d\u\!\rd}){/}\sqrt{2}$. This state will also be important in the dynamics and we denote it by $\ket{\Phi}$.

\begin{figure}[t!]
	\centering
	\includegraphics[width=\linewidth]{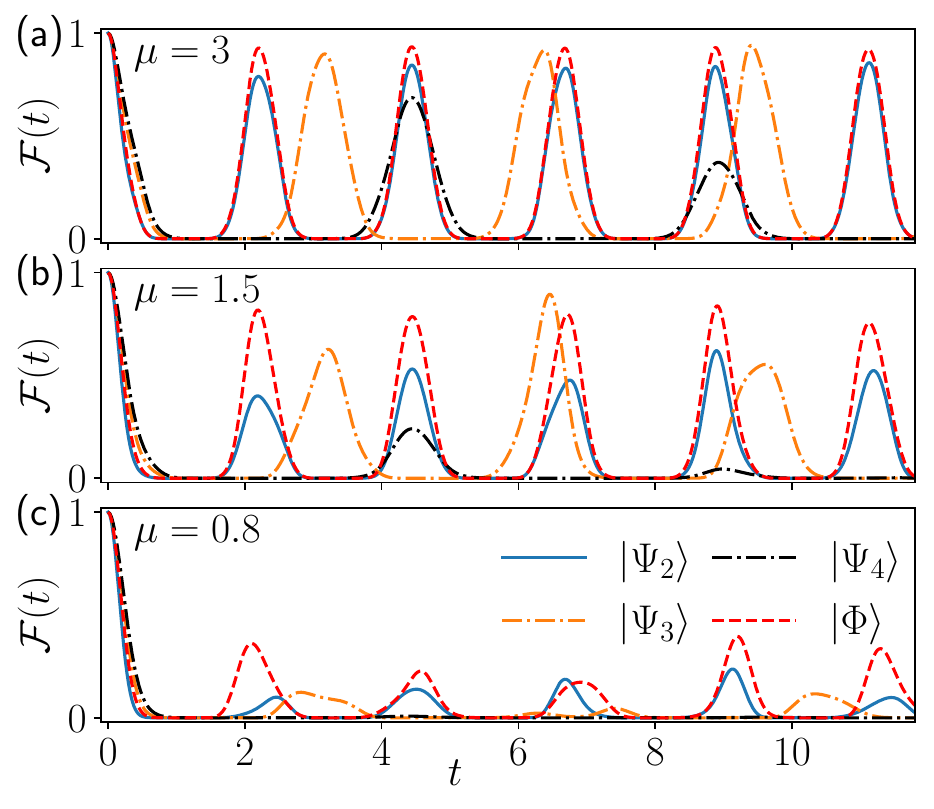}
    \caption{Quenches from different initial states for (a) $\mu=h=3$, (b) $\mu=h=1.5$ and (c) $\mu=h=0.8$. The system size used is $L=20$ for all states except for $\ket{\Psi_3}$ where we use $L=18$ instead. While all states show good revivals for large values of $\mu=h$, this is no longer true when these two parameters are close to 1. 
	}\label{fig:CDW_revs}
\end{figure}

\begin{figure}[tb]
	\centering
	\includegraphics[width=\linewidth]{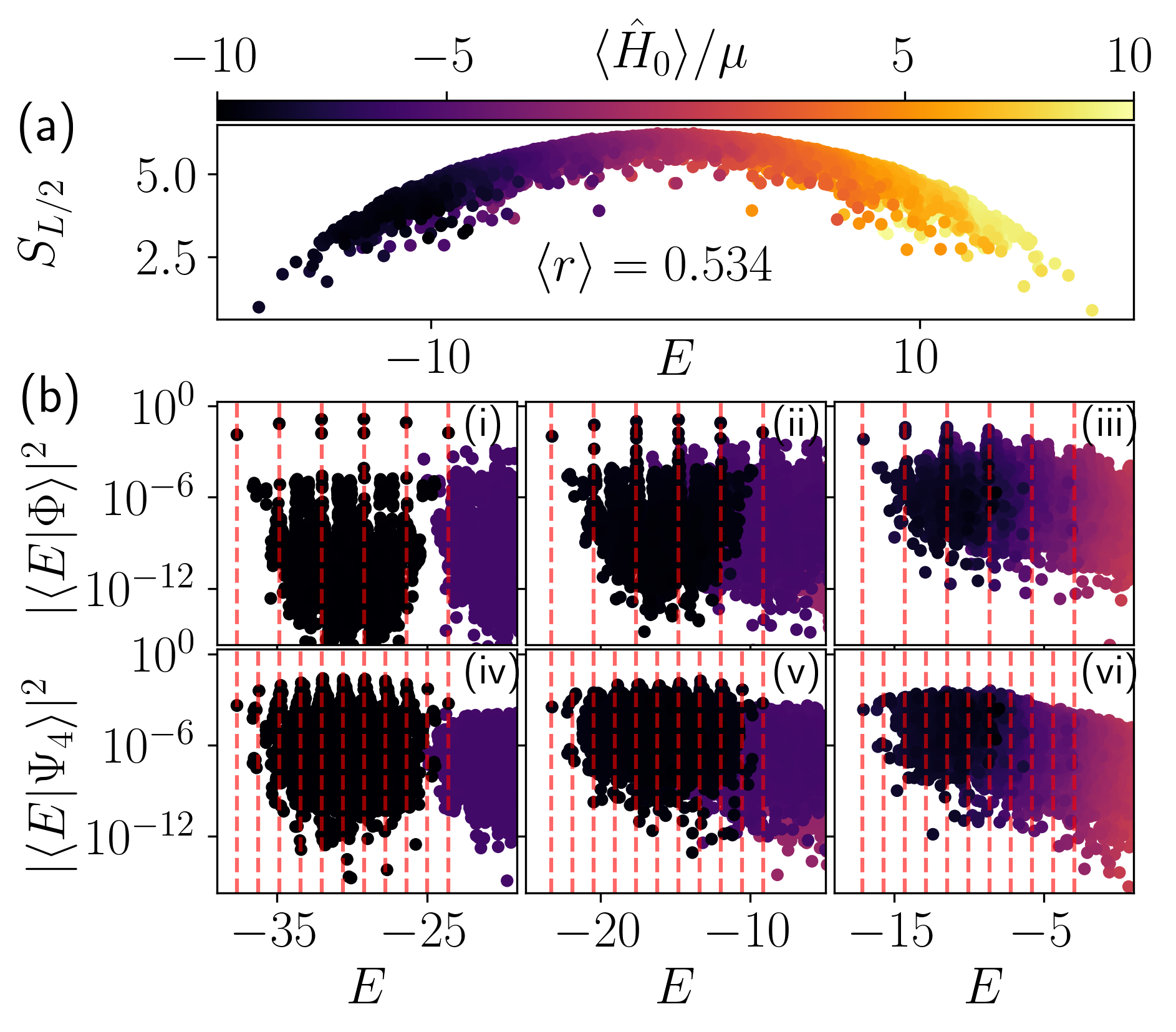}
    \caption{Properties of eigenstates for $L=20$ and $10$ domain-walls. The color indicates the expectation value of $\Hn/\mu$. (a) Bipartite entanglement entropy for $\mu=h=0.8$ in the fully symmetric sector. The entanglement entropy profile and the mean energy-level spacing~\cite{Oganesyan07,Atas13} are typical of chaotic systems 
    (b) Overlap with (i)-(iii) $\ket{\Phi}$ and (iv)-(vi) $\ket{\Psi_4}$. The parameters are (i), (iv) $\mu=h=3.0$, (ii), (v) $\mu=h=1.5$ and (iii), (vi) $\mu=h=0.8$. The red dashed lines denote the exact energies at which the scarred eigenstates are expected according to the spin-$1$ picture. While the scarred towers of state persist in panel (iii) for $\ket{\Phi}$, this is not the case for $\ket{\Psi_4}$ in panel (vi).
	}\label{fig:olap_phi}
\end{figure}

\begin{figure}[tb]
	\centering
	\includegraphics[width=\linewidth]{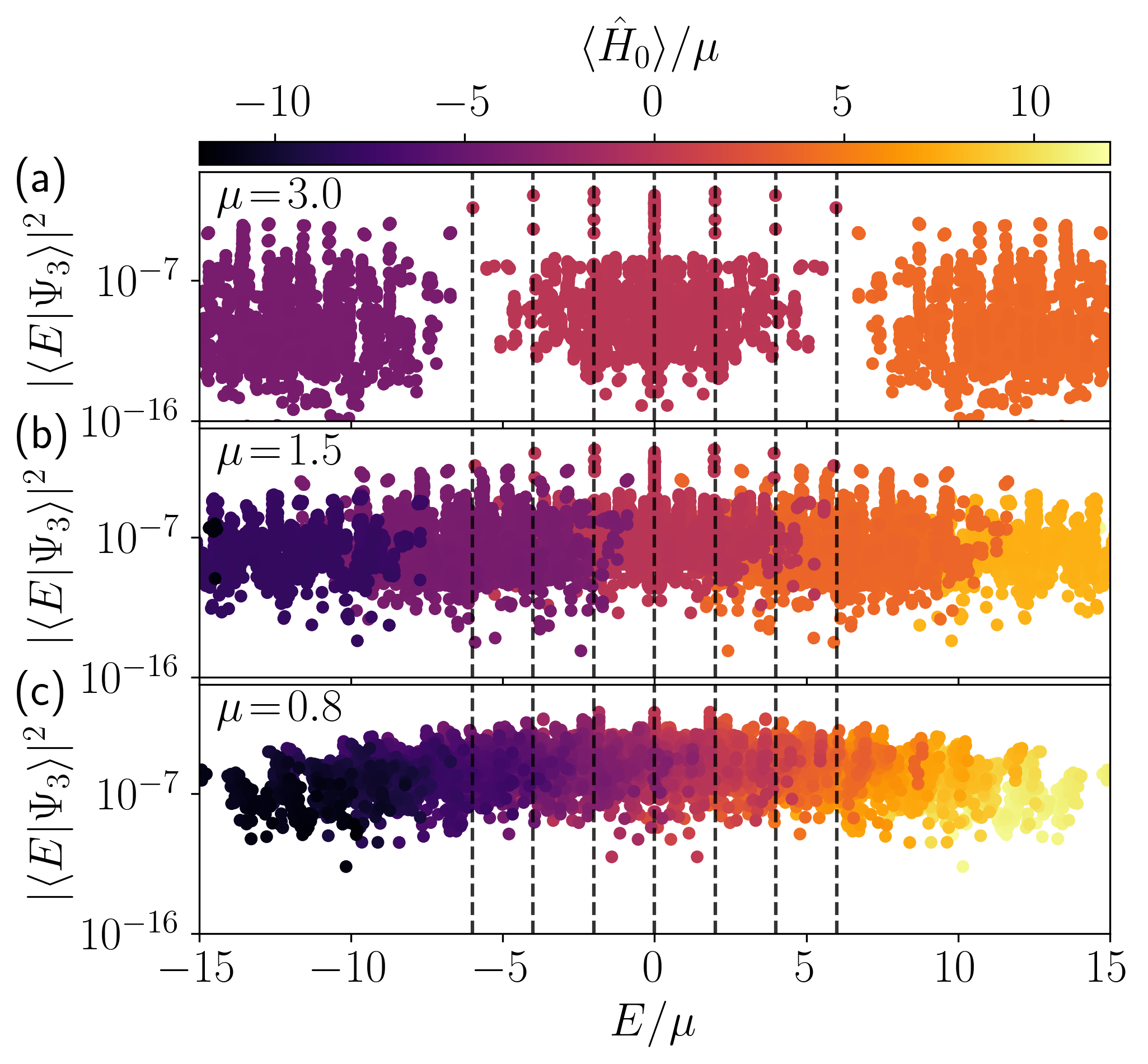}
\caption{Overlap of $\ket{\Psi_3}$ with the eigenstates of the Hamiltonian for $L=18$ and $\mu=h=0.8$, $1.5$ and $3$. The color indicates the expectation value of $\Hn/\mu$ with respect to each eigenstate. The black dashed lines denote the exact energies at which the scarred eigenstates are expected according to the spin-$1/2$ picture. As $\mu=h$ is decreased, the various $\Hn$ sectors rapidly merge together.}\label{fig:olap_psi3}
\end{figure}

Figure~\ref{fig:CDW_revs} shows the fidelity $\mathcal{F}(t)=|\braket{\psi(0)|\psi(t)}|^2$ after a quench from these four states at different values of $\mu=h$. For values as small as $\mu=h=3$, clear oscillations stemming from the integrable limit are visible for all initial states considered. As we further lower $\mu$ and $h$, we expect the oscillations to disappear as ergodicity sets in for $\mu=h\approx 1$. However, this is not what we observe. Instead, while $\ket{\Psi_4}$ thermalizes as expected, $\ket{\Phi}$ and $\ket{\Psi_2}$ show oscillations even at $\mu=h=0.8$. This is surprising, as for these values of the parameters the system is not separated into sectors of $\Hn$ and is chaotic as demonstrated by the entanglement entropy of eigenstates and mean energy spacing in Fig.~\ref{fig:olap_phi}(a).
Nonetheless, Fig.~\ref{fig:olap_phi} (b) demonstrates that the overlap of these eigenstates with $\ket{\Phi}$ still shows regular structures consistent with the spin-$1$ mapping valid for large $\mu=h$. This becomes clearer as $\mu=h$ is increased to $1.5$, along with the different sectors starting to separate. Meanwhile, for the $\ket{\Psi_4}$ state while these structures in the overlap are visible for larger values of $\mu=h$, they disappear as these parameters get smaller.

We briefly mention that the addition of the mass term does not destroy the exact QMBS that were found in Ref.~\cite{Iadecola2020quantum}, see Appendix~\ref{app:QMBS} for more details.  
But while the model in Eq.~(\ref{eq:H-ZXZ}) still has these scars, they cannot be responsible for the non-ergodicity we witness here. Indeed, as we are restricting to a single sector of $\hat{N}_\mathrm{DW}$, there is only one of the exact scars present, which means it cannot be causing the oscillatory dynamics we observe. 
We also note that detuning between $\mu$ and $h$ does not destroy the scarring but allows to tune its period, as discussed in more detail in Appendix~\ref{app:det}

Similar to the case of $\ket{\Psi_4}$, the regular structure in the overlap with eigenstates stemming from the spin-$1/2$ picture disappear as $\mu=h$ when considering the $\ket{\Psi_3}$ state. This is shown in Fig.~\ref{fig:olap_psi3} and stems from the various sectors of $\Hn$ merging with each other. This is the generic behavior expected and explains why revivals from the $\ket{\Psi_3}$ taper off as the parameters are taken towards the ergodic regime.
One of the reasons why this does not happen for the $\Hn$ sector of $\ket{\Phi}$ is relatively simple. While the $\Hn$ sector of $\ket{\Psi_3}$ is centered around the average $\Hn$ value, the sector of the other states is at the lowest $\Hn$ value. This implies that there is no other sector at a lower energy to merge with, protecting the spin-1 structure in part of the sector. 
However, the same should be true for $\ket{\Psi_4}$, so this does not elucidate the  reason why $\ket{\Phi}$ and $\ket{\Psi_4}$ behave so unalike as $\mu=h$ is lowered.

\begin{figure}[tb]
	\centering
	\includegraphics[width=\linewidth]{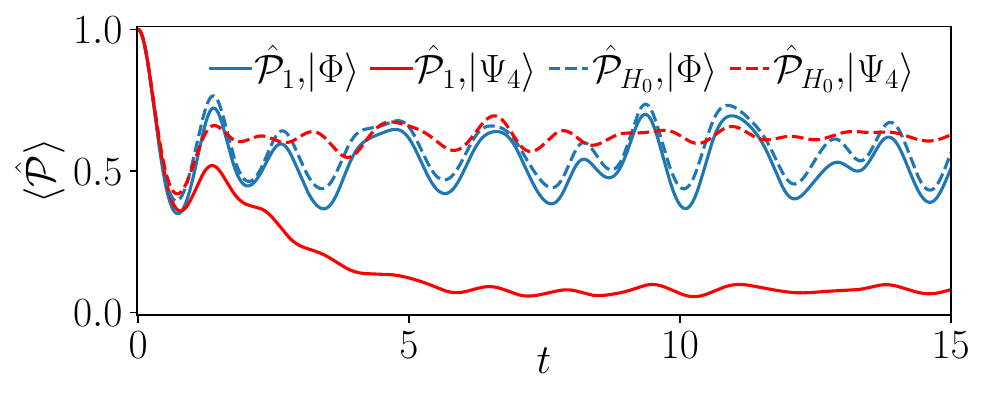}
    \caption{ Expectation value of projectors after quenches for $L=16$ and $\mu=h=1$. For both $\ket{\Psi_4}$ and $\ket{\Phi}$, the amount of the wave function remaining in the ``correct'' sector of $\Hn$ is similar, but this is not at all true for the spin-$1$ sector. This means that the main difference is to be found in the effective dynamics in the $\Hn$ sector.  
	}\label{fig:comp_proj}
\end{figure}

Indeed, since both states live in the same $\Hn$ sector (and even in the same spin-1 subsector), we expect the influence of other $\Hn$ sectors to be similar. While this is already visible in Fig.~\ref{fig:olap_phi}, we can test it during the dynamics. We define the operator $\hat{\mathcal{P}}_{H_0}$, which projects onto all states with the same expectation value of $\Hn$ as $\ket{\Phi}$ (or, correspondingly, as $\ket{\Psi_2}$ or $\ket{\Psi_4}$). We also define $\hat{\mathcal{P}}_1$ as the projector onto the spin-$1$ sector in which these states live . We recall that the relevant spin-1 sector is a subset of the states with the same expectation value of $\Hn$ and that it corresponds to the set of states connected to the initial states of interest by $\Heff$ in Eq.~(\ref{eq:Heff}).
Figure~\ref{fig:comp_proj} clearly shows that $\hat{\mathcal{P}}_{H_0}$ is similar for both $\ket{\Phi}$ and $\ket{\Psi_4}$, but that $\hat{\mathcal{P}}_1$ is not. This implies that the difference between these states does \emph{not} result from connection to other sectors of $\Hn$, but rather stems from the intra-sector dynamics generated by higher-order terms of the Schrieffer-Wolff transformation.

\subsection{Beyond first order}
While $\Heff$ provides an exact description of the dynamics for infinite $\mu=h$, higher-order terms in the SW transformation are needed for finite values of these two parameters. The term at second order reads
\begin{equation}\label{eq:SW2}
    \begin{aligned}
        H_\mathrm{eff}^{(2)}&=\frac{J^2}{4\mu}\Big[\sum_{j \ {\rm even}} \Pu_{j-1}(\hat{\sigma}^+_{j}\hat{\sigma}^-_{j+1}+\hat{\sigma}^-_{j}\hat{\sigma}^+_{j+1})\Pu_{j+2}\\
        &{-}\sum_{j \ {\rm odd}} \Pd_{j-1}(\hat{\sigma}^+_{j}\hat{\sigma}^-_{j+1}+\hat{\sigma}^-_{j}\hat{\sigma}^+_{j+1})\Pd_{j+2}\\
&{+}\sum_{j \ {\rm even}} \Pu_{j-1}\hat{Z}_{j}\Pd_{j+1}{+}\sum_{j \ {\rm odd}} \Pd_{j-1}\hat{Z}_{j}\Pu_{j+1}\Big].
    \end{aligned}
\end{equation}
While the third-order term is nonzero (see Appendix~\ref{app:SW3}), its prefactor is $J^3/(32\mu^2)$; thus, even if $J/\mu{\approx} 1$, it is already highly suppressed. As such, our analysis will concentrate on the second-order term. We define a new Hamiltonian $\hat{H}^\prime=\Heff+\Heff^{(2)}$ which captures the first two leading orders.

$\Heff^{(2)}$ does not commute with the $\hat{O}_j$ operators. It can for example take the state $\ket{\rd\! \d \u \! \rd \rd\rd}$ to $\ket{\rd \rd \rd\! \u \d\!\rd}$, thus changing the value of $\hat{O}_2=\Pd_2\Pd_3$ and $\hat{O}_4=\Pd_4\Pd_5$. So we expect the resulting model to be fully ergodic, but this depends on the filling factor. 

Indeed, we find that $\Heff^{(2)}$ has surprisingly no effect on $\ket{\Psi_3}$. Due to the frozen blocks alternating between $\u \u $ and $\d \d$, the dynamical terms are always annihilated and $\Heff^{(2)}$ does not create leakage out of the spin-$1/2$ subspace. So the Hilbert space \emph{still} admits three isolated spin-$1/2$ sectors of dimension $2^{L/3}$ which contain $\ket{\Psi_3}$ and its translated counterparts. In addition, while the diagonal term of $\Heff^{(2)}$ is not constant in the spin-$1/2$ subspace, it is always equal to zero along the dynamical trajectory of $\ket{\Psi_3}$~\footnote{This is because unfrozen spins on odd sites are always in the opposite state as unfrozen spins on even sites. When an even site $j$ is $\d$, this allows the term $\Pd_j\Z_{j+1}\Pu_{j+2}$ to be $+1$. But at the same time, the odd site $k=j+3$ is $\u$, allowing $\Pu_k\Z_{k+1}\Pd_{k+2}$ to be $-1$. The same can be seen when $j$ is $\u$ and $k$ is $\d$, but with the terms  $\Pd_{j-1}\Z_{j-1}\Pu_{j}$ and $\Pu_{k-2}\Z_{k-1}\Pd_{k}$.}. 
Thus $\Heff^{(2)}$ has absolutely no impact on the revivals of the $\ket{\Psi_3}$ state, which explains why the revivals are still visible for values of $\mu=h\approx 1$, despite the influence of the other $\Hn$ sectors.

\begin{figure}[t]
\centering
    \includegraphics[width=\linewidth]{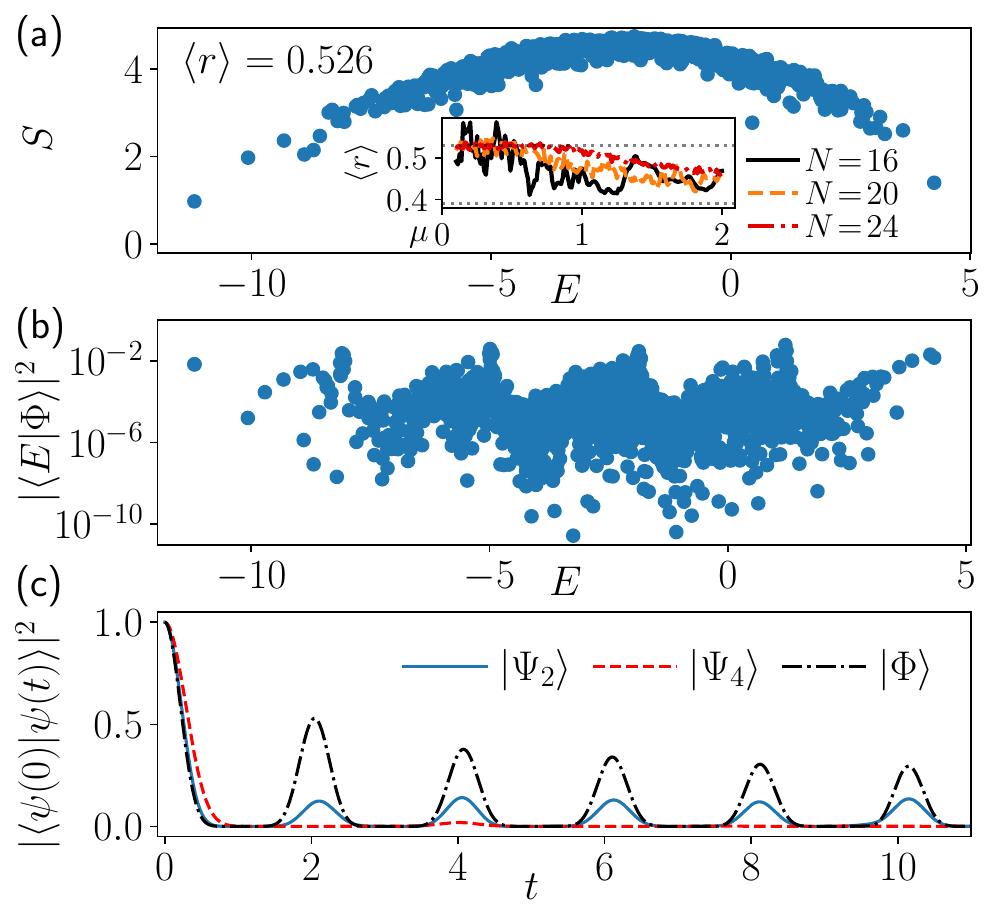}       
\caption{Properties of $\hat{H}^\prime=\Heff+\Heff^{(2)}$ for $\mu=h=0.5$ and $L=20$ in the $(\hat{N}_\mathrm{DW},\Hn)$ sector containing $\ket{\Psi_2}$, $\ket{\Psi_4}$ and $\ket{\Phi}$. The entanglement entropy and mean level spacing value in panel (a) are for the symmetry sector with zero momentum and symmetric under spatial reflection. The inset shows the level spacing value for a wider range of values of $\mu=h$ and for various system sizes. The entanglement entropy of eigenstates is seemingly thermal while the spectrum still holds some nontrivial structure as indicated in the overlap with the $\ket{\Phi}$ state in panel (b). The approximately equal peaks lead to the revivals observed in panel (c).}
\label{fig:all_mu_comp_SW2}
\end{figure}

For the $(\hat{N}_\mathrm{DW},\Hn)$ sector containing $\ket{\Psi_2}$, $\ket{\Psi_4}$ and $\ket{\Phi}$, the picture is radically different. We find that the Hamiltonian is instead fully connected in the computational basis. The level statistics also match Wigner-Dyson predictions for $\mu{>}0.5$, demonstrating that $\Heff^{(2)}$ completely destroys any apparent structure of $\Heff$. This is shown in the inset of Fig.~\ref{fig:all_mu_comp_SW2} (a). We also check that $\hat{H}^\prime$ captures the scarring by checking the overlap with the states of interest and the dynamics. This is shown in Fig.~\ref{fig:all_mu_comp_SW2} (b) and (c) and we find clear signatures of QMBS in both.

\begin{figure}[t!]
	\centering
	\includegraphics[width=\linewidth]{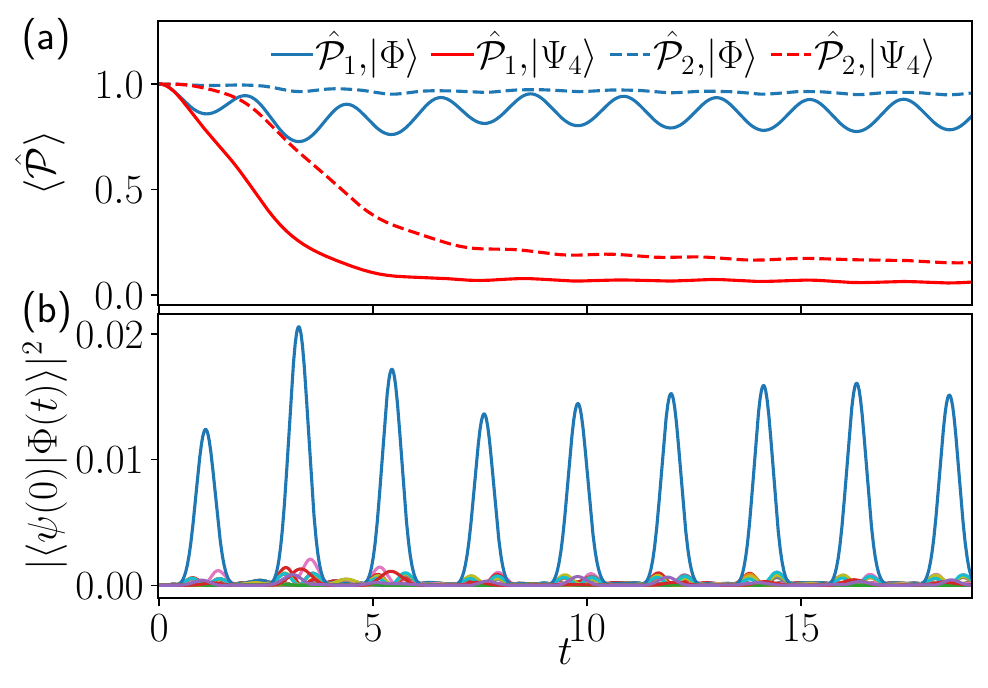}
    \caption{ Dynamics after a quench from $\hat{H}^\prime$ for $L=20$ and $\mu=h=1$. 
    (a) For $\ket{\Phi}$ the wave function escaping the spin-$1$ sector is almost entirely going to the ``neighbor states''. Meanwhile, for $\ket{\Psi_4}$ it spreads into the entire $\Hn$ sector. 
    (b) Overlap of the wave function with all 135 (after resolution of symmetries) neighbor states after a quench from $\ket{\Phi}$. The wave function is concentrated on a single state. We denote the symmetric superposition of that state and its equivalents under symmetries by $\kb$.
	}\label{fig:comp_proj_eff}
\end{figure}

Now that we have shown that $\hat{H}^\prime$ is chaotic, scarred, and behaves similarly to the full Hamiltonian, we can start looking into the origin of the robust scarring of $\ket{\Phi}$.
While the off-diagonal terms of $\Heff^{(2)}$ do not connect any pair of states within the spin-$1$ sector, they lead to leakage out of it. This is reminiscent of QMBS observed in Refs.~\cite{Desaules2021TFH,Zhang2023Many-body}. In these works, there is also an effective spin-$1$ or spin-$1/2$ sector with additional terms creating leakage. In both cases, there are only revivals from the states that do not directly leak out of the regular sector. The dynamics is then state transfer between these two special states. In our model, there is a single state with no leakage caused by $\Heff^{(2)}$, which is $\ket{\Psi_2}$. However, it leads to state transfer to the state $\ket{\Phi}$, which has \textit{maximum} leakage in the spin-$1$ sector. The fact that this does not prevent $\ket{\Psi_2}$ from reviving and that $\ket{\Phi}$ has even better revivals means that this simple picture of leakage is not enough to understand the dynamics.

\begin{figure}[tb]
	\centering
	\includegraphics[width=0.9\linewidth]{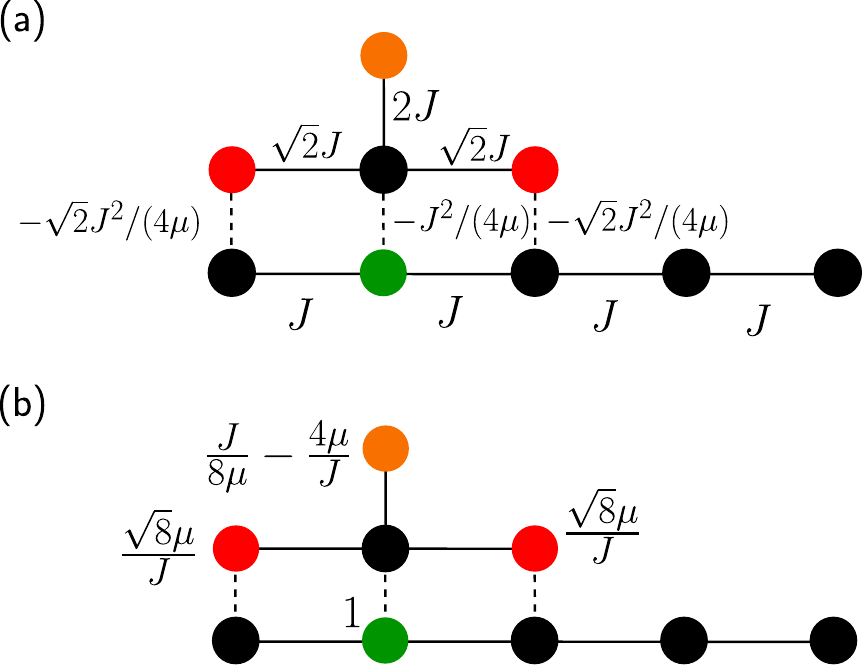}
    \caption{Graph of the Hilbert space (symmetrized under translation and spatial inversion) for $L=8$ in the Krylov subspace generated by $\Heff$ and by a \emph{single} application of $\Heff^{(2)}$. The orange vertex is $\ket{\Psi_2}$, the superposition of the red vertices is $\ket{\Phi}$ and the green vertex $\kb$. Since it is not fully symmetric, the $\ket{\Psi_4}$ state itself does not correspond to a full vertex but has overlap with the rightmost red vertex. Full edges represent transitions allowed by $\Heff$ while dashed edges are transitions enabled by $\Heff^{(2)}$. Transition amplitudes are shown in panel (a). The regular structure of the graph allows atypical eigenstates which are mostly localized in the upper part of the graph. An example of such an exact eigenstate is shown in panel (b), with only four states with nonzero weight (whose unnormalized values are printed next to the corresponding vertices).
	}\label{fig:graph_eff}
\end{figure}

We cannot simply consider the rest of the Hilbert space as a kind of ``reservoir'' from which no information comes back. Instead, we have to look at where $\Heff^{(2)}$ takes it. Importantly, for $\mu{\geq}0.5$, we find that instead of spreading into the full Hilbert space, the wave function remains close to the spin-$1$ subsector. Let us denote by ``neighbor states'' all computational basis states in the Krylov subspaces of $\Heff$ that can be reached from the spin-$1$ subsector with a \textit{single} application of $\Heff^{(2)}$.
We find that when evolving $\ket{\Phi}$ with $\hat{H}^\prime$, even if the wave function leaves the spin-$1$ subsector, it essentially entirely remains within the ``neighbor states''. To show this, we can define the projector $\hat{\mathcal{P}}_2$ onto ``neighbor states'' and states in the spin-$1$ subsector. Figure~\ref{fig:comp_proj_eff}(a) shows that the expectation value of this projector essentially remains at $1$ for $\ket{\Phi}$, even as $\hat{\mathcal{P}}_1$ itself oscillates. In fact, we find that within the ``neighbor states'' the wave function concentrates into a small subset of states. These take the form of $\ket{\rd\! \u \u \!\rd\! \cdots \!\rd\! \u \u \!\rd \rd \rd \rd\! \u \d \u \u \!\rd}$ and its equivalents under all possible translations (by an even number of sites) and spatial inversion. This is shown very clearly in Fig.~\ref{fig:comp_proj_eff} (b). We will denote by $\kb$ the symmetric superposition of these $L/2$ states. An important property of $\kb$ (and of all states it is a superposition of) is that applying $\Heff^{(2)}$ to it only leads to the spin-$1$ subsector and does not lead to leakage further away from it.

We can now ask \emph{why} the wave function does not spread further into the ``neighbor states''. For that, we focus on the simpler case $L=8$. This is also representative of larger system sizes, as $\Heff^{(2)}$ only affects two neighboring cells while the others stay as effective spins-$1$. In Fig.~\ref{fig:graph_eff}(a), we plot the graph representing the Hamiltonian action in the Hilbert space limited to the Krylov subspaces of interest. For simplicity, we only consider the off-diagonal parts of $\Heff^{(2)}$.

As the graph has a very regular structure, destructive interference of transition amplitudes can lead to 
eigenstates with zero overlap on certain Fock states. For example, for all values of $\mu$ there is a zero mode (i.e., a state annihilated by the off-diagonal part of $\hat{H}^\prime=\Heff+\Heff^{(2)}$) comprised only of the $\kb$ state and states in the original Krylov subspace. The (unnormalized) amplitudes of this state are shown in Fig.~\ref{fig:graph_eff}(c). That this state is annihilated by $\hat{H}^\prime$ can be seen using the transition amplitudes shown in Fig.~\ref{fig:graph_eff}(a). In presence of the diagonal terms of $\Heff^{(2)}$, the eigenstates generally do not have exactly zero overlap with the other ``neighbor states''. Nonetheless, these other states still have very low participation in many eigenstates when $\mu$ is $\geq 0.5$.

\begin{figure}[t]
	\centering
	\includegraphics[width=\linewidth]{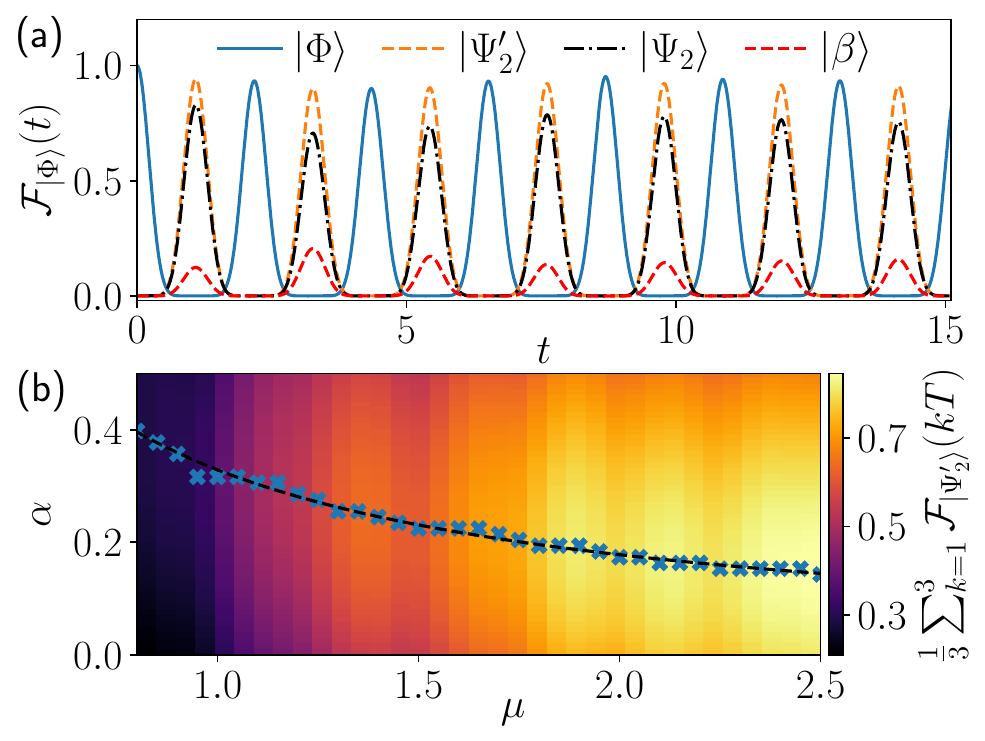}
    \caption{Deformation of the $\ket{\Phi}$ trajectory. (a) Overlap of the wave function on several states after evolving $\ket{\Phi}$ with $\hat{H}^\prime$ for $L=20$ and $\mu=h=0.8$. The $\ket{\Psi_2^\prime}$ state with $\alpha=0.408$ captures the wave function after a half-period with very high accuracy. (b) Average fidelity of the first three revivals for different values of $\alpha$ and $\mu=h$ after evolving $\ket{\Psi_2^\prime}$ with $\hat{H}$ for $L=16$. The blue crosses show the optimal value of $\alpha$ for each $\mu$ while he black dashed line is the fit to it $1/(0.46+2.58\mu)$.
	}\label{fig:psi2_prime}
\end{figure}

Of course, this destructive interference occurs only for states that occupy the graph symmetrically. This is true for the $\ket{\Phi}$ state, as well as when evolving the $\ket{\Psi_2}$ state, but it is not the case for $\ket{\Psi_4}$, as in one cell we only have $\ket{\rd\! \d \u \!\rd}$ but not its symmetric partner $\ket{\rd \!\u \d \!\rd}$. $\ket{\Psi_4}$ then has a strong overlap with the eigenstates of Fig.~\ref{fig:graph_eff} that are spread evenly over the whole graph. This leads to leakage beyond the ``neighbor states'' and to thermalization.

\begin{figure}[htb]
	\centering
	\includegraphics[width=\linewidth]{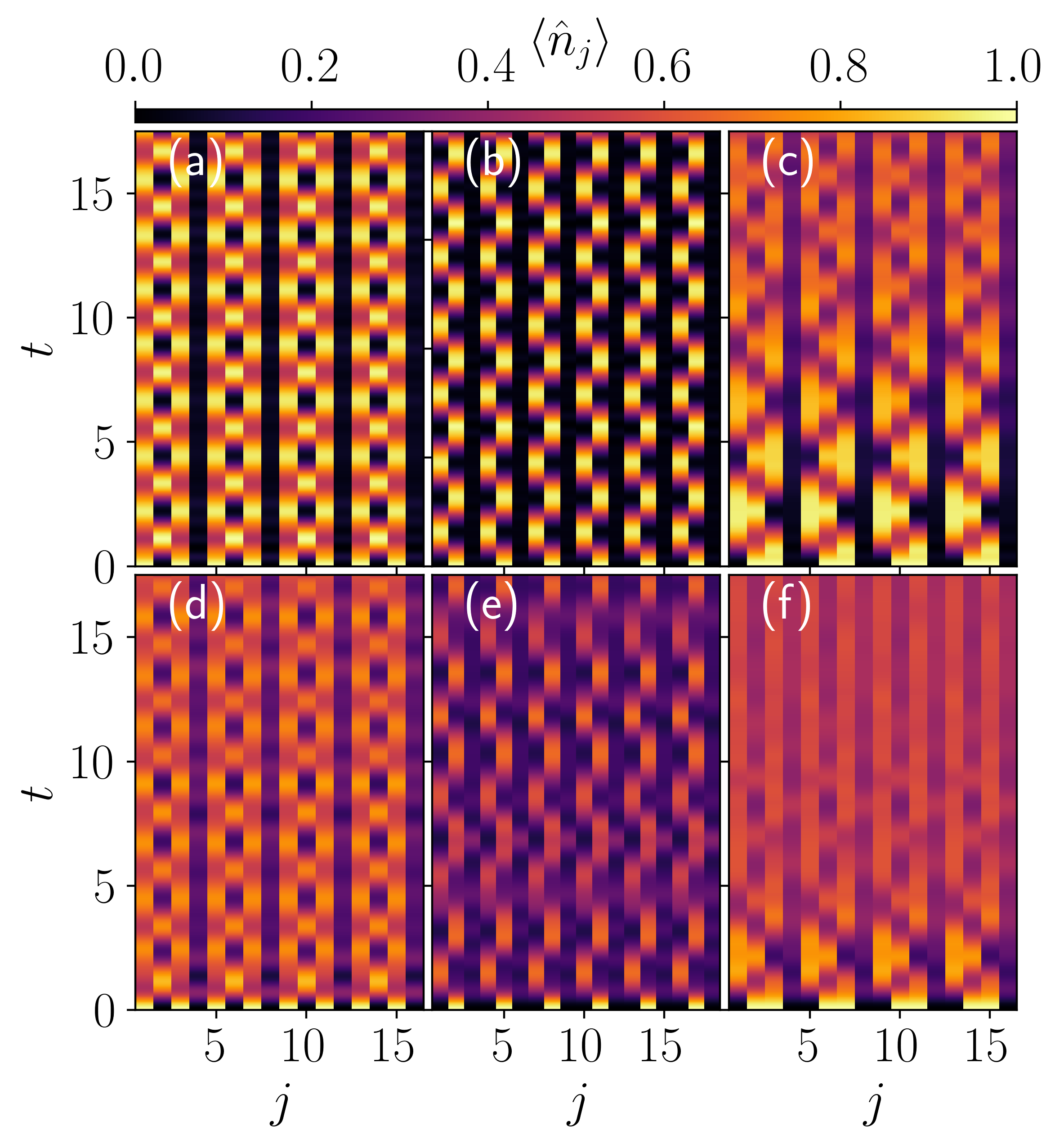}
    \caption{Dynamics of the fermion density after a quench from (a), (d) $\ket{\Psi_2}$, (b), (e) $\ket{\Psi_3}$ and (c), (f) $\ket{\Psi_4}$. Panels (a)-(c) are for $\mu=h=2$ while panels (d)-(f) are for $\mu=h=0.8$. Panels (a), (c), (d) and (f) are for $L=16$ while panels (b) and (e) are for $L=18$.
	}\label{fig:matter_dyn20}
\end{figure}

\begin{figure}[tb]
	\centering
	\includegraphics[width=\linewidth]{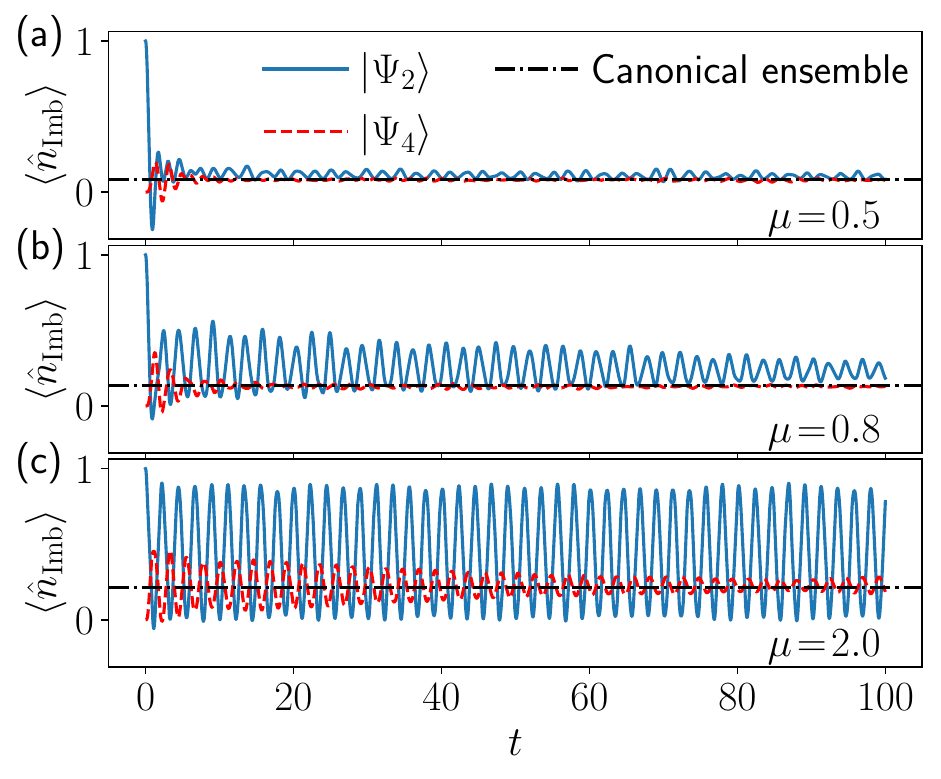}
    \caption{Dynamics of the fermion imbalance after a quench from $\ket{\Psi_2}$ and $\ket{\Psi_4}$ for various values of $\mu=h$ and $L=16$. Despite being at the same energy, both states show very different thermalization patterns.
	}\label{fig:matter_imb}
\end{figure}

These results help us understand the effect of $\Heff^{(2)}$ as deforming the scarred trajectory from $\ket{\Phi}$ instead of destroying it as for $\ket{\Psi_4}$. Indeed, in the limit $\mu=h\to \infty$ the $\ket{\Phi}$ trajectory has no overlap with $\kb$---as it is not in the spin-1 sector---while this overlap is non-negligible for $\mu=h\approx 1$, as shown in Fig.~\ref{fig:psi2_prime}(a). We can characterize this deformation using the state at half the period. This state goes from $\ket{\Psi_2}$ (at $\mu{=}h{=}\infty)$ to $\ket{\Psi_2^\prime}{=}(\ket{\Psi_2}{-}\alpha\kb){/}\sqrt{1{+}\alpha^2}$, with $\alpha$ a smooth function of $\mu$. Fig.~\ref{fig:psi2_prime}(a) shows that this new state captures the deformation almost perfectly when evolving the $\ket{\Phi}$ state with $\hat{H}^\prime$. 
Using the full Hamiltonian in Eq.~\eqref{eq:H-ZXZ}, we also find an increased revival fidelity when quenching from $\ket{\Psi_2^\prime}$ instead of $\ket{\Psi_2}$, with an optimal $\alpha$ of $\alpha^\star(\mu){\approx} 1{/}(0.46{+}2.58\mu)$ as shown on Fig.~\ref{fig:psi2_prime}(b). This type of scaling is expected, as the matrix elements of $\Heff^{(2)}$ connecting the spin-$1$ sector and $\kb$ are of order $1/\mu$.

\subsection{Matter dynamics}
Finally, we briefly discuss the scarred dynamics from the point of view of the fermions of the original LGT. As a reminder, the fermion density simply corresponds to the domain wall density in the spin model as $\n_j\equiv (1-\Z_j\Z_{j+1})/2$. In Fig,~\ref{fig:matter_dyn20}, we show the dynamics of the fermions after quenches from the states $\ket{\Psi_2}$, $\ket{\Psi_3}$ and $\ket{\Psi_4}$. As expected, for $\mu=h=2$ as shown in panels (a)-(c) the dynamics is essentially that of multiple independent cells separated by frozen holes. However, for $\ket{\Psi_4}$ we already see the effects of interactions between cells at later times. 
We also show the same dynamics at $\mu=h=0.8$ in panels (d)-(f). The empty sites between active cells are no longer fully frozen but show oscillations for $\ket{\Psi_2}$ and $\ket{\Psi_3}$. This is a sign that the cells are no longer isolated but instead interact. Importantly, these oscillations of the previously frozen sites show no clear damping which would be indicative of equilibration. On the other hand, for $\ket{\Psi_4}$ we see that the sites between cells gradually fill up and that the dynamics leads to thermalization.

Nonetheless, the equilibrated expectation value seemingly still shows imbalance between the occupation of odd and even matter sites. However, this does not denote ergodicity breaking as the Hamiltonian itself is only invariant under translations by two sites. To test if this state indeed thermalizes, we look at the fermion imbalance defined as 
\begin{equation}
    \hat{n}_\mathrm{Imb}=\frac{2}{L}\left(\sum_{j \ \mathrm{odd}}\hat{n}_j -\sum_{j \ \mathrm{even}}\hat{n}_j \right).
\end{equation}
In Fig.~\ref{fig:matter_imb}, we show the dynamics of this quantity after quenches from the $\ket{\Psi_2}$ and $\ket{\Psi_4}$ states. We do not show it for $\ket{\Psi_3}$ as it is always zero for symmetry reasons. As expected, we see that $\ket{\Psi_2}$ exhibits oscillations in the imbalance that are much more pronounced and long-lived than for $\ket{\Psi_4}$. Perhaps more surprisingly, we see that the center of the oscillations and the late-time values also differ between the two states. For $\ket{\Psi_4}$, both of them agree well with the prediction of the canonical ensemble. Meanwhile, for $\ket{\Psi_2}$ this is only the case when $\mu=h$ is very small. Even for $\mu=h=0.5$, there is still a noticeable difference between the equilibration value of the imbalance and the canonical ensemble prediction. This further highlights the clear difference between $\ket{\Psi_2}$ and $\ket{\Psi_4}$. While the latter only shows non-thermal dynamics at shorter time due to the integrability of $\Heff$, the former shows actual many-body scarring because of the interplay of $\Heff$ with the higher-order terms.

\section{Summary and outlook} In this work, we have explored adding a resonant mass term to a strongly confined $\mathbb{Z}_2$ LGT. The mass term has the counterintuitive effect of inducing \textit{local} deconfinement such that a single hole within a fully packed matter background becomes mobile. In the extreme limit where only a single hole is present, information can propagate across the system; otherwise, the holes remain spatially confined to local regions of varying sizes, each of which constitutes an independent tight-binding chain. For certain CDW matter configurations where these tight-binding chains have commensurate spectra, this structure induces periodic fidelity revivals. These features are sharp in the limit $\mu{=}h{=}\infty$, where the model becomes integrable. Remarkably, when integrability breaking perturbative corrections are included that weaken these effects, a particular period-$2$ CDW state continues to exhibit robust revivals even for modest $\mu{\sim} h{\sim} J$. We trace these revivals back to destructive interferences along the Hilbert space trajectory of the initial state, made possible by the intricate interplay between the first and second terms of the effective Hamiltonian. This leads to the periodic trajectory of the integrable limit being smoothly deformed instead of breaking down as $\mu/J=h/J$ is decreased. This is unlike previous examples of QMBS surviving from an integrable regime, where the regular dynamics relied simply on the trajectory being mostly annihilated by higher-order corrections. This provides a new mechanism for QMBS in the $\mathbb{Z}_2$ LGT and in models with an integrable limit.

Signatures of these QMBS, including oscillations of the matter imbalance, are readily observable in present-day quantum simulators. The model in Eq.~\eqref{eq:H-ZXZ}, including the mass term, can be realized as a limit of the mixed-field Ising model~\cite{Bastianello22,aramthottil2022scar}, making trapped-ion and Rydberg-atom platforms particularly well suited. Furthermore, it is interesting to ask whether the local deconfinement mechanism explored in this paper is generic to $\mathbb Z_2$ LGTs in higher dimensions, or whether it arises for other gauge groups. Indeed, the absence of this mechanism in the $\mathrm{U}(1)$ quantum link model~\cite{Desaules2024ergodicitybreaking} begs the question of what are the necessary and sufficient conditions for this behavior to occur. Understanding this will shed light on the nature of confinement and on  mechanisms to avoid it.

\section*{Data availability}
The data supporting the results within this paper is available at \cite{RData}.

\begin{acknowledgments}
The authors are grateful to Fiona Burnell, Gaurav Gyawali, Zlatko Papi\'c, Elliot Rosenberg, Pedram Roushan, Michael Schecter and Una \v{S}lanka for insightful discussions. J.-Y.D.~acknowledges funding from the European Union’s Horizon 2020 research and innovation programme under the Marie Sk\l odowska-Curie Grant Agreement No.~101034413. T.I.~acknowledges support from the National Science Foundation under Grant No.~DMR-2143635. 
J.C.H. acknowledges funding by  the Emmy Noether Programme of the German Research Foundation (DFG) under Grant No. HA 8206/1-1.s, the Max Planck Society, the Deutsche Forschungsgemeinschaft (DFG, German Research Foundation) under Germany’s Excellence Strategy – EXC-2111 – 390814868, and the European Research Council (ERC) under the European Union’s Horizon Europe research and innovation program (Grant Agreement No.~101165667)—ERC Starting Grant QuSiGauge. This work is part of the Quantum Computing for High-Energy Physics (QC4HEP) working group.
\end{acknowledgments}

\appendix

\section{Model and symmetries}\label{app:syms}

\begin{figure}[bt]
	\centering
	\includegraphics[width=\linewidth]{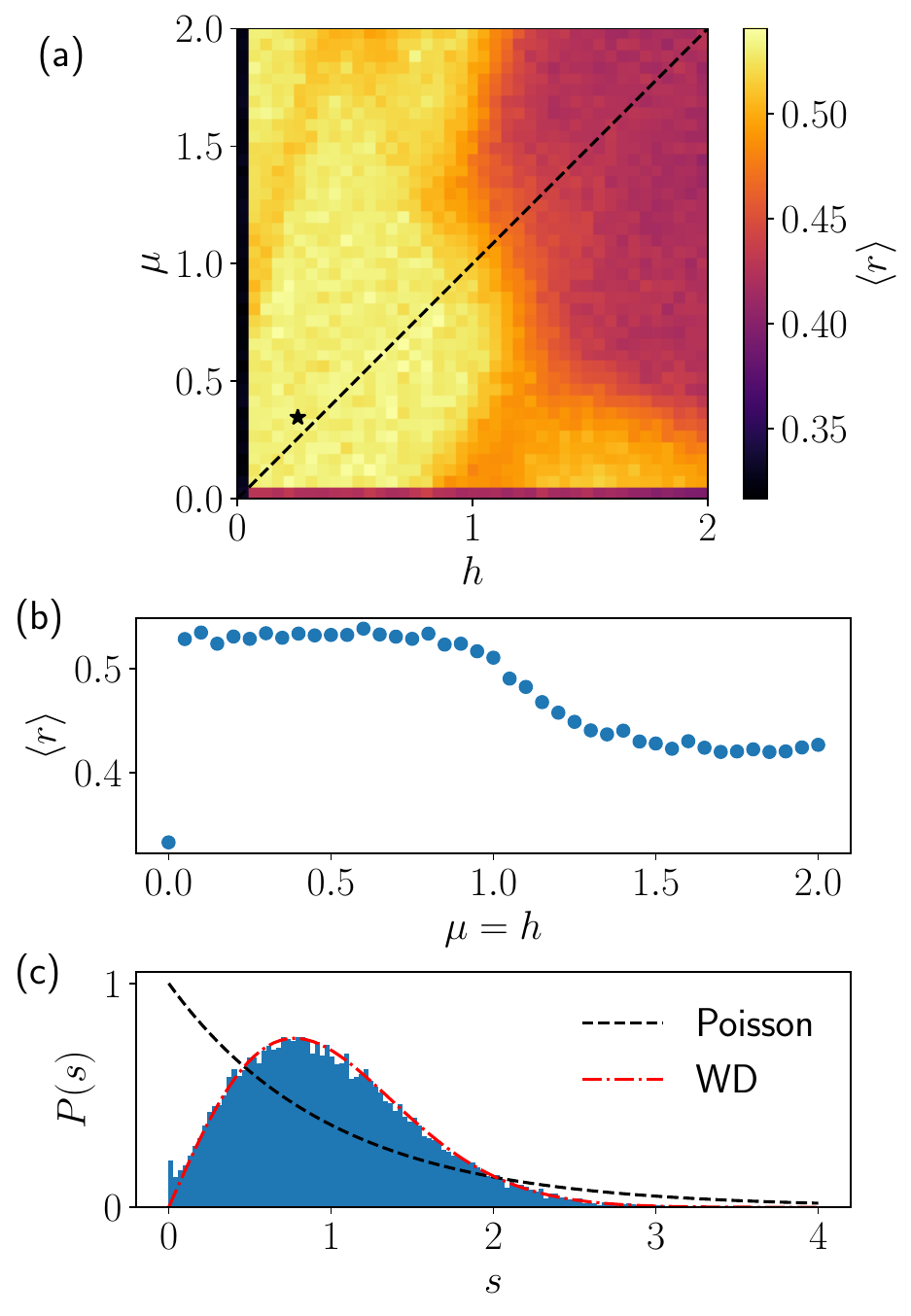}
\caption{Level spacing statistics in the sector with $2\lfloor L/4\rfloor$ domain walls, zero momentum, and eigenvalue $+1$ under spatial reflection. (a) Mean level spacing ratio~\cite{Oganesyan07} for a wide range of values of $\mu$ and $h$ for $L=20$. In a broad range of values, we have $\langle r \rangle \approx 0.53$, as expected for a chaotic system~\cite{Atas13}. The dashed line indicates $\mu=h$, for which the data is plotted in panel (b). The black star indicates the parameter used in panel (c). (b) Mean level spacing ratio for $\mu=h$ for $L=20$. The system is chaotic until around $\mu=h=1$. (c) Full distribution of level spacings after unfolding for $L=22$ with $\mu=\sqrt{3}/5\approx 0.346$ and $h=\sqrt{8}/11\approx 0.257$. There is clear agreement with the Wigner-Dyson distribution for the Gaussian orthogonal ensemble, indicating that the model is chaotic.
	}\label{fig:level_stats}
\end{figure}

In this section we briefly review the properties of the Hamiltonian in Eq.~(\ref{eq:H-ZXZ}). It conserves the number of domain walls $\hat{N}_\mathrm{DW}=\sum_j \left(1-\Z_j \Z_{j+1}\right)/2$.
This splits the system into $L/2$ sectors for $L$ even with periodic boundary conditions (PBC). In addition, for even $L$ the system is also invariant under spatial reflection and translation by two sites. We observe that applying a translation by a single site maps to the same model but with $\mu \rightarrow -\mu$. In a similar way, performing a particle-hole exchange $\sum_j \X_j$ only changes $h$ to $-h$ while leaving the other parameters invariant. This means that we can focus solely on the case where $\mu, h\geq 0$, as the other three sectors are identical up to a unitary transformation.

To verify that the Hamiltonian is chaotic, we check the energy level statistics, and in particular the mean energy spacing ratio~\cite{Oganesyan07}. This is plotted in Fig.~\ref{fig:level_stats}(a) and shows $\langle r \rangle\approx 0.53$ in a relatively broad range of parameters. This is the value expected from the Wigner-Dyson distribution which is relevant for a real-valued random matrix~\cite{Atas13}. For $\mu=0$ the model has additional symmetries (translation by one instead of two sites) while for $h=0$ the model is integrable via a mapping to free fermions~\cite{Iadecola2020quantum,Khor23}.  Along the resonant regime $\mu=h$ that we explore in the main text, the level spacing statistics indicate that the model is chaotic until $\mu=h\approx J$ for $L=20$, as shown in Fig.~\ref{fig:level_stats}(b). Additionally, we check the full distribution of level spacings for incommensurate values of $h$ and $\mu$ in Fig.~\ref{fig:level_stats}(c), once again finding good agreement with the Wigner-Dyson distribution.

\section{Detuning}\label{app:det}
In this section, we consider the case  where $h$ and $\mu$ are not exactly equal. In particular, we will set $h=\mu+\delta$ where $\delta$ is the detuning. As long as $\delta$ is small compared to $h$ and $\mu$, we can  perform the Schrieffer-Wolff transformation with the same $\Hn$ where $h=\mu$, but with an additional term $\delta\sum_j \hat{Z}_j$ in $\hat{V}$. This additional term will directly appear in $\Heff$ at order 1. It will not influence the terms at second order but will have an effect at third order. Indeed, there will be new terms appearing at third order with a prefactor of $J^2\delta/\mu^2$. These new terms can actually be simply described as $-\frac{\delta}{2\mu}\Heff^{(2)}$, as they correspond to inserting the diagonal field $\delta$ in the process described by $\Heff^{(2)}$.

If we only look at the first order term, since the detuning term is a simple Z-field it will not change the connectivity. So the Hilbert space fragmentation is not affected. In each disconnected part of the chain, moving the hole is done by flipping a spin up or down depending on the parity of the site. This means that the effective particle will feel a staggered Z-field. Since this just adds local Z-terms, it does not kill the mapping to free-fermions and each part of the chain remains integrable.

\begin{figure}[tb]
	\centering
	\includegraphics[width=\linewidth]{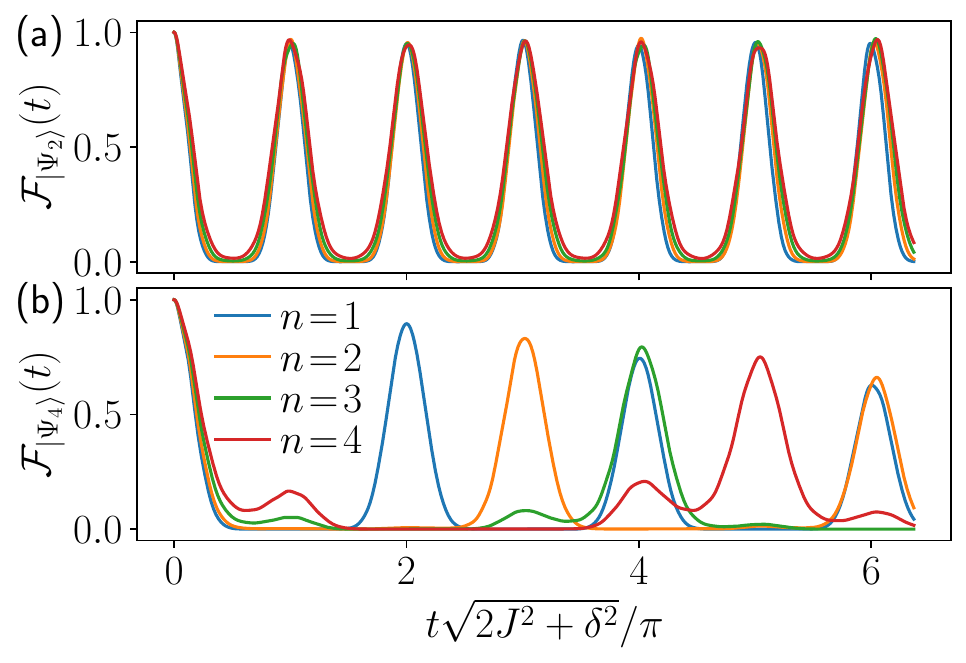}
\caption{Fidelity after a quench from the states $\ket{\Psi_2}$ (top) and $\ket{\Psi_4}$ (bottom) with $L=16$ and  $\mu=5$. Each curve corresponds to a different detuning $\frac{(n-1)}{\sqrt{2n}}$. For $\ket{\Psi_2}$, only the period and fidelity minimum changes with detuning. Meanwhile, for $\ket{\Psi_4}$ the period is directly affected by detuning.	}\label{fig:revs_det}
\end{figure}

\begin{figure}[tb]
	\centering
	\includegraphics[width=\linewidth]{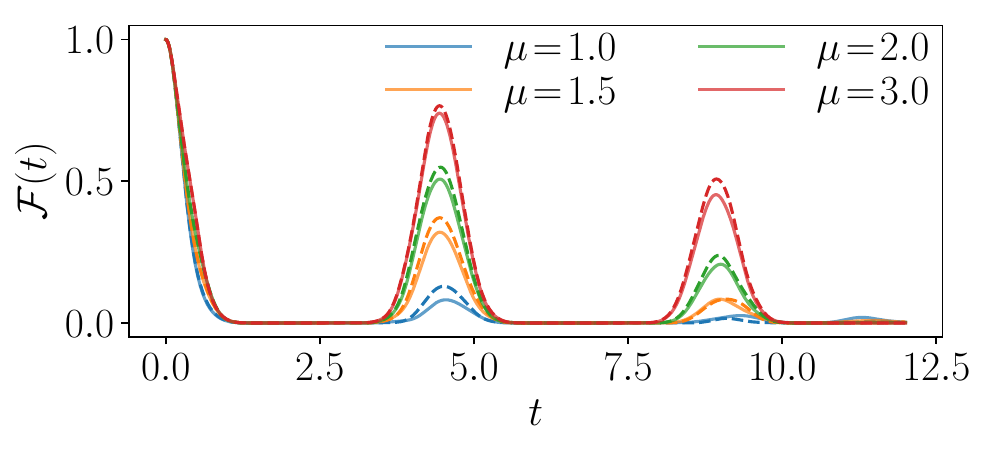}
\caption{Fidelity after a quench from $\ket{\Psi_4}$ with $L=16$ close to the resonant point $\mu=h$. Each color corresponds to a different value of $\mu$. Solid curves are with no detuning while dashed lines are with $\delta=\frac{1}{8\mu}\frac{1}{1+1/(16\mu^2)}$. While the effect is small, the latter consistently shows better revivals.}\label{fig:revs_det_HF2}
\end{figure}

We can also focus on the effective spin-$1$ sector. Let us ignore the action of the higher-order Schrieffer-Wolff terms for the moment. In that case, the Z-field will change the effective Hamiltonian of each spin-$1$ as
\begin{equation}\label{eq:HF_det}
\begin{pmatrix}
    0& J & 0 \\
    J & 0 & J \\
    0 & J & 0
\end{pmatrix}
\to 
\begin{pmatrix}
    -2\delta & J & 0 \\
    J & 0 & J \\
    0 & J & -2\delta
\end{pmatrix},
\end{equation}
so that the eigenvalues then go from $-\sqrt{2}|J|$, $0$ and $\sqrt{2}|J|$ to $-\delta-\sqrt{2J^2+\delta^2}$, $-2\delta$ and $-\delta+\sqrt{2J^2+\delta^2}$. This has important consequences. If we quench from $\ket{\Psi_2}$, then only the highest and lowest eigenvalues are involved. This means that there is a single energy spacing of $2\sqrt{2J^2+\delta^2}$ so that revivals still occur but with a renormalized period of $\pi/\sqrt{2J^2+\delta^2}$. Meanwhile, if we quench from $\ket{\Psi_4}$ then all three eigenvalues are involved and we get two different energy spacings. We can only get perfect revivals if these are commensurate, meaning if $\delta+\sqrt{2J^2+\delta^2}=n\left(\sqrt{2J^2+\delta^2}-\delta\right)$, with $n$ a rational number. We can turn that expression into $\delta=\pm \frac{(n-1)|J|}{\sqrt{2n}}$. For $n$ integer, the revival then has a period of $(n+1)\pi/\sqrt{2J^2+\delta^2}$. The case with no detuning corresponds to $n=1$ and we recover that $\ket{\Psi_4}$ revives with twice the period of $\ket{\Psi_2}$. This dependence on detuning is shown in Fig.~\ref{fig:revs_det} for various integer $n$.

We can also ask if there is any interesting interaction between the higher order Schrieffer-Wolff terms and the detuning. Interestingly, in each spin-$1$ cell the effect of the detuning is the opposite of that of the diagonal part of $\Heff^{(2)}$. Indeed, the action of the latter in the spin-$1$ sector is 
\begin{equation}\label{eq:HF2_diag}
\begin{pmatrix}
    0& J & 0 \\
    J & 0 & J \\
    0 & J & 0
\end{pmatrix}
\to 
\begin{pmatrix}
    -1/(4\mu) & J & 0 \\
    J & -1/(2\mu) & J \\
    0 & J & -1/(4\mu)
\end{pmatrix}.
\end{equation}
Looking at both (\ref{eq:HF2_diag}) and Eq.~(\ref{eq:HF_det}), if we combine their action we get that the diagonal term becomes $-2\delta-1/(4\mu)$, $-1/(2\mu)$ and $-2\delta-1/(4\mu)$. Choosing $\delta=1/(8\mu)$ makes all these terms equal to $-1/(2\mu)$ and so restores the equal spacing of the three energy levels.  For added accuracy, we can also consider the effect of the third order term. With detuning, its diagonal contribution is equal to $-\frac{\delta}{2\mu}$ times that of $\Heff^{(2)}$. This means we can just multiply the diagonal term of $\Heff^{(2)}$ by a factor of $1-\delta/(2\mu)$. This changes the optimal $\delta$ from $\frac{1}{8\mu}$ to $\frac{1}{8\mu}\frac{1}{1+1/(16\mu^2)}$. While adding this detuning has little effect on revivals from $\ket{\Psi_2}$, it actually improves revivals from $\ket{\Psi_4}$. This is shown in Fig.~\ref{fig:revs_det_HF2}. The effect is visible but small, as most of the revival decay is caused by leakage which is not corrected by the detuning.

\begin{figure}[t]
	\centering
	\includegraphics[width=\linewidth]{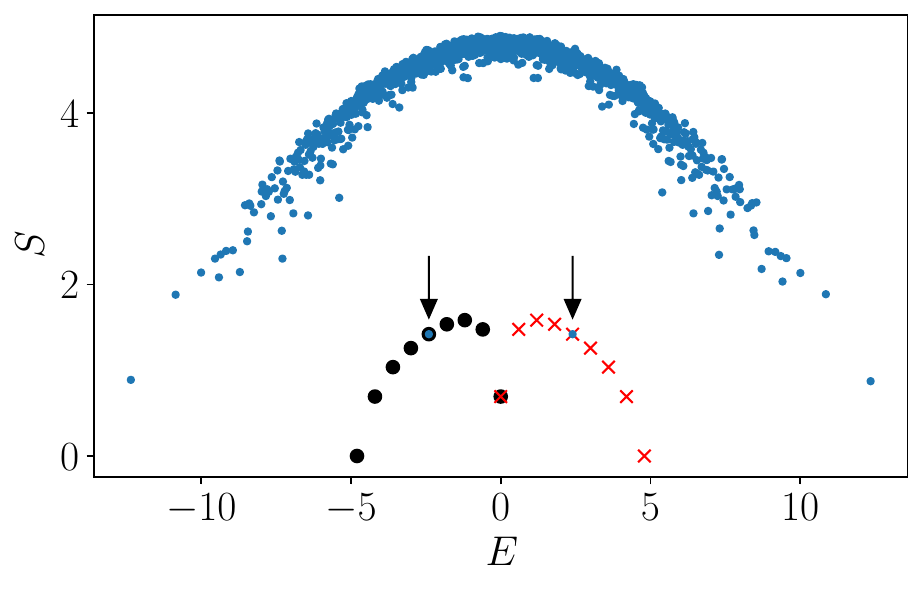}
\caption{Entanglement entropy of eigenstates for $L=16$ with $h=0.3$ and $\mu=0.74$. The blue dots are eigenstates in the sector with 8 domain walls, 0 momentum, and eigenvalue +1 under spatial inversion. Black and red markers indicates the two families of exact scars found in Ref.~\cite{Iadecola2020quantum}, which have an energy spacing of $2h$. The arrows show which states among them live in the same symmetry sector as the rest of the data.
	}\label{fig:stag_scars}
\end{figure}

\section{Exact quantum many-body scars}\label{app:QMBS}
In this section, we discuss the presence of exact quantum many-body scars (QMBSs) in the model.
If we consider the same model as in the main text but without the staggering in the ZZ term, then it has been shown to hosts two families of exact QMBS~\cite{Iadecola2020quantum}. When the ZZ term is staggered, the same states are still present, as shown in Fig.~\ref{fig:stag_scars}. However, their energy now only depends on the field $h$ and not on $\mu$, as $E_n=h(-L+2n)$ with $n=0$ to $L+1$. Indeed, scarred eigenstates of the first kind are symmetric superpositions of all Fock states with a fixed number of up-spins but none of them next to each other. These isolated up-spins do not change the energy of the mass term, as they create two domain walls but on even-odd and odd-even sites. Thus their contribution will cancel each other due to the staggering. The second family of scarred states is the same up to a global spin flip. This means that they are symmetric superpositions of all Fock states with a fixed number of down-spins but none of them next to each other. The same argument holds as to why their energy only depends on $h$. 

Importantly, these $N+1$ scarred states are all eigenstates of $\Hn$. This means that they will not get split up into different eigenstates as $\mu=h$ is increased. Instead, they will be exact eigenstates of \textit{all} the higher order Schrieffer-Wolff terms independently. Nonetheless, we emphasize that they do not contribute at all to the scarring we observe from the $\ket{\Phi}$ and $\ket{\Psi_2}$ states. First of all, as both of these states have a definite number of up-spins they could have overlap with at most one of the exact scarred states. Thus, the equal spacing with the other QMBSs would be irrelevant for the dynamics. On top of that, we can show that these QMBSs have exactly zero overlap with $\ket{\Psi_2}$ and $\ket{\Phi}$. For the former this is easy to see, as $\ket{\Psi_2}$ has both consecutive down-spins and consecutive up-spins. For the latter, we need to take into the account the phases in the states composing the exact QMBSs. Having an up-spin on an odd site leads to a phase factor of $-1$ while it is $+1$ if the up-spin is on an even site. So we can write the relevant part of the scarred state as a tensor product of 4-site cells where each cell is $(\ket{\d \u \d \d }-\ket{\d \d \u \d })/\sqrt{2}$. As for the $\ket{\Phi}$ state, each cell is in the state
$(\ket{\d \u \d \d }+\ket{\d \d \u \d })/\sqrt{2}$, and we immediately see that they are orthogonal.

\section{Third order SW term}\label{app:SW3}
We can go beyond second order and look at the third order SW term:
\begin{equation}\label{eq:SW3}
    \begin{aligned}
    H_\mathrm{eff}^{(3)}&{=}\frac{J^3}{32\mu^2}\Big[
\sum_{j\ {\rm odd}}\Big(Q_{j-1}\sigma^+_j\sigma^+_{j+1}\sigma^-_{j+2}Q_{j+3}+\text{h.c.}\\
&\quad \quad \quad +P_{j-1}\sigma^+_j\sigma^-_{j+1}\sigma^-_{j+2}P_{j+3}+\text{h.c.}\Big)\\
&{+}\sum_{j\ {\rm even}}\Big(P_{j-1}\sigma^+_j\sigma^+_{j+1}\sigma^-_{j+2}P_{j+3}+\text{h.c.}\\
&\quad \quad \quad +Q_{j-1}\sigma^+_j\sigma^-_{j+1}\sigma^-_{j+2}Q_{j+3}+\text{h.c.}\Big)\\
&{-}\sum_{j\ {\rm even}} \Big(\Pd_{j-1}\hat{X}_{j}\Pu_{j+1}\Pu_{j+2}+\Pd_{j-2}\Pd_{j-1}\hat{X}_{j}\Pu_{j+1}\Big)\\
&{-}\sum_{j\ {\rm odd}}\Big( \Pu_{j-1}\hat{X}_{j}\Pd_{j+1}\Pd_{j+2}{+} \Pu_{j-2}\Pu_{j-1}\hat{X}_{j}\Pd_{j+1}\Big)\Big]
\end{aligned}
\end{equation}
The terms on the two last lines are the same as in the effective Hamiltonian $\Heff$ but with additional projectors on sites $j+2$ or $j-2$, and so do not cause any leakage out of the integrable subspaces. However, the terms in the first four lines do. For example, they act on $\ket{\Psi_2}=\ket{\rd \!\u\u\!\rd\rd\!\u\u\!\rd}$, taking it to $\ket{\u\!\rd\rd\rd\rd\!\u\u\d}$, $\ket{\rd\!\u\u\d\u\!\rd\rd\rd}$, $\ket{\rd\rd\rd\!\u\d\u\u\!\rd}$ and $\ket{\d\u\u\!\rd\rd\rd\rd\!\u}$.
We note that $\Heff^{(3)}$ has a prefactor of $J^3/(32\mu^2)$, so for $\mu~\approx J$ its contribution will be small compared to that of $\Heff$ and $\Heff^{(2)}$.

\pagebreak

\bibliography{biblio}

\begin{thebibliography}{86}%
\makeatletter
\providecommand \@ifxundefined [1]{%
 \@ifx{#1\undefined}
}%
\providecommand \@ifnum [1]{%
 \ifnum #1\expandafter \@firstoftwo
 \else \expandafter \@secondoftwo
 \fi
}%
\providecommand \@ifx [1]{%
 \ifx #1\expandafter \@firstoftwo
 \else \expandafter \@secondoftwo
 \fi
}%
\providecommand \natexlab [1]{#1}%
\providecommand \enquote  [1]{``#1''}%
\providecommand \bibnamefont  [1]{#1}%
\providecommand \bibfnamefont [1]{#1}%
\providecommand \citenamefont [1]{#1}%
\providecommand \href@noop [0]{\@secondoftwo}%
\providecommand \href [0]{\begingroup \@sanitize@url \@href}%
\providecommand \@href[1]{\@@startlink{#1}\@@href}%
\providecommand \@@href[1]{\endgroup#1\@@endlink}%
\providecommand \@sanitize@url [0]{\catcode `\\12\catcode `\$12\catcode
  `\&12\catcode `\#12\catcode `\^12\catcode `\_12\catcode `\%12\relax}%
\providecommand \@@startlink[1]{}%
\providecommand \@@endlink[0]{}%
\providecommand \url  [0]{\begingroup\@sanitize@url \@url }%
\providecommand \@url [1]{\endgroup\@href {#1}{\urlprefix }}%
\providecommand \urlprefix  [0]{URL }%
\providecommand \Eprint [0]{\href }%
\providecommand \doibase [0]{http://dx.doi.org/}%
\providecommand \selectlanguage [0]{\@gobble}%
\providecommand \bibinfo  [0]{\@secondoftwo}%
\providecommand \bibfield  [0]{\@secondoftwo}%
\providecommand \translation [1]{[#1]}%
\providecommand \BibitemOpen [0]{}%
\providecommand \bibitemStop [0]{}%
\providecommand \bibitemNoStop [0]{.\EOS\space}%
\providecommand \EOS [0]{\spacefactor3000\relax}%
\providecommand \BibitemShut  [1]{\csname bibitem#1\endcsname}%
\let\auto@bib@innerbib\@empty
\bibitem [{\citenamefont {Wilson}(1974)}]{Wilson1974confinement}%
  \BibitemOpen
  \bibfield  {author} {\bibinfo {author} {\bibfnamefont {Kenneth~G.}\
  \bibnamefont {Wilson}},\ }\bibfield  {title} {\enquote {\bibinfo {title}
  {Confinement of quarks},}\ }\href {\doibase 10.1103/PhysRevD.10.2445}
  {\bibfield  {journal} {\bibinfo  {journal} {Phys. Rev. D}\ }\textbf {\bibinfo
  {volume} {10}},\ \bibinfo {pages} {2445--2459} (\bibinfo {year}
  {1974})}\BibitemShut {NoStop}%
\bibitem [{\citenamefont {Rothe}(2005)}]{Rothe_book}%
  \BibitemOpen
  \bibfield  {author} {\bibinfo {author} {\bibfnamefont {H.J.}\ \bibnamefont
  {Rothe}},\ }\href {https://books.google.de/books?id=U1hBLG-\_WxAC} {\emph
  {\bibinfo {title} {Lattice Gauge Theories: An Introduction}}},\ EBSCO ebook
  academic collection\ (\bibinfo  {publisher} {World Scientific},\ \bibinfo
  {year} {2005})\BibitemShut {NoStop}%
\bibitem [{\citenamefont {Dalmonte}\ and\ \citenamefont
  {Montangero}(2016)}]{Dalmonte_review}%
  \BibitemOpen
  \bibfield  {author} {\bibinfo {author} {\bibfnamefont {M.}~\bibnamefont
  {Dalmonte}}\ and\ \bibinfo {author} {\bibfnamefont {S.}~\bibnamefont
  {Montangero}},\ }\bibfield  {title} {\enquote {\bibinfo {title} {Lattice
  gauge theory simulations in the quantum information era},}\ }\href {\doibase
  10.1080/00107514.2016.1151199} {\bibfield  {journal} {\bibinfo  {journal}
  {Contemporary Physics}\ }\textbf {\bibinfo {volume} {57}},\ \bibinfo {pages}
  {388--412} (\bibinfo {year} {2016})}\BibitemShut {NoStop}%
\bibitem [{\citenamefont {Ba{\~n}uls}\ \emph {et~al.}(2020)\citenamefont
  {Ba{\~n}uls}, \citenamefont {Blatt}, \citenamefont {Catani}, \citenamefont
  {Celi}, \citenamefont {Cirac}, \citenamefont {Dalmonte}, \citenamefont
  {Fallani}, \citenamefont {Jansen}, \citenamefont {Lewenstein}, \citenamefont
  {Montangero}, \citenamefont {Muschik}, \citenamefont {Reznik}, \citenamefont
  {Rico}, \citenamefont {Tagliacozzo}, \citenamefont {Van~Acoleyen},
  \citenamefont {Verstraete}, \citenamefont {Wiese}, \citenamefont {Wingate},
  \citenamefont {Zakrzewski},\ and\ \citenamefont {Zoller}}]{Pasquans_review}%
  \BibitemOpen
  \bibfield  {author} {\bibinfo {author} {\bibfnamefont {Mari~Carmen}\
  \bibnamefont {Ba{\~n}uls}}, \bibinfo {author} {\bibfnamefont {Rainer}\
  \bibnamefont {Blatt}}, \bibinfo {author} {\bibfnamefont {Jacopo}\
  \bibnamefont {Catani}}, \bibinfo {author} {\bibfnamefont {Alessio}\
  \bibnamefont {Celi}}, \bibinfo {author} {\bibfnamefont {Juan~Ignacio}\
  \bibnamefont {Cirac}}, \bibinfo {author} {\bibfnamefont {Marcello}\
  \bibnamefont {Dalmonte}}, \bibinfo {author} {\bibfnamefont {Leonardo}\
  \bibnamefont {Fallani}}, \bibinfo {author} {\bibfnamefont {Karl}\
  \bibnamefont {Jansen}}, \bibinfo {author} {\bibfnamefont {Maciej}\
  \bibnamefont {Lewenstein}}, \bibinfo {author} {\bibfnamefont {Simone}\
  \bibnamefont {Montangero}}, \bibinfo {author} {\bibfnamefont {Christine~A.}\
  \bibnamefont {Muschik}}, \bibinfo {author} {\bibfnamefont {Benni}\
  \bibnamefont {Reznik}}, \bibinfo {author} {\bibfnamefont {Enrique}\
  \bibnamefont {Rico}}, \bibinfo {author} {\bibfnamefont {Luca}\ \bibnamefont
  {Tagliacozzo}}, \bibinfo {author} {\bibfnamefont {Karel}\ \bibnamefont
  {Van~Acoleyen}}, \bibinfo {author} {\bibfnamefont {Frank}\ \bibnamefont
  {Verstraete}}, \bibinfo {author} {\bibfnamefont {Uwe-Jens}\ \bibnamefont
  {Wiese}}, \bibinfo {author} {\bibfnamefont {Matthew}\ \bibnamefont
  {Wingate}}, \bibinfo {author} {\bibfnamefont {Jakub}\ \bibnamefont
  {Zakrzewski}}, \ and\ \bibinfo {author} {\bibfnamefont {Peter}\ \bibnamefont
  {Zoller}},\ }\bibfield  {title} {\enquote {\bibinfo {title} {Simulating
  lattice gauge theories within quantum technologies},}\ }\href {\doibase
  10.1140/epjd/e2020-100571-8} {\bibfield  {journal} {\bibinfo  {journal} {The
  European Physical Journal D}\ }\textbf {\bibinfo {volume} {74}},\ \bibinfo
  {pages} {165} (\bibinfo {year} {2020})}\BibitemShut {NoStop}%
\bibitem [{\citenamefont {Zohar}\ \emph {et~al.}(2015)\citenamefont {Zohar},
  \citenamefont {Cirac},\ and\ \citenamefont {Reznik}}]{Zohar_review}%
  \BibitemOpen
  \bibfield  {author} {\bibinfo {author} {\bibfnamefont {Erez}\ \bibnamefont
  {Zohar}}, \bibinfo {author} {\bibfnamefont {J~Ignacio}\ \bibnamefont
  {Cirac}}, \ and\ \bibinfo {author} {\bibfnamefont {Benni}\ \bibnamefont
  {Reznik}},\ }\bibfield  {title} {\enquote {\bibinfo {title} {Quantum
  simulations of lattice gauge theories using ultracold atoms in optical
  lattices},}\ }\href {\doibase 10.1088/0034-4885/79/1/014401} {\bibfield
  {journal} {\bibinfo  {journal} {Reports on Progress in Physics}\ }\textbf
  {\bibinfo {volume} {79}},\ \bibinfo {pages} {014401} (\bibinfo {year}
  {2015})}\BibitemShut {NoStop}%
\bibitem [{\citenamefont {Alexeev}\ \emph {et~al.}(2021)\citenamefont
  {Alexeev}, \citenamefont {Bacon}, \citenamefont {Brown}, \citenamefont
  {Calderbank}, \citenamefont {Carr}, \citenamefont {Chong}, \citenamefont
  {DeMarco}, \citenamefont {Englund}, \citenamefont {Farhi}, \citenamefont
  {Fefferman}, \citenamefont {Gorshkov}, \citenamefont {Houck}, \citenamefont
  {Kim}, \citenamefont {Kimmel}, \citenamefont {Lange}, \citenamefont {Lloyd},
  \citenamefont {Lukin}, \citenamefont {Maslov}, \citenamefont {Maunz},
  \citenamefont {Monroe}, \citenamefont {Preskill}, \citenamefont {Roetteler},
  \citenamefont {Savage},\ and\ \citenamefont {Thompson}}]{Alexeev_review}%
  \BibitemOpen
  \bibfield  {author} {\bibinfo {author} {\bibfnamefont {Yuri}\ \bibnamefont
  {Alexeev}}, \bibinfo {author} {\bibfnamefont {Dave}\ \bibnamefont {Bacon}},
  \bibinfo {author} {\bibfnamefont {Kenneth~R.}\ \bibnamefont {Brown}},
  \bibinfo {author} {\bibfnamefont {Robert}\ \bibnamefont {Calderbank}},
  \bibinfo {author} {\bibfnamefont {Lincoln~D.}\ \bibnamefont {Carr}}, \bibinfo
  {author} {\bibfnamefont {Frederic~T.}\ \bibnamefont {Chong}}, \bibinfo
  {author} {\bibfnamefont {Brian}\ \bibnamefont {DeMarco}}, \bibinfo {author}
  {\bibfnamefont {Dirk}\ \bibnamefont {Englund}}, \bibinfo {author}
  {\bibfnamefont {Edward}\ \bibnamefont {Farhi}}, \bibinfo {author}
  {\bibfnamefont {Bill}\ \bibnamefont {Fefferman}}, \bibinfo {author}
  {\bibfnamefont {Alexey~V.}\ \bibnamefont {Gorshkov}}, \bibinfo {author}
  {\bibfnamefont {Andrew}\ \bibnamefont {Houck}}, \bibinfo {author}
  {\bibfnamefont {Jungsang}\ \bibnamefont {Kim}}, \bibinfo {author}
  {\bibfnamefont {Shelby}\ \bibnamefont {Kimmel}}, \bibinfo {author}
  {\bibfnamefont {Michael}\ \bibnamefont {Lange}}, \bibinfo {author}
  {\bibfnamefont {Seth}\ \bibnamefont {Lloyd}}, \bibinfo {author}
  {\bibfnamefont {Mikhail~D.}\ \bibnamefont {Lukin}}, \bibinfo {author}
  {\bibfnamefont {Dmitri}\ \bibnamefont {Maslov}}, \bibinfo {author}
  {\bibfnamefont {Peter}\ \bibnamefont {Maunz}}, \bibinfo {author}
  {\bibfnamefont {Christopher}\ \bibnamefont {Monroe}}, \bibinfo {author}
  {\bibfnamefont {John}\ \bibnamefont {Preskill}}, \bibinfo {author}
  {\bibfnamefont {Martin}\ \bibnamefont {Roetteler}}, \bibinfo {author}
  {\bibfnamefont {Martin~J.}\ \bibnamefont {Savage}}, \ and\ \bibinfo {author}
  {\bibfnamefont {Jeff}\ \bibnamefont {Thompson}},\ }\bibfield  {title}
  {\enquote {\bibinfo {title} {Quantum computer systems for scientific
  discovery},}\ }\href {\doibase 10.1103/PRXQuantum.2.017001} {\bibfield
  {journal} {\bibinfo  {journal} {PRX Quantum}\ }\textbf {\bibinfo {volume}
  {2}},\ \bibinfo {pages} {017001} (\bibinfo {year} {2021})}\BibitemShut
  {NoStop}%
\bibitem [{\citenamefont {Aidelsburger}\ \emph {et~al.}(2022)\citenamefont
  {Aidelsburger}, \citenamefont {Barbiero}, \citenamefont {Bermudez},
  \citenamefont {Chanda}, \citenamefont {Dauphin}, \citenamefont
  {González-Cuadra}, \citenamefont {Grzybowski}, \citenamefont {Hands},
  \citenamefont {Jendrzejewski}, \citenamefont {Jünemann}, \citenamefont
  {Juzeliūnas}, \citenamefont {Kasper}, \citenamefont {Piga}, \citenamefont
  {Ran}, \citenamefont {Rizzi}, \citenamefont {Sierra}, \citenamefont
  {Tagliacozzo}, \citenamefont {Tirrito}, \citenamefont {Zache}, \citenamefont
  {Zakrzewski}, \citenamefont {Zohar},\ and\ \citenamefont
  {Lewenstein}}]{aidelsburger2021cold}%
  \BibitemOpen
  \bibfield  {author} {\bibinfo {author} {\bibfnamefont {Monika}\ \bibnamefont
  {Aidelsburger}}, \bibinfo {author} {\bibfnamefont {Luca}\ \bibnamefont
  {Barbiero}}, \bibinfo {author} {\bibfnamefont {Alejandro}\ \bibnamefont
  {Bermudez}}, \bibinfo {author} {\bibfnamefont {Titas}\ \bibnamefont
  {Chanda}}, \bibinfo {author} {\bibfnamefont {Alexandre}\ \bibnamefont
  {Dauphin}}, \bibinfo {author} {\bibfnamefont {Daniel}\ \bibnamefont
  {González-Cuadra}}, \bibinfo {author} {\bibfnamefont {Przemysław~R.}\
  \bibnamefont {Grzybowski}}, \bibinfo {author} {\bibfnamefont {Simon}\
  \bibnamefont {Hands}}, \bibinfo {author} {\bibfnamefont {Fred}\ \bibnamefont
  {Jendrzejewski}}, \bibinfo {author} {\bibfnamefont {Johannes}\ \bibnamefont
  {Jünemann}}, \bibinfo {author} {\bibfnamefont {Gediminas}\ \bibnamefont
  {Juzeliūnas}}, \bibinfo {author} {\bibfnamefont {Valentin}\ \bibnamefont
  {Kasper}}, \bibinfo {author} {\bibfnamefont {Angelo}\ \bibnamefont {Piga}},
  \bibinfo {author} {\bibfnamefont {Shi-Ju}\ \bibnamefont {Ran}}, \bibinfo
  {author} {\bibfnamefont {Matteo}\ \bibnamefont {Rizzi}}, \bibinfo {author}
  {\bibfnamefont {Germán}\ \bibnamefont {Sierra}}, \bibinfo {author}
  {\bibfnamefont {Luca}\ \bibnamefont {Tagliacozzo}}, \bibinfo {author}
  {\bibfnamefont {Emanuele}\ \bibnamefont {Tirrito}}, \bibinfo {author}
  {\bibfnamefont {Torsten~V.}\ \bibnamefont {Zache}}, \bibinfo {author}
  {\bibfnamefont {Jakub}\ \bibnamefont {Zakrzewski}}, \bibinfo {author}
  {\bibfnamefont {Erez}\ \bibnamefont {Zohar}}, \ and\ \bibinfo {author}
  {\bibfnamefont {Maciej}\ \bibnamefont {Lewenstein}},\ }\bibfield  {title}
  {\enquote {\bibinfo {title} {Cold atoms meet lattice gauge theory},}\ }\href
  {\doibase 10.1098/rsta.2021.0064} {\bibfield  {journal} {\bibinfo  {journal}
  {Philosophical Transactions of the Royal Society A: Mathematical, Physical
  and Engineering Sciences}\ }\textbf {\bibinfo {volume} {380}},\ \bibinfo
  {pages} {20210064} (\bibinfo {year} {2022})}\BibitemShut {NoStop}%
\bibitem [{\citenamefont {Zohar}(2022)}]{zohar2021quantum}%
  \BibitemOpen
  \bibfield  {author} {\bibinfo {author} {\bibfnamefont {Erez}\ \bibnamefont
  {Zohar}},\ }\bibfield  {title} {\enquote {\bibinfo {title} {Quantum
  simulation of lattice gauge theories in more than one space dimension:
  requirements, challenges and methods},}\ }\href {\doibase
  10.1098/rsta.2021.0069} {\bibfield  {journal} {\bibinfo  {journal}
  {Philosophical Transactions of the Royal Society A: Mathematical, Physical
  and Engineering Sciences}\ }\textbf {\bibinfo {volume} {380}},\ \bibinfo
  {pages} {20210069} (\bibinfo {year} {2022})}\BibitemShut {NoStop}%
\bibitem [{\citenamefont {Klco}\ \emph {et~al.}(2022)\citenamefont {Klco},
  \citenamefont {Roggero},\ and\ \citenamefont {Savage}}]{klco2021standard}%
  \BibitemOpen
  \bibfield  {author} {\bibinfo {author} {\bibfnamefont {Natalie}\ \bibnamefont
  {Klco}}, \bibinfo {author} {\bibfnamefont {Alessandro}\ \bibnamefont
  {Roggero}}, \ and\ \bibinfo {author} {\bibfnamefont {Martin~J}\ \bibnamefont
  {Savage}},\ }\bibfield  {title} {\enquote {\bibinfo {title} {Standard model
  physics and the digital quantum revolution: thoughts about the interface},}\
  }\href {\doibase 10.1088/1361-6633/ac58a4} {\bibfield  {journal} {\bibinfo
  {journal} {Reports on Progress in Physics}\ }\textbf {\bibinfo {volume}
  {85}},\ \bibinfo {pages} {064301} (\bibinfo {year} {2022})}\BibitemShut
  {NoStop}%
\bibitem [{\citenamefont {Bauer}\ \emph {et~al.}(2023)\citenamefont {Bauer},
  \citenamefont {Davoudi}, \citenamefont {Balantekin}, \citenamefont
  {Bhattacharya}, \citenamefont {Carena}, \citenamefont {de~Jong},
  \citenamefont {Draper}, \citenamefont {El-Khadra}, \citenamefont {Gemelke},
  \citenamefont {Hanada}, \citenamefont {Kharzeev}, \citenamefont {Lamm},
  \citenamefont {Li}, \citenamefont {Liu}, \citenamefont {Lukin}, \citenamefont
  {Meurice}, \citenamefont {Monroe}, \citenamefont {Nachman}, \citenamefont
  {Pagano}, \citenamefont {Preskill}, \citenamefont {Rinaldi}, \citenamefont
  {Roggero}, \citenamefont {Santiago}, \citenamefont {Savage}, \citenamefont
  {Siddiqi}, \citenamefont {Siopsis}, \citenamefont {Van~Zanten}, \citenamefont
  {Wiebe}, \citenamefont {Yamauchi}, \citenamefont {Yeter-Aydeniz},\ and\
  \citenamefont {Zorzetti}}]{Bauer_review}%
  \BibitemOpen
  \bibfield  {author} {\bibinfo {author} {\bibfnamefont {Christian~W.}\
  \bibnamefont {Bauer}}, \bibinfo {author} {\bibfnamefont {Zohreh}\
  \bibnamefont {Davoudi}}, \bibinfo {author} {\bibfnamefont {A.~Baha}\
  \bibnamefont {Balantekin}}, \bibinfo {author} {\bibfnamefont {Tanmoy}\
  \bibnamefont {Bhattacharya}}, \bibinfo {author} {\bibfnamefont {Marcela}\
  \bibnamefont {Carena}}, \bibinfo {author} {\bibfnamefont {Wibe~A.}\
  \bibnamefont {de~Jong}}, \bibinfo {author} {\bibfnamefont {Patrick}\
  \bibnamefont {Draper}}, \bibinfo {author} {\bibfnamefont {Aida}\ \bibnamefont
  {El-Khadra}}, \bibinfo {author} {\bibfnamefont {Nate}\ \bibnamefont
  {Gemelke}}, \bibinfo {author} {\bibfnamefont {Masanori}\ \bibnamefont
  {Hanada}}, \bibinfo {author} {\bibfnamefont {Dmitri}\ \bibnamefont
  {Kharzeev}}, \bibinfo {author} {\bibfnamefont {Henry}\ \bibnamefont {Lamm}},
  \bibinfo {author} {\bibfnamefont {Ying-Ying}\ \bibnamefont {Li}}, \bibinfo
  {author} {\bibfnamefont {Junyu}\ \bibnamefont {Liu}}, \bibinfo {author}
  {\bibfnamefont {Mikhail}\ \bibnamefont {Lukin}}, \bibinfo {author}
  {\bibfnamefont {Yannick}\ \bibnamefont {Meurice}}, \bibinfo {author}
  {\bibfnamefont {Christopher}\ \bibnamefont {Monroe}}, \bibinfo {author}
  {\bibfnamefont {Benjamin}\ \bibnamefont {Nachman}}, \bibinfo {author}
  {\bibfnamefont {Guido}\ \bibnamefont {Pagano}}, \bibinfo {author}
  {\bibfnamefont {John}\ \bibnamefont {Preskill}}, \bibinfo {author}
  {\bibfnamefont {Enrico}\ \bibnamefont {Rinaldi}}, \bibinfo {author}
  {\bibfnamefont {Alessandro}\ \bibnamefont {Roggero}}, \bibinfo {author}
  {\bibfnamefont {David~I.}\ \bibnamefont {Santiago}}, \bibinfo {author}
  {\bibfnamefont {Martin~J.}\ \bibnamefont {Savage}}, \bibinfo {author}
  {\bibfnamefont {Irfan}\ \bibnamefont {Siddiqi}}, \bibinfo {author}
  {\bibfnamefont {George}\ \bibnamefont {Siopsis}}, \bibinfo {author}
  {\bibfnamefont {David}\ \bibnamefont {Van~Zanten}}, \bibinfo {author}
  {\bibfnamefont {Nathan}\ \bibnamefont {Wiebe}}, \bibinfo {author}
  {\bibfnamefont {Yukari}\ \bibnamefont {Yamauchi}}, \bibinfo {author}
  {\bibfnamefont {K\"ubra}\ \bibnamefont {Yeter-Aydeniz}}, \ and\ \bibinfo
  {author} {\bibfnamefont {Silvia}\ \bibnamefont {Zorzetti}},\ }\bibfield
  {title} {\enquote {\bibinfo {title} {Quantum simulation for high-energy
  physics},}\ }\href {\doibase 10.1103/PRXQuantum.4.027001} {\bibfield
  {journal} {\bibinfo  {journal} {PRX Quantum}\ }\textbf {\bibinfo {volume}
  {4}},\ \bibinfo {pages} {027001} (\bibinfo {year} {2023})}\BibitemShut
  {NoStop}%
\bibitem [{\citenamefont {Di~Meglio}\ \emph {et~al.}(2024)\citenamefont
  {Di~Meglio}, \citenamefont {Jansen}, \citenamefont {Tavernelli},
  \citenamefont {Alexandrou}, \citenamefont {Arunachalam}, \citenamefont
  {Bauer}, \citenamefont {Borras}, \citenamefont {Carrazza}, \citenamefont
  {Crippa}, \citenamefont {Croft}, \citenamefont {de~Putter}, \citenamefont
  {Delgado}, \citenamefont {Dunjko}, \citenamefont {Egger}, \citenamefont
  {Fern\'andez-Combarro}, \citenamefont {Fuchs}, \citenamefont {Funcke},
  \citenamefont {Gonz\'alez-Cuadra}, \citenamefont {Grossi}, \citenamefont
  {Halimeh}, \citenamefont {Holmes}, \citenamefont {K\"uhn}, \citenamefont
  {Lacroix}, \citenamefont {Lewis}, \citenamefont {Lucchesi}, \citenamefont
  {Martinez}, \citenamefont {Meloni}, \citenamefont {Mezzacapo}, \citenamefont
  {Montangero}, \citenamefont {Nagano}, \citenamefont {Pascuzzi}, \citenamefont
  {Radescu}, \citenamefont {Ortega}, \citenamefont {Roggero}, \citenamefont
  {Schuhmacher}, \citenamefont {Seixas}, \citenamefont {Silvi}, \citenamefont
  {Spentzouris}, \citenamefont {Tacchino}, \citenamefont {Temme}, \citenamefont
  {Terashi}, \citenamefont {Tura}, \citenamefont {T\"uys\"uz}, \citenamefont
  {Vallecorsa}, \citenamefont {Wiese}, \citenamefont {Yoo},\ and\ \citenamefont
  {Zhang}}]{dimeglio2023quantum}%
  \BibitemOpen
  \bibfield  {author} {\bibinfo {author} {\bibfnamefont {Alberto}\ \bibnamefont
  {Di~Meglio}}, \bibinfo {author} {\bibfnamefont {Karl}\ \bibnamefont
  {Jansen}}, \bibinfo {author} {\bibfnamefont {Ivano}\ \bibnamefont
  {Tavernelli}}, \bibinfo {author} {\bibfnamefont {Constantia}\ \bibnamefont
  {Alexandrou}}, \bibinfo {author} {\bibfnamefont {Srinivasan}\ \bibnamefont
  {Arunachalam}}, \bibinfo {author} {\bibfnamefont {Christian~W.}\ \bibnamefont
  {Bauer}}, \bibinfo {author} {\bibfnamefont {Kerstin}\ \bibnamefont {Borras}},
  \bibinfo {author} {\bibfnamefont {Stefano}\ \bibnamefont {Carrazza}},
  \bibinfo {author} {\bibfnamefont {Arianna}\ \bibnamefont {Crippa}}, \bibinfo
  {author} {\bibfnamefont {Vincent}\ \bibnamefont {Croft}}, \bibinfo {author}
  {\bibfnamefont {Roland}\ \bibnamefont {de~Putter}}, \bibinfo {author}
  {\bibfnamefont {Andrea}\ \bibnamefont {Delgado}}, \bibinfo {author}
  {\bibfnamefont {Vedran}\ \bibnamefont {Dunjko}}, \bibinfo {author}
  {\bibfnamefont {Daniel~J.}\ \bibnamefont {Egger}}, \bibinfo {author}
  {\bibfnamefont {Elias}\ \bibnamefont {Fern\'andez-Combarro}}, \bibinfo
  {author} {\bibfnamefont {Elina}\ \bibnamefont {Fuchs}}, \bibinfo {author}
  {\bibfnamefont {Lena}\ \bibnamefont {Funcke}}, \bibinfo {author}
  {\bibfnamefont {Daniel}\ \bibnamefont {Gonz\'alez-Cuadra}}, \bibinfo {author}
  {\bibfnamefont {Michele}\ \bibnamefont {Grossi}}, \bibinfo {author}
  {\bibfnamefont {Jad~C.}\ \bibnamefont {Halimeh}}, \bibinfo {author}
  {\bibfnamefont {Zo\"e}\ \bibnamefont {Holmes}}, \bibinfo {author}
  {\bibfnamefont {Stefan}\ \bibnamefont {K\"uhn}}, \bibinfo {author}
  {\bibfnamefont {Denis}\ \bibnamefont {Lacroix}}, \bibinfo {author}
  {\bibfnamefont {Randy}\ \bibnamefont {Lewis}}, \bibinfo {author}
  {\bibfnamefont {Donatella}\ \bibnamefont {Lucchesi}}, \bibinfo {author}
  {\bibfnamefont {Miriam~Lucio}\ \bibnamefont {Martinez}}, \bibinfo {author}
  {\bibfnamefont {Federico}\ \bibnamefont {Meloni}}, \bibinfo {author}
  {\bibfnamefont {Antonio}\ \bibnamefont {Mezzacapo}}, \bibinfo {author}
  {\bibfnamefont {Simone}\ \bibnamefont {Montangero}}, \bibinfo {author}
  {\bibfnamefont {Lento}\ \bibnamefont {Nagano}}, \bibinfo {author}
  {\bibfnamefont {Vincent~R.}\ \bibnamefont {Pascuzzi}}, \bibinfo {author}
  {\bibfnamefont {Voica}\ \bibnamefont {Radescu}}, \bibinfo {author}
  {\bibfnamefont {Enrique~Rico}\ \bibnamefont {Ortega}}, \bibinfo {author}
  {\bibfnamefont {Alessandro}\ \bibnamefont {Roggero}}, \bibinfo {author}
  {\bibfnamefont {Julian}\ \bibnamefont {Schuhmacher}}, \bibinfo {author}
  {\bibfnamefont {Joao}\ \bibnamefont {Seixas}}, \bibinfo {author}
  {\bibfnamefont {Pietro}\ \bibnamefont {Silvi}}, \bibinfo {author}
  {\bibfnamefont {Panagiotis}\ \bibnamefont {Spentzouris}}, \bibinfo {author}
  {\bibfnamefont {Francesco}\ \bibnamefont {Tacchino}}, \bibinfo {author}
  {\bibfnamefont {Kristan}\ \bibnamefont {Temme}}, \bibinfo {author}
  {\bibfnamefont {Koji}\ \bibnamefont {Terashi}}, \bibinfo {author}
  {\bibfnamefont {Jordi}\ \bibnamefont {Tura}}, \bibinfo {author}
  {\bibfnamefont {Cenk}\ \bibnamefont {T\"uys\"uz}}, \bibinfo {author}
  {\bibfnamefont {Sofia}\ \bibnamefont {Vallecorsa}}, \bibinfo {author}
  {\bibfnamefont {Uwe-Jens}\ \bibnamefont {Wiese}}, \bibinfo {author}
  {\bibfnamefont {Shinjae}\ \bibnamefont {Yoo}}, \ and\ \bibinfo {author}
  {\bibfnamefont {Jinglei}\ \bibnamefont {Zhang}},\ }\bibfield  {title}
  {\enquote {\bibinfo {title} {Quantum computing for high-energy physics: State
  of the art and challenges},}\ }\href {\doibase 10.1103/PRXQuantum.5.037001}
  {\bibfield  {journal} {\bibinfo  {journal} {PRX Quantum}\ }\textbf {\bibinfo
  {volume} {5}},\ \bibinfo {pages} {037001} (\bibinfo {year}
  {2024})}\BibitemShut {NoStop}%
\bibitem [{\citenamefont {Halimeh}\ \emph {et~al.}(2025)\citenamefont
  {Halimeh}, \citenamefont {Aidelsburger}, \citenamefont {Grusdt},
  \citenamefont {Hauke},\ and\ \citenamefont {Yang}}]{halimeh2023coldatom}%
  \BibitemOpen
  \bibfield  {author} {\bibinfo {author} {\bibfnamefont {Jad~C}\ \bibnamefont
  {Halimeh}}, \bibinfo {author} {\bibfnamefont {Monika}\ \bibnamefont
  {Aidelsburger}}, \bibinfo {author} {\bibfnamefont {Fabian}\ \bibnamefont
  {Grusdt}}, \bibinfo {author} {\bibfnamefont {Philipp}\ \bibnamefont {Hauke}},
  \ and\ \bibinfo {author} {\bibfnamefont {Bing}\ \bibnamefont {Yang}},\
  }\bibfield  {title} {\enquote {\bibinfo {title} {Cold-atom quantum simulators
  of gauge theories},}\ }\href {\doibase 10.1038/s41567-024-02721-8} {\bibfield
   {journal} {\bibinfo  {journal} {Nature Physics}\ }\textbf {\bibinfo {volume}
  {21}},\ \bibinfo {pages} {25--36} (\bibinfo {year} {2025})}\BibitemShut
  {NoStop}%
\bibitem [{\citenamefont {Cheng}\ and\ \citenamefont
  {Zhai}(2024)}]{cheng2024emergent}%
  \BibitemOpen
  \bibfield  {author} {\bibinfo {author} {\bibfnamefont {Yanting}\ \bibnamefont
  {Cheng}}\ and\ \bibinfo {author} {\bibfnamefont {Hui}\ \bibnamefont {Zhai}},\
  }\bibfield  {title} {\enquote {\bibinfo {title} {Emergent {U(1)} lattice
  gauge theory in rydberg atom arrays},}\ }\href {\doibase
  10.1038/s42254-024-00749-6} {\bibfield  {journal} {\bibinfo  {journal}
  {Nature Reviews Physics}\ }\textbf {\bibinfo {volume} {6}},\ \bibinfo {pages}
  {566--576} (\bibinfo {year} {2024})}\BibitemShut {NoStop}%
\bibitem [{\citenamefont {Martinez}\ \emph {et~al.}(2016)\citenamefont
  {Martinez}, \citenamefont {Muschik}, \citenamefont {Schindler}, \citenamefont
  {Nigg}, \citenamefont {Erhard}, \citenamefont {Heyl}, \citenamefont {Hauke},
  \citenamefont {Dalmonte}, \citenamefont {Monz}, \citenamefont {Zoller},\ and\
  \citenamefont {Blatt}}]{Martinez2016}%
  \BibitemOpen
  \bibfield  {author} {\bibinfo {author} {\bibfnamefont {Esteban~A.}\
  \bibnamefont {Martinez}}, \bibinfo {author} {\bibfnamefont {Christine~A.}\
  \bibnamefont {Muschik}}, \bibinfo {author} {\bibfnamefont {Philipp}\
  \bibnamefont {Schindler}}, \bibinfo {author} {\bibfnamefont {Daniel}\
  \bibnamefont {Nigg}}, \bibinfo {author} {\bibfnamefont {Alexander}\
  \bibnamefont {Erhard}}, \bibinfo {author} {\bibfnamefont {Markus}\
  \bibnamefont {Heyl}}, \bibinfo {author} {\bibfnamefont {Philipp}\
  \bibnamefont {Hauke}}, \bibinfo {author} {\bibfnamefont {Marcello}\
  \bibnamefont {Dalmonte}}, \bibinfo {author} {\bibfnamefont {Thomas}\
  \bibnamefont {Monz}}, \bibinfo {author} {\bibfnamefont {Peter}\ \bibnamefont
  {Zoller}}, \ and\ \bibinfo {author} {\bibfnamefont {Rainer}\ \bibnamefont
  {Blatt}},\ }\bibfield  {title} {\enquote {\bibinfo {title} {Real-time
  dynamics of lattice gauge theories with a few-qubit quantum computer},}\
  }\href {\doibase 10.1038/nature18318} {\bibfield  {journal} {\bibinfo
  {journal} {Nature}\ }\textbf {\bibinfo {volume} {534}},\ \bibinfo {pages}
  {516--519} (\bibinfo {year} {2016})}\BibitemShut {NoStop}%
\bibitem [{\citenamefont {Klco}\ \emph {et~al.}(2018)\citenamefont {Klco},
  \citenamefont {Dumitrescu}, \citenamefont {McCaskey}, \citenamefont {Morris},
  \citenamefont {Pooser}, \citenamefont {Sanz}, \citenamefont {Solano},
  \citenamefont {Lougovski},\ and\ \citenamefont {Savage}}]{Klco2018}%
  \BibitemOpen
  \bibfield  {author} {\bibinfo {author} {\bibfnamefont {N.}~\bibnamefont
  {Klco}}, \bibinfo {author} {\bibfnamefont {E.~F.}\ \bibnamefont
  {Dumitrescu}}, \bibinfo {author} {\bibfnamefont {A.~J.}\ \bibnamefont
  {McCaskey}}, \bibinfo {author} {\bibfnamefont {T.~D.}\ \bibnamefont
  {Morris}}, \bibinfo {author} {\bibfnamefont {R.~C.}\ \bibnamefont {Pooser}},
  \bibinfo {author} {\bibfnamefont {M.}~\bibnamefont {Sanz}}, \bibinfo {author}
  {\bibfnamefont {E.}~\bibnamefont {Solano}}, \bibinfo {author} {\bibfnamefont
  {P.}~\bibnamefont {Lougovski}}, \ and\ \bibinfo {author} {\bibfnamefont
  {M.~J.}\ \bibnamefont {Savage}},\ }\bibfield  {title} {\enquote {\bibinfo
  {title} {Quantum-classical computation of {Schwinger} model dynamics using
  quantum computers},}\ }\href {\doibase 10.1103/PhysRevA.98.032331} {\bibfield
   {journal} {\bibinfo  {journal} {Phys. Rev. A}\ }\textbf {\bibinfo {volume}
  {98}},\ \bibinfo {pages} {032331} (\bibinfo {year} {2018})}\BibitemShut
  {NoStop}%
\bibitem [{\citenamefont {G{\"o}rg}\ \emph {et~al.}(2019)\citenamefont
  {G{\"o}rg}, \citenamefont {Sandholzer}, \citenamefont {Minguzzi},
  \citenamefont {Desbuquois}, \citenamefont {Messer},\ and\ \citenamefont
  {Esslinger}}]{Goerg2019}%
  \BibitemOpen
  \bibfield  {author} {\bibinfo {author} {\bibfnamefont {Frederik}\
  \bibnamefont {G{\"o}rg}}, \bibinfo {author} {\bibfnamefont {Kilian}\
  \bibnamefont {Sandholzer}}, \bibinfo {author} {\bibfnamefont {Joaqu{\'\i}n}\
  \bibnamefont {Minguzzi}}, \bibinfo {author} {\bibfnamefont {R{\'e}mi}\
  \bibnamefont {Desbuquois}}, \bibinfo {author} {\bibfnamefont {Michael}\
  \bibnamefont {Messer}}, \ and\ \bibinfo {author} {\bibfnamefont {Tilman}\
  \bibnamefont {Esslinger}},\ }\bibfield  {title} {\enquote {\bibinfo {title}
  {Realization of density-dependent {Peierls} phases to engineer quantized
  gauge fields coupled to ultracold matter},}\ }\href {\doibase
  10.1038/s41567-019-0615-4} {\bibfield  {journal} {\bibinfo  {journal} {Nature
  Physics}\ }\textbf {\bibinfo {volume} {15}},\ \bibinfo {pages} {1161--1167}
  (\bibinfo {year} {2019})}\BibitemShut {NoStop}%
\bibitem [{\citenamefont {Schweizer}\ \emph {et~al.}(2019)\citenamefont
  {Schweizer}, \citenamefont {Grusdt}, \citenamefont {Berngruber},
  \citenamefont {Barbiero}, \citenamefont {Demler}, \citenamefont {Goldman},
  \citenamefont {Bloch},\ and\ \citenamefont {Aidelsburger}}]{Schweizer2019}%
  \BibitemOpen
  \bibfield  {author} {\bibinfo {author} {\bibfnamefont {Christian}\
  \bibnamefont {Schweizer}}, \bibinfo {author} {\bibfnamefont {Fabian}\
  \bibnamefont {Grusdt}}, \bibinfo {author} {\bibfnamefont {Moritz}\
  \bibnamefont {Berngruber}}, \bibinfo {author} {\bibfnamefont {Luca}\
  \bibnamefont {Barbiero}}, \bibinfo {author} {\bibfnamefont {Eugene}\
  \bibnamefont {Demler}}, \bibinfo {author} {\bibfnamefont {Nathan}\
  \bibnamefont {Goldman}}, \bibinfo {author} {\bibfnamefont {Immanuel}\
  \bibnamefont {Bloch}}, \ and\ \bibinfo {author} {\bibfnamefont {Monika}\
  \bibnamefont {Aidelsburger}},\ }\bibfield  {title} {\enquote {\bibinfo
  {title} {Floquet approach to $\mathbb{Z}$2 lattice gauge theories with
  ultracold atoms in optical lattices},}\ }\href {\doibase
  10.1038/s41567-019-0649-7} {\bibfield  {journal} {\bibinfo  {journal} {Nature
  Physics}\ }\textbf {\bibinfo {volume} {15}},\ \bibinfo {pages} {1168--1173}
  (\bibinfo {year} {2019})}\BibitemShut {NoStop}%
\bibitem [{\citenamefont {Mil}\ \emph {et~al.}(2020)\citenamefont {Mil},
  \citenamefont {Zache}, \citenamefont {Hegde}, \citenamefont {Xia},
  \citenamefont {Bhatt}, \citenamefont {Oberthaler}, \citenamefont {Hauke},
  \citenamefont {Berges},\ and\ \citenamefont {Jendrzejewski}}]{Mil2020}%
  \BibitemOpen
  \bibfield  {author} {\bibinfo {author} {\bibfnamefont {Alexander}\
  \bibnamefont {Mil}}, \bibinfo {author} {\bibfnamefont {Torsten~V.}\
  \bibnamefont {Zache}}, \bibinfo {author} {\bibfnamefont {Apoorva}\
  \bibnamefont {Hegde}}, \bibinfo {author} {\bibfnamefont {Andy}\ \bibnamefont
  {Xia}}, \bibinfo {author} {\bibfnamefont {Rohit~P.}\ \bibnamefont {Bhatt}},
  \bibinfo {author} {\bibfnamefont {Markus~K.}\ \bibnamefont {Oberthaler}},
  \bibinfo {author} {\bibfnamefont {Philipp}\ \bibnamefont {Hauke}}, \bibinfo
  {author} {\bibfnamefont {J{\"u}rgen}\ \bibnamefont {Berges}}, \ and\ \bibinfo
  {author} {\bibfnamefont {Fred}\ \bibnamefont {Jendrzejewski}},\ }\bibfield
  {title} {\enquote {\bibinfo {title} {A scalable realization of local {U(1)}
  gauge invariance in cold atomic mixtures},}\ }\href {\doibase
  10.1126/science.aaz5312} {\bibfield  {journal} {\bibinfo  {journal}
  {Science}\ }\textbf {\bibinfo {volume} {367}},\ \bibinfo {pages} {1128--1130}
  (\bibinfo {year} {2020})}\BibitemShut {NoStop}%
\bibitem [{\citenamefont {Yang}\ \emph
  {et~al.}(2020{\natexlab{a}})\citenamefont {Yang}, \citenamefont {Sun},
  \citenamefont {Ott}, \citenamefont {Wang}, \citenamefont {Zache},
  \citenamefont {Halimeh}, \citenamefont {Yuan}, \citenamefont {Hauke},\ and\
  \citenamefont {Pan}}]{Yang2020}%
  \BibitemOpen
  \bibfield  {author} {\bibinfo {author} {\bibfnamefont {Bing}\ \bibnamefont
  {Yang}}, \bibinfo {author} {\bibfnamefont {Hui}\ \bibnamefont {Sun}},
  \bibinfo {author} {\bibfnamefont {Robert}\ \bibnamefont {Ott}}, \bibinfo
  {author} {\bibfnamefont {Han-Yi}\ \bibnamefont {Wang}}, \bibinfo {author}
  {\bibfnamefont {Torsten~V.}\ \bibnamefont {Zache}}, \bibinfo {author}
  {\bibfnamefont {Jad~C.}\ \bibnamefont {Halimeh}}, \bibinfo {author}
  {\bibfnamefont {Zhen-Sheng}\ \bibnamefont {Yuan}}, \bibinfo {author}
  {\bibfnamefont {Philipp}\ \bibnamefont {Hauke}}, \ and\ \bibinfo {author}
  {\bibfnamefont {Jian-Wei}\ \bibnamefont {Pan}},\ }\bibfield  {title}
  {\enquote {\bibinfo {title} {Observation of gauge invariance in a 71-site
  {Bose--Hubbard} quantum simulator},}\ }\href {\doibase
  10.1038/s41586-020-2910-8} {\bibfield  {journal} {\bibinfo  {journal}
  {Nature}\ }\textbf {\bibinfo {volume} {587}},\ \bibinfo {pages} {392--396}
  (\bibinfo {year} {2020}{\natexlab{a}})}\BibitemShut {NoStop}%
\bibitem [{\citenamefont {Wang}\ \emph {et~al.}(2022)\citenamefont {Wang},
  \citenamefont {Ge}, \citenamefont {Xiang}, \citenamefont {Song},
  \citenamefont {Huang}, \citenamefont {Song}, \citenamefont {Guo},
  \citenamefont {Su}, \citenamefont {Xu}, \citenamefont {Zheng},\ and\
  \citenamefont {Fan}}]{Wang2021}%
  \BibitemOpen
  \bibfield  {author} {\bibinfo {author} {\bibfnamefont {Zhan}\ \bibnamefont
  {Wang}}, \bibinfo {author} {\bibfnamefont {Zi-Yong}\ \bibnamefont {Ge}},
  \bibinfo {author} {\bibfnamefont {Zhongcheng}\ \bibnamefont {Xiang}},
  \bibinfo {author} {\bibfnamefont {Xiaohui}\ \bibnamefont {Song}}, \bibinfo
  {author} {\bibfnamefont {Rui-Zhen}\ \bibnamefont {Huang}}, \bibinfo {author}
  {\bibfnamefont {Pengtao}\ \bibnamefont {Song}}, \bibinfo {author}
  {\bibfnamefont {Xue-Yi}\ \bibnamefont {Guo}}, \bibinfo {author}
  {\bibfnamefont {Luhong}\ \bibnamefont {Su}}, \bibinfo {author} {\bibfnamefont
  {Kai}\ \bibnamefont {Xu}}, \bibinfo {author} {\bibfnamefont {Dongning}\
  \bibnamefont {Zheng}}, \ and\ \bibinfo {author} {\bibfnamefont {Heng}\
  \bibnamefont {Fan}},\ }\bibfield  {title} {\enquote {\bibinfo {title}
  {Observation of emergent {${\mathbb{Z}}_{2}$} gauge invariance in a
  superconducting circuit},}\ }\href {\doibase
  10.1103/PhysRevResearch.4.L022060} {\bibfield  {journal} {\bibinfo  {journal}
  {Phys. Rev. Research}\ }\textbf {\bibinfo {volume} {4}},\ \bibinfo {pages}
  {L022060} (\bibinfo {year} {2022})}\BibitemShut {NoStop}%
\bibitem [{\citenamefont {Zhou}\ \emph {et~al.}(2022)\citenamefont {Zhou},
  \citenamefont {Su}, \citenamefont {Halimeh}, \citenamefont {Ott},
  \citenamefont {Sun}, \citenamefont {Hauke}, \citenamefont {Yang},
  \citenamefont {Yuan}, \citenamefont {Berges},\ and\ \citenamefont
  {Pan}}]{Zhou2022}%
  \BibitemOpen
  \bibfield  {author} {\bibinfo {author} {\bibfnamefont {Zhao-Yu}\ \bibnamefont
  {Zhou}}, \bibinfo {author} {\bibfnamefont {Guo-Xian}\ \bibnamefont {Su}},
  \bibinfo {author} {\bibfnamefont {Jad~C.}\ \bibnamefont {Halimeh}}, \bibinfo
  {author} {\bibfnamefont {Robert}\ \bibnamefont {Ott}}, \bibinfo {author}
  {\bibfnamefont {Hui}\ \bibnamefont {Sun}}, \bibinfo {author} {\bibfnamefont
  {Philipp}\ \bibnamefont {Hauke}}, \bibinfo {author} {\bibfnamefont {Bing}\
  \bibnamefont {Yang}}, \bibinfo {author} {\bibfnamefont {Zhen-Sheng}\
  \bibnamefont {Yuan}}, \bibinfo {author} {\bibfnamefont {Jürgen}\
  \bibnamefont {Berges}}, \ and\ \bibinfo {author} {\bibfnamefont {Jian-Wei}\
  \bibnamefont {Pan}},\ }\bibfield  {title} {\enquote {\bibinfo {title}
  {Thermalization dynamics of a gauge theory on a quantum simulator},}\ }\href
  {\doibase 10.1126/science.abl6277} {\bibfield  {journal} {\bibinfo  {journal}
  {Science}\ }\textbf {\bibinfo {volume} {377}},\ \bibinfo {pages} {311--314}
  (\bibinfo {year} {2022})}\BibitemShut {NoStop}%
\bibitem [{\citenamefont {Mildenberger}\ \emph {et~al.}(2025)\citenamefont
  {Mildenberger}, \citenamefont {Mruczkiewicz}, \citenamefont {Halimeh},
  \citenamefont {Jiang},\ and\ \citenamefont {Hauke}}]{Mildenberger2022}%
  \BibitemOpen
  \bibfield  {author} {\bibinfo {author} {\bibfnamefont {Julius}\ \bibnamefont
  {Mildenberger}}, \bibinfo {author} {\bibfnamefont {Wojciech}\ \bibnamefont
  {Mruczkiewicz}}, \bibinfo {author} {\bibfnamefont {Jad~C.}\ \bibnamefont
  {Halimeh}}, \bibinfo {author} {\bibfnamefont {Zhang}\ \bibnamefont {Jiang}},
  \ and\ \bibinfo {author} {\bibfnamefont {Philipp}\ \bibnamefont {Hauke}},\
  }\bibfield  {title} {\enquote {\bibinfo {title} {Confinement in a
  {$\mathbb{Z}_2$} lattice gauge theory on a quantum computer},}\ }\href
  {\doibase 10.1038/s41567-024-02723-6} {\bibfield  {journal} {\bibinfo
  {journal} {Nature Physics}\ }\textbf {\bibinfo {volume} {21}},\ \bibinfo
  {pages} {312--317} (\bibinfo {year} {2025})}\BibitemShut {NoStop}%
\bibitem [{\citenamefont {Zhang}\ \emph {et~al.}(2025)\citenamefont {Zhang},
  \citenamefont {Liu}, \citenamefont {Cheng}, \citenamefont {He}, \citenamefont
  {Wang}, \citenamefont {Wang}, \citenamefont {Zhu}, \citenamefont {Su},
  \citenamefont {Zhou}, \citenamefont {Zheng}, \citenamefont {Sun},
  \citenamefont {Yang}, \citenamefont {Hauke}, \citenamefont {Zheng},
  \citenamefont {Halimeh}, \citenamefont {Yuan},\ and\ \citenamefont
  {Pan}}]{Zhang2023observation}%
  \BibitemOpen
  \bibfield  {author} {\bibinfo {author} {\bibfnamefont {Wei-Yong}\
  \bibnamefont {Zhang}}, \bibinfo {author} {\bibfnamefont {Ying}\ \bibnamefont
  {Liu}}, \bibinfo {author} {\bibfnamefont {Yanting}\ \bibnamefont {Cheng}},
  \bibinfo {author} {\bibfnamefont {Ming-Gen}\ \bibnamefont {He}}, \bibinfo
  {author} {\bibfnamefont {Han-Yi}\ \bibnamefont {Wang}}, \bibinfo {author}
  {\bibfnamefont {Tian-Yi}\ \bibnamefont {Wang}}, \bibinfo {author}
  {\bibfnamefont {Zi-Hang}\ \bibnamefont {Zhu}}, \bibinfo {author}
  {\bibfnamefont {Guo-Xian}\ \bibnamefont {Su}}, \bibinfo {author}
  {\bibfnamefont {Zhao-Yu}\ \bibnamefont {Zhou}}, \bibinfo {author}
  {\bibfnamefont {Yong-Guang}\ \bibnamefont {Zheng}}, \bibinfo {author}
  {\bibfnamefont {Hui}\ \bibnamefont {Sun}}, \bibinfo {author} {\bibfnamefont
  {Bing}\ \bibnamefont {Yang}}, \bibinfo {author} {\bibfnamefont {Philipp}\
  \bibnamefont {Hauke}}, \bibinfo {author} {\bibfnamefont {Wei}\ \bibnamefont
  {Zheng}}, \bibinfo {author} {\bibfnamefont {Jad~C.}\ \bibnamefont {Halimeh}},
  \bibinfo {author} {\bibfnamefont {Zhen-Sheng}\ \bibnamefont {Yuan}}, \ and\
  \bibinfo {author} {\bibfnamefont {Jian-Wei}\ \bibnamefont {Pan}},\ }\bibfield
   {title} {\enquote {\bibinfo {title} {Observation of microscopic confinement
  dynamics by a tunable topological {$\theta$}-angle},}\ }\href {\doibase
  10.1038/s41567-024-02702-x} {\bibfield  {journal} {\bibinfo  {journal}
  {Nature Physics}\ }\textbf {\bibinfo {volume} {21}},\ \bibinfo {pages}
  {155--160} (\bibinfo {year} {2025})}\BibitemShut {NoStop}%
\bibitem [{\citenamefont {Farrell}\ \emph {et~al.}(2024)\citenamefont
  {Farrell}, \citenamefont {Illa}, \citenamefont {Ciavarella},\ and\
  \citenamefont {Savage}}]{farrell2023scalable}%
  \BibitemOpen
  \bibfield  {author} {\bibinfo {author} {\bibfnamefont {Roland~C.}\
  \bibnamefont {Farrell}}, \bibinfo {author} {\bibfnamefont {Marc}\
  \bibnamefont {Illa}}, \bibinfo {author} {\bibfnamefont {Anthony~N.}\
  \bibnamefont {Ciavarella}}, \ and\ \bibinfo {author} {\bibfnamefont
  {Martin~J.}\ \bibnamefont {Savage}},\ }\bibfield  {title} {\enquote {\bibinfo
  {title} {Scalable circuits for preparing ground states on digital quantum
  computers: The {Schwinger} model vacuum on 100 qubits},}\ }\href {\doibase
  10.1103/PRXQuantum.5.020315} {\bibfield  {journal} {\bibinfo  {journal} {PRX
  Quantum}\ }\textbf {\bibinfo {volume} {5}},\ \bibinfo {pages} {020315}
  (\bibinfo {year} {2024})}\BibitemShut {NoStop}%
\bibitem [{\citenamefont {Angelides}\ \emph {et~al.}(2025)\citenamefont
  {Angelides}, \citenamefont {Naredi}, \citenamefont {Crippa}, \citenamefont
  {Jansen}, \citenamefont {K{\"u}hn}, \citenamefont {Tavernelli},\ and\
  \citenamefont {Wang}}]{angelides2023firstorder}%
  \BibitemOpen
  \bibfield  {author} {\bibinfo {author} {\bibfnamefont {Takis}\ \bibnamefont
  {Angelides}}, \bibinfo {author} {\bibfnamefont {Pranay}\ \bibnamefont
  {Naredi}}, \bibinfo {author} {\bibfnamefont {Arianna}\ \bibnamefont
  {Crippa}}, \bibinfo {author} {\bibfnamefont {Karl}\ \bibnamefont {Jansen}},
  \bibinfo {author} {\bibfnamefont {Stefan}\ \bibnamefont {K{\"u}hn}}, \bibinfo
  {author} {\bibfnamefont {Ivano}\ \bibnamefont {Tavernelli}}, \ and\ \bibinfo
  {author} {\bibfnamefont {Derek~S.}\ \bibnamefont {Wang}},\ }\bibfield
  {title} {\enquote {\bibinfo {title} {First-order phase transition of the
  {Schwinger} model with a quantum computer},}\ }\href {\doibase
  10.1038/s41534-024-00950-6} {\bibfield  {journal} {\bibinfo  {journal} {npj
  Quantum Information}\ }\textbf {\bibinfo {volume} {11}},\ \bibinfo {pages}
  {6} (\bibinfo {year} {2025})}\BibitemShut {NoStop}%
\bibitem [{\citenamefont {Charles}\ \emph {et~al.}(2024)\citenamefont
  {Charles}, \citenamefont {Gustafson}, \citenamefont {Hardt}, \citenamefont
  {Herren}, \citenamefont {Hogan}, \citenamefont {Lamm}, \citenamefont
  {Starecheski}, \citenamefont {Van~de Water},\ and\ \citenamefont
  {Wagman}}]{charles2023simulating}%
  \BibitemOpen
  \bibfield  {author} {\bibinfo {author} {\bibfnamefont {Clement}\ \bibnamefont
  {Charles}}, \bibinfo {author} {\bibfnamefont {Erik~J.}\ \bibnamefont
  {Gustafson}}, \bibinfo {author} {\bibfnamefont {Elizabeth}\ \bibnamefont
  {Hardt}}, \bibinfo {author} {\bibfnamefont {Florian}\ \bibnamefont {Herren}},
  \bibinfo {author} {\bibfnamefont {Norman}\ \bibnamefont {Hogan}}, \bibinfo
  {author} {\bibfnamefont {Henry}\ \bibnamefont {Lamm}}, \bibinfo {author}
  {\bibfnamefont {Sara}\ \bibnamefont {Starecheski}}, \bibinfo {author}
  {\bibfnamefont {Ruth~S.}\ \bibnamefont {Van~de Water}}, \ and\ \bibinfo
  {author} {\bibfnamefont {Michael~L.}\ \bibnamefont {Wagman}},\ }\bibfield
  {title} {\enquote {\bibinfo {title} {Simulating {${\mathbb{Z}}_{2}$} lattice
  gauge theory on a quantum computer},}\ }\href {\doibase
  10.1103/PhysRevE.109.015307} {\bibfield  {journal} {\bibinfo  {journal}
  {Phys. Rev. E}\ }\textbf {\bibinfo {volume} {109}},\ \bibinfo {pages}
  {015307} (\bibinfo {year} {2024})}\BibitemShut {NoStop}%
\bibitem [{\citenamefont {Halimeh}\ \emph {et~al.}(2022)\citenamefont
  {Halimeh}, \citenamefont {McCulloch}, \citenamefont {Yang},\ and\
  \citenamefont {Hauke}}]{Halimeh2022tuning}%
  \BibitemOpen
  \bibfield  {author} {\bibinfo {author} {\bibfnamefont {Jad~C.}\ \bibnamefont
  {Halimeh}}, \bibinfo {author} {\bibfnamefont {Ian~P.}\ \bibnamefont
  {McCulloch}}, \bibinfo {author} {\bibfnamefont {Bing}\ \bibnamefont {Yang}},
  \ and\ \bibinfo {author} {\bibfnamefont {Philipp}\ \bibnamefont {Hauke}},\
  }\bibfield  {title} {\enquote {\bibinfo {title} {Tuning the topological
  $\ensuremath{\theta}$-angle in cold-atom quantum simulators of gauge
  theories},}\ }\href {\doibase 10.1103/PRXQuantum.3.040316} {\bibfield
  {journal} {\bibinfo  {journal} {PRX Quantum}\ }\textbf {\bibinfo {volume}
  {3}},\ \bibinfo {pages} {040316} (\bibinfo {year} {2022})}\BibitemShut
  {NoStop}%
\bibitem [{\citenamefont {Cheng}\ \emph {et~al.}(2022)\citenamefont {Cheng},
  \citenamefont {Liu}, \citenamefont {Zheng}, \citenamefont {Zhang},\ and\
  \citenamefont {Zhai}}]{Cheng2022tunable}%
  \BibitemOpen
  \bibfield  {author} {\bibinfo {author} {\bibfnamefont {Yanting}\ \bibnamefont
  {Cheng}}, \bibinfo {author} {\bibfnamefont {Shang}\ \bibnamefont {Liu}},
  \bibinfo {author} {\bibfnamefont {Wei}\ \bibnamefont {Zheng}}, \bibinfo
  {author} {\bibfnamefont {Pengfei}\ \bibnamefont {Zhang}}, \ and\ \bibinfo
  {author} {\bibfnamefont {Hui}\ \bibnamefont {Zhai}},\ }\bibfield  {title}
  {\enquote {\bibinfo {title} {Tunable confinement-deconfinement transition in
  an ultracold-atom quantum simulator},}\ }\href {\doibase
  10.1103/PRXQuantum.3.040317} {\bibfield  {journal} {\bibinfo  {journal} {PRX
  Quantum}\ }\textbf {\bibinfo {volume} {3}},\ \bibinfo {pages} {040317}
  (\bibinfo {year} {2022})}\BibitemShut {NoStop}%
\bibitem [{\citenamefont {Bernien}\ \emph {et~al.}(2017)\citenamefont
  {Bernien}, \citenamefont {Schwartz}, \citenamefont {Keesling}, \citenamefont
  {Levine}, \citenamefont {Omran}, \citenamefont {Pichler}, \citenamefont
  {Choi}, \citenamefont {Zibrov}, \citenamefont {Endres}, \citenamefont
  {Greiner}, \citenamefont {Vuleti{\'c}},\ and\ \citenamefont
  {Lukin}}]{Bernien2017}%
  \BibitemOpen
  \bibfield  {author} {\bibinfo {author} {\bibfnamefont {Hannes}\ \bibnamefont
  {Bernien}}, \bibinfo {author} {\bibfnamefont {Sylvain}\ \bibnamefont
  {Schwartz}}, \bibinfo {author} {\bibfnamefont {Alexander}\ \bibnamefont
  {Keesling}}, \bibinfo {author} {\bibfnamefont {Harry}\ \bibnamefont
  {Levine}}, \bibinfo {author} {\bibfnamefont {Ahmed}\ \bibnamefont {Omran}},
  \bibinfo {author} {\bibfnamefont {Hannes}\ \bibnamefont {Pichler}}, \bibinfo
  {author} {\bibfnamefont {Soonwon}\ \bibnamefont {Choi}}, \bibinfo {author}
  {\bibfnamefont {Alexander~S.}\ \bibnamefont {Zibrov}}, \bibinfo {author}
  {\bibfnamefont {Manuel}\ \bibnamefont {Endres}}, \bibinfo {author}
  {\bibfnamefont {Markus}\ \bibnamefont {Greiner}}, \bibinfo {author}
  {\bibfnamefont {Vladan}\ \bibnamefont {Vuleti{\'c}}}, \ and\ \bibinfo
  {author} {\bibfnamefont {Mikhail~D.}\ \bibnamefont {Lukin}},\ }\bibfield
  {title} {\enquote {\bibinfo {title} {Probing many-body dynamics on a 51-atom
  quantum simulator},}\ }\href {\doibase 10.1038/nature24622} {\bibfield
  {journal} {\bibinfo  {journal} {Nature}\ }\textbf {\bibinfo {volume} {551}},\
  \bibinfo {pages} {579--584} (\bibinfo {year} {2017})}\BibitemShut {NoStop}%
\bibitem [{\citenamefont {Turner}\ \emph {et~al.}(2018)\citenamefont {Turner},
  \citenamefont {Michailidis}, \citenamefont {Abanin}, \citenamefont {Serbyn},\
  and\ \citenamefont {Papi{\'c}}}]{Turner2018}%
  \BibitemOpen
  \bibfield  {author} {\bibinfo {author} {\bibfnamefont {C.~J.}\ \bibnamefont
  {Turner}}, \bibinfo {author} {\bibfnamefont {A.~A.}\ \bibnamefont
  {Michailidis}}, \bibinfo {author} {\bibfnamefont {D.~A.}\ \bibnamefont
  {Abanin}}, \bibinfo {author} {\bibfnamefont {M.}~\bibnamefont {Serbyn}}, \
  and\ \bibinfo {author} {\bibfnamefont {Z.}~\bibnamefont {Papi{\'c}}},\
  }\bibfield  {title} {\enquote {\bibinfo {title} {Weak ergodicity breaking
  from quantum many-body scars},}\ }\href {\doibase 10.1038/s41567-018-0137-5}
  {\bibfield  {journal} {\bibinfo  {journal} {Nature Physics}\ }\textbf
  {\bibinfo {volume} {14}},\ \bibinfo {pages} {745--749} (\bibinfo {year}
  {2018})}\BibitemShut {NoStop}%
\bibitem [{\citenamefont {Moudgalya}\ \emph
  {et~al.}(2018{\natexlab{a}})\citenamefont {Moudgalya}, \citenamefont
  {Rachel}, \citenamefont {Bernevig},\ and\ \citenamefont
  {Regnault}}]{Moudgalya2018}%
  \BibitemOpen
  \bibfield  {author} {\bibinfo {author} {\bibfnamefont {Sanjay}\ \bibnamefont
  {Moudgalya}}, \bibinfo {author} {\bibfnamefont {Stephan}\ \bibnamefont
  {Rachel}}, \bibinfo {author} {\bibfnamefont {B.~Andrei}\ \bibnamefont
  {Bernevig}}, \ and\ \bibinfo {author} {\bibfnamefont {Nicolas}\ \bibnamefont
  {Regnault}},\ }\bibfield  {title} {\enquote {\bibinfo {title} {Exact excited
  states of nonintegrable models},}\ }\href {\doibase
  10.1103/PhysRevB.98.235155} {\bibfield  {journal} {\bibinfo  {journal} {Phys.
  Rev. B}\ }\textbf {\bibinfo {volume} {98}},\ \bibinfo {pages} {235155}
  (\bibinfo {year} {2018}{\natexlab{a}})}\BibitemShut {NoStop}%
\bibitem [{\citenamefont {Zhao}\ \emph {et~al.}(2020)\citenamefont {Zhao},
  \citenamefont {Vovrosh}, \citenamefont {Mintert},\ and\ \citenamefont
  {Knolle}}]{Zhao2020}%
  \BibitemOpen
  \bibfield  {author} {\bibinfo {author} {\bibfnamefont {Hongzheng}\
  \bibnamefont {Zhao}}, \bibinfo {author} {\bibfnamefont {Joseph}\ \bibnamefont
  {Vovrosh}}, \bibinfo {author} {\bibfnamefont {Florian}\ \bibnamefont
  {Mintert}}, \ and\ \bibinfo {author} {\bibfnamefont {Johannes}\ \bibnamefont
  {Knolle}},\ }\bibfield  {title} {\enquote {\bibinfo {title} {Quantum
  many-body scars in optical lattices},}\ }\href {\doibase
  10.1103/PhysRevLett.124.160604} {\bibfield  {journal} {\bibinfo  {journal}
  {Phys. Rev. Lett.}\ }\textbf {\bibinfo {volume} {124}},\ \bibinfo {pages}
  {160604} (\bibinfo {year} {2020})}\BibitemShut {NoStop}%
\bibitem [{\citenamefont {Jepsen}\ \emph {et~al.}(2022)\citenamefont {Jepsen},
  \citenamefont {Lee}, \citenamefont {Lin}, \citenamefont {Dimitrova},
  \citenamefont {Margalit}, \citenamefont {Ho},\ and\ \citenamefont
  {Ketterle}}]{Jepsen2021}%
  \BibitemOpen
  \bibfield  {author} {\bibinfo {author} {\bibfnamefont {Paul~Niklas}\
  \bibnamefont {Jepsen}}, \bibinfo {author} {\bibfnamefont {Yoo
  Kyung~`Eunice'}\ \bibnamefont {Lee}}, \bibinfo {author} {\bibfnamefont
  {Hanzhen}\ \bibnamefont {Lin}}, \bibinfo {author} {\bibfnamefont {Ivana}\
  \bibnamefont {Dimitrova}}, \bibinfo {author} {\bibfnamefont {Yair}\
  \bibnamefont {Margalit}}, \bibinfo {author} {\bibfnamefont {Wen~Wei}\
  \bibnamefont {Ho}}, \ and\ \bibinfo {author} {\bibfnamefont {Wolfgang}\
  \bibnamefont {Ketterle}},\ }\bibfield  {title} {\enquote {\bibinfo {title}
  {Long-lived phantom helix states in {Heisenberg} quantum magnets},}\ }\href
  {\doibase 10.1038/s41567-022-01651-7} {\bibfield  {journal} {\bibinfo
  {journal} {Nature Physics}\ }\textbf {\bibinfo {volume} {18}},\ \bibinfo
  {pages} {899--904} (\bibinfo {year} {2022})}\BibitemShut {NoStop}%
\bibitem [{\citenamefont {Serbyn}\ \emph {et~al.}(2021)\citenamefont {Serbyn},
  \citenamefont {Abanin},\ and\ \citenamefont {Papi{\'c}}}]{Serbyn2020}%
  \BibitemOpen
  \bibfield  {author} {\bibinfo {author} {\bibfnamefont {Maksym}\ \bibnamefont
  {Serbyn}}, \bibinfo {author} {\bibfnamefont {Dmitry~A.}\ \bibnamefont
  {Abanin}}, \ and\ \bibinfo {author} {\bibfnamefont {Zlatko}\ \bibnamefont
  {Papi{\'c}}},\ }\bibfield  {title} {\enquote {\bibinfo {title} {Quantum
  many-body scars and weak breaking of ergodicity},}\ }\href {\doibase
  10.1038/s41567-021-01230-2} {\bibfield  {journal} {\bibinfo  {journal}
  {Nature Physics}\ }\textbf {\bibinfo {volume} {17}},\ \bibinfo {pages}
  {675--685} (\bibinfo {year} {2021})}\BibitemShut {NoStop}%
\bibitem [{\citenamefont {Moudgalya}\ \emph {et~al.}(2022)\citenamefont
  {Moudgalya}, \citenamefont {Bernevig},\ and\ \citenamefont
  {Regnault}}]{Moudgalya_review}%
  \BibitemOpen
  \bibfield  {author} {\bibinfo {author} {\bibfnamefont {Sanjay}\ \bibnamefont
  {Moudgalya}}, \bibinfo {author} {\bibfnamefont {B~Andrei}\ \bibnamefont
  {Bernevig}}, \ and\ \bibinfo {author} {\bibfnamefont {Nicolas}\ \bibnamefont
  {Regnault}},\ }\bibfield  {title} {\enquote {\bibinfo {title} {Quantum
  many-body scars and {Hilbert} space fragmentation: a review of exact
  results},}\ }\href {\doibase 10.1088/1361-6633/ac73a0} {\bibfield  {journal}
  {\bibinfo  {journal} {Reports on Progress in Physics}\ }\textbf {\bibinfo
  {volume} {85}},\ \bibinfo {pages} {086501} (\bibinfo {year}
  {2022})}\BibitemShut {NoStop}%
\bibitem [{\citenamefont {Chandran}\ \emph {et~al.}(2023)\citenamefont
  {Chandran}, \citenamefont {Iadecola}, \citenamefont {Khemani},\ and\
  \citenamefont {Moessner}}]{Chandran_review}%
  \BibitemOpen
  \bibfield  {author} {\bibinfo {author} {\bibfnamefont {Anushya}\ \bibnamefont
  {Chandran}}, \bibinfo {author} {\bibfnamefont {Thomas}\ \bibnamefont
  {Iadecola}}, \bibinfo {author} {\bibfnamefont {Vedika}\ \bibnamefont
  {Khemani}}, \ and\ \bibinfo {author} {\bibfnamefont {Roderich}\ \bibnamefont
  {Moessner}},\ }\bibfield  {title} {\enquote {\bibinfo {title} {Quantum
  many-body scars: A quasiparticle perspective},}\ }\href {\doibase
  10.1146/annurev-conmatphys-031620-101617} {\bibfield  {journal} {\bibinfo
  {journal} {Annual Review of Condensed Matter Physics}\ }\textbf {\bibinfo
  {volume} {14}},\ \bibinfo {pages} {443--469} (\bibinfo {year}
  {2023})}\BibitemShut {NoStop}%
\bibitem [{\citenamefont {Moudgalya}\ \emph
  {et~al.}(2018{\natexlab{b}})\citenamefont {Moudgalya}, \citenamefont
  {Regnault},\ and\ \citenamefont {Bernevig}}]{BernevigEnt}%
  \BibitemOpen
  \bibfield  {author} {\bibinfo {author} {\bibfnamefont {Sanjay}\ \bibnamefont
  {Moudgalya}}, \bibinfo {author} {\bibfnamefont {Nicolas}\ \bibnamefont
  {Regnault}}, \ and\ \bibinfo {author} {\bibfnamefont {B.~Andrei}\
  \bibnamefont {Bernevig}},\ }\bibfield  {title} {\enquote {\bibinfo {title}
  {Entanglement of exact excited states of {Affleck-Kennedy-Lieb-Tasaki}
  models: Exact results, many-body scars, and violation of the strong
  eigenstate thermalization hypothesis},}\ }\href {\doibase
  10.1103/PhysRevB.98.235156} {\bibfield  {journal} {\bibinfo  {journal} {Phys.
  Rev. B}\ }\textbf {\bibinfo {volume} {98}},\ \bibinfo {pages} {235156}
  (\bibinfo {year} {2018}{\natexlab{b}})}\BibitemShut {NoStop}%
\bibitem [{\citenamefont {Schecter}\ and\ \citenamefont
  {Iadecola}(2019)}]{Schecter2019}%
  \BibitemOpen
  \bibfield  {author} {\bibinfo {author} {\bibfnamefont {Michael}\ \bibnamefont
  {Schecter}}\ and\ \bibinfo {author} {\bibfnamefont {Thomas}\ \bibnamefont
  {Iadecola}},\ }\bibfield  {title} {\enquote {\bibinfo {title} {Weak
  ergodicity breaking and quantum many-body scars in spin-1 {$XY$} magnets},}\
  }\href {\doibase 10.1103/PhysRevLett.123.147201} {\bibfield  {journal}
  {\bibinfo  {journal} {Phys. Rev. Lett.}\ }\textbf {\bibinfo {volume} {123}},\
  \bibinfo {pages} {147201} (\bibinfo {year} {2019})}\BibitemShut {NoStop}%
\bibitem [{\citenamefont {Lin}\ and\ \citenamefont
  {Motrunich}(2019)}]{lin2018exact}%
  \BibitemOpen
  \bibfield  {author} {\bibinfo {author} {\bibfnamefont {Cheng-Ju}\
  \bibnamefont {Lin}}\ and\ \bibinfo {author} {\bibfnamefont {Olexei~I.}\
  \bibnamefont {Motrunich}},\ }\bibfield  {title} {\enquote {\bibinfo {title}
  {Exact quantum many-body scar states in the {Rydberg}-blockaded atom
  chain},}\ }\href {\doibase 10.1103/PhysRevLett.122.173401} {\bibfield
  {journal} {\bibinfo  {journal} {Phys. Rev. Lett.}\ }\textbf {\bibinfo
  {volume} {122}},\ \bibinfo {pages} {173401} (\bibinfo {year}
  {2019})}\BibitemShut {NoStop}%
\bibitem [{\citenamefont {Bluvstein}\ \emph {et~al.}(2021)\citenamefont
  {Bluvstein}, \citenamefont {Omran}, \citenamefont {Levine}, \citenamefont
  {Keesling}, \citenamefont {Semeghini}, \citenamefont {Ebadi}, \citenamefont
  {Wang}, \citenamefont {Michailidis}, \citenamefont {Maskara}, \citenamefont
  {Ho}, \citenamefont {Choi}, \citenamefont {Serbyn}, \citenamefont {Greiner},
  \citenamefont {Vuletić},\ and\ \citenamefont {Lukin}}]{Bluvstein2021}%
  \BibitemOpen
  \bibfield  {author} {\bibinfo {author} {\bibfnamefont {D.}~\bibnamefont
  {Bluvstein}}, \bibinfo {author} {\bibfnamefont {A.}~\bibnamefont {Omran}},
  \bibinfo {author} {\bibfnamefont {H.}~\bibnamefont {Levine}}, \bibinfo
  {author} {\bibfnamefont {A.}~\bibnamefont {Keesling}}, \bibinfo {author}
  {\bibfnamefont {G.}~\bibnamefont {Semeghini}}, \bibinfo {author}
  {\bibfnamefont {S.}~\bibnamefont {Ebadi}}, \bibinfo {author} {\bibfnamefont
  {T.~T.}\ \bibnamefont {Wang}}, \bibinfo {author} {\bibfnamefont {A.~A.}\
  \bibnamefont {Michailidis}}, \bibinfo {author} {\bibfnamefont
  {N.}~\bibnamefont {Maskara}}, \bibinfo {author} {\bibfnamefont {W.~W.}\
  \bibnamefont {Ho}}, \bibinfo {author} {\bibfnamefont {S.}~\bibnamefont
  {Choi}}, \bibinfo {author} {\bibfnamefont {M.}~\bibnamefont {Serbyn}},
  \bibinfo {author} {\bibfnamefont {M.}~\bibnamefont {Greiner}}, \bibinfo
  {author} {\bibfnamefont {V.}~\bibnamefont {Vuletić}}, \ and\ \bibinfo
  {author} {\bibfnamefont {M.~D.}\ \bibnamefont {Lukin}},\ }\bibfield  {title}
  {\enquote {\bibinfo {title} {Controlling quantum many-body dynamics in driven
  {Rydberg} atom arrays},}\ }\href {\doibase 10.1126/science.abg2530}
  {\bibfield  {journal} {\bibinfo  {journal} {Science}\ }\textbf {\bibinfo
  {volume} {371}},\ \bibinfo {pages} {1355--1359} (\bibinfo {year}
  {2021})}\BibitemShut {NoStop}%
\bibitem [{\citenamefont {Bluvstein}\ \emph {et~al.}(2022)\citenamefont
  {Bluvstein}, \citenamefont {Levine}, \citenamefont {Semeghini}, \citenamefont
  {Wang}, \citenamefont {Ebadi}, \citenamefont {Kalinowski}, \citenamefont
  {Keesling}, \citenamefont {Maskara}, \citenamefont {Pichler}, \citenamefont
  {Greiner}, \citenamefont {Vuleti{\'c}},\ and\ \citenamefont
  {Lukin}}]{Bluvstein2022quantum}%
  \BibitemOpen
  \bibfield  {author} {\bibinfo {author} {\bibfnamefont {Dolev}\ \bibnamefont
  {Bluvstein}}, \bibinfo {author} {\bibfnamefont {Harry}\ \bibnamefont
  {Levine}}, \bibinfo {author} {\bibfnamefont {Giulia}\ \bibnamefont
  {Semeghini}}, \bibinfo {author} {\bibfnamefont {Tout~T.}\ \bibnamefont
  {Wang}}, \bibinfo {author} {\bibfnamefont {Sepehr}\ \bibnamefont {Ebadi}},
  \bibinfo {author} {\bibfnamefont {Marcin}\ \bibnamefont {Kalinowski}},
  \bibinfo {author} {\bibfnamefont {Alexander}\ \bibnamefont {Keesling}},
  \bibinfo {author} {\bibfnamefont {Nishad}\ \bibnamefont {Maskara}}, \bibinfo
  {author} {\bibfnamefont {Hannes}\ \bibnamefont {Pichler}}, \bibinfo {author}
  {\bibfnamefont {Markus}\ \bibnamefont {Greiner}}, \bibinfo {author}
  {\bibfnamefont {Vladan}\ \bibnamefont {Vuleti{\'c}}}, \ and\ \bibinfo
  {author} {\bibfnamefont {Mikhail~D.}\ \bibnamefont {Lukin}},\ }\bibfield
  {title} {\enquote {\bibinfo {title} {A quantum processor based on coherent
  transport of entangled atom arrays},}\ }\href {\doibase
  10.1038/s41586-022-04592-6} {\bibfield  {journal} {\bibinfo  {journal}
  {Nature}\ }\textbf {\bibinfo {volume} {604}},\ \bibinfo {pages} {451--456}
  (\bibinfo {year} {2022})}\BibitemShut {NoStop}%
\bibitem [{\citenamefont {Su}\ \emph {et~al.}(2023)\citenamefont {Su},
  \citenamefont {Sun}, \citenamefont {Hudomal}, \citenamefont {Desaules},
  \citenamefont {Zhou}, \citenamefont {Yang}, \citenamefont {Halimeh},
  \citenamefont {Yuan}, \citenamefont {Papi\ifmmode~\acute{c}\else
  \'{c}\fi{}},\ and\ \citenamefont {Pan}}]{Su2022}%
  \BibitemOpen
  \bibfield  {author} {\bibinfo {author} {\bibfnamefont {Guo-Xian}\
  \bibnamefont {Su}}, \bibinfo {author} {\bibfnamefont {Hui}\ \bibnamefont
  {Sun}}, \bibinfo {author} {\bibfnamefont {Ana}\ \bibnamefont {Hudomal}},
  \bibinfo {author} {\bibfnamefont {Jean-Yves}\ \bibnamefont {Desaules}},
  \bibinfo {author} {\bibfnamefont {Zhao-Yu}\ \bibnamefont {Zhou}}, \bibinfo
  {author} {\bibfnamefont {Bing}\ \bibnamefont {Yang}}, \bibinfo {author}
  {\bibfnamefont {Jad~C.}\ \bibnamefont {Halimeh}}, \bibinfo {author}
  {\bibfnamefont {Zhen-Sheng}\ \bibnamefont {Yuan}}, \bibinfo {author}
  {\bibfnamefont {Zlatko}\ \bibnamefont {Papi\ifmmode~\acute{c}\else
  \'{c}\fi{}}}, \ and\ \bibinfo {author} {\bibfnamefont {Jian-Wei}\
  \bibnamefont {Pan}},\ }\bibfield  {title} {\enquote {\bibinfo {title}
  {Observation of many-body scarring in a {Bose-Hubbard} quantum simulator},}\
  }\href {\doibase 10.1103/PhysRevResearch.5.023010} {\bibfield  {journal}
  {\bibinfo  {journal} {Phys. Rev. Res.}\ }\textbf {\bibinfo {volume} {5}},\
  \bibinfo {pages} {023010} (\bibinfo {year} {2023})}\BibitemShut {NoStop}%
\bibitem [{\citenamefont {Zhang}\ \emph {et~al.}(2023)\citenamefont {Zhang},
  \citenamefont {Dong}, \citenamefont {Gao}, \citenamefont {Zhao},
  \citenamefont {Hao}, \citenamefont {Desaules}, \citenamefont {Guo},
  \citenamefont {Chen}, \citenamefont {Deng}, \citenamefont {Liu},
  \citenamefont {Ren}, \citenamefont {Yao}, \citenamefont {Zhang},
  \citenamefont {Xu}, \citenamefont {Wang}, \citenamefont {Jin}, \citenamefont
  {Zhu}, \citenamefont {Zhang}, \citenamefont {Li}, \citenamefont {Song},
  \citenamefont {Wang}, \citenamefont {Liu}, \citenamefont {Papi{\'c}},
  \citenamefont {Ying}, \citenamefont {Wang},\ and\ \citenamefont
  {Lai}}]{Zhang2023Many-body}%
  \BibitemOpen
  \bibfield  {author} {\bibinfo {author} {\bibfnamefont {Pengfei}\ \bibnamefont
  {Zhang}}, \bibinfo {author} {\bibfnamefont {Hang}\ \bibnamefont {Dong}},
  \bibinfo {author} {\bibfnamefont {Yu}~\bibnamefont {Gao}}, \bibinfo {author}
  {\bibfnamefont {Liangtian}\ \bibnamefont {Zhao}}, \bibinfo {author}
  {\bibfnamefont {Jie}\ \bibnamefont {Hao}}, \bibinfo {author} {\bibfnamefont
  {Jean-Yves}\ \bibnamefont {Desaules}}, \bibinfo {author} {\bibfnamefont
  {Qiujiang}\ \bibnamefont {Guo}}, \bibinfo {author} {\bibfnamefont {Jiachen}\
  \bibnamefont {Chen}}, \bibinfo {author} {\bibfnamefont {Jinfeng}\
  \bibnamefont {Deng}}, \bibinfo {author} {\bibfnamefont {Bobo}\ \bibnamefont
  {Liu}}, \bibinfo {author} {\bibfnamefont {Wenhui}\ \bibnamefont {Ren}},
  \bibinfo {author} {\bibfnamefont {Yunyan}\ \bibnamefont {Yao}}, \bibinfo
  {author} {\bibfnamefont {Xu}~\bibnamefont {Zhang}}, \bibinfo {author}
  {\bibfnamefont {Shibo}\ \bibnamefont {Xu}}, \bibinfo {author} {\bibfnamefont
  {Ke}~\bibnamefont {Wang}}, \bibinfo {author} {\bibfnamefont {Feitong}\
  \bibnamefont {Jin}}, \bibinfo {author} {\bibfnamefont {Xuhao}\ \bibnamefont
  {Zhu}}, \bibinfo {author} {\bibfnamefont {Bing}\ \bibnamefont {Zhang}},
  \bibinfo {author} {\bibfnamefont {Hekang}\ \bibnamefont {Li}}, \bibinfo
  {author} {\bibfnamefont {Chao}\ \bibnamefont {Song}}, \bibinfo {author}
  {\bibfnamefont {Zhen}\ \bibnamefont {Wang}}, \bibinfo {author} {\bibfnamefont
  {Fangli}\ \bibnamefont {Liu}}, \bibinfo {author} {\bibfnamefont {Zlatko}\
  \bibnamefont {Papi{\'c}}}, \bibinfo {author} {\bibfnamefont {Lei}\
  \bibnamefont {Ying}}, \bibinfo {author} {\bibfnamefont {H.}~\bibnamefont
  {Wang}}, \ and\ \bibinfo {author} {\bibfnamefont {Ying-Cheng}\ \bibnamefont
  {Lai}},\ }\bibfield  {title} {\enquote {\bibinfo {title} {Many-body {Hilbert}
  space scarring on a superconducting processor},}\ }\href {\doibase
  10.1038/s41567-022-01784-9} {\bibfield  {journal} {\bibinfo  {journal}
  {Nature Physics}\ }\textbf {\bibinfo {volume} {19}},\ \bibinfo {pages}
  {120--125} (\bibinfo {year} {2023})}\BibitemShut {NoStop}%
\bibitem [{\citenamefont {Dong}\ \emph {et~al.}(2023)\citenamefont {Dong},
  \citenamefont {Desaules}, \citenamefont {Gao}, \citenamefont {Wang},
  \citenamefont {Guo}, \citenamefont {Chen}, \citenamefont {Zou}, \citenamefont
  {Jin}, \citenamefont {Zhu}, \citenamefont {Zhang}, \citenamefont {Li},
  \citenamefont {Wang}, \citenamefont {Guo}, \citenamefont {Zhang},
  \citenamefont {Ying},\ and\ \citenamefont {Papić}}]{Dong2023Disorder}%
  \BibitemOpen
  \bibfield  {author} {\bibinfo {author} {\bibfnamefont {Hang}\ \bibnamefont
  {Dong}}, \bibinfo {author} {\bibfnamefont {Jean-Yves}\ \bibnamefont
  {Desaules}}, \bibinfo {author} {\bibfnamefont {Yu}~\bibnamefont {Gao}},
  \bibinfo {author} {\bibfnamefont {Ning}\ \bibnamefont {Wang}}, \bibinfo
  {author} {\bibfnamefont {Zexian}\ \bibnamefont {Guo}}, \bibinfo {author}
  {\bibfnamefont {Jiachen}\ \bibnamefont {Chen}}, \bibinfo {author}
  {\bibfnamefont {Yiren}\ \bibnamefont {Zou}}, \bibinfo {author} {\bibfnamefont
  {Feitong}\ \bibnamefont {Jin}}, \bibinfo {author} {\bibfnamefont {Xuhao}\
  \bibnamefont {Zhu}}, \bibinfo {author} {\bibfnamefont {Pengfei}\ \bibnamefont
  {Zhang}}, \bibinfo {author} {\bibfnamefont {Hekang}\ \bibnamefont {Li}},
  \bibinfo {author} {\bibfnamefont {Zhen}\ \bibnamefont {Wang}}, \bibinfo
  {author} {\bibfnamefont {Qiujiang}\ \bibnamefont {Guo}}, \bibinfo {author}
  {\bibfnamefont {Junxiang}\ \bibnamefont {Zhang}}, \bibinfo {author}
  {\bibfnamefont {Lei}\ \bibnamefont {Ying}}, \ and\ \bibinfo {author}
  {\bibfnamefont {Zlatko}\ \bibnamefont {Papić}},\ }\bibfield  {title}
  {\enquote {\bibinfo {title} {Disorder-tunable entanglement at infinite
  temperature},}\ }\href {\doibase 10.1126/sciadv.adj3822} {\bibfield
  {journal} {\bibinfo  {journal} {Science Advances}\ }\textbf {\bibinfo
  {volume} {9}},\ \bibinfo {pages} {eadj3822} (\bibinfo {year}
  {2023})}\BibitemShut {NoStop}%
\bibitem [{\citenamefont {Surace}\ \emph {et~al.}(2020)\citenamefont {Surace},
  \citenamefont {Mazza}, \citenamefont {Giudici}, \citenamefont {Lerose},
  \citenamefont {Gambassi},\ and\ \citenamefont {Dalmonte}}]{Surace2020}%
  \BibitemOpen
  \bibfield  {author} {\bibinfo {author} {\bibfnamefont {Federica~M.}\
  \bibnamefont {Surace}}, \bibinfo {author} {\bibfnamefont {Paolo~P.}\
  \bibnamefont {Mazza}}, \bibinfo {author} {\bibfnamefont {Giuliano}\
  \bibnamefont {Giudici}}, \bibinfo {author} {\bibfnamefont {Alessio}\
  \bibnamefont {Lerose}}, \bibinfo {author} {\bibfnamefont {Andrea}\
  \bibnamefont {Gambassi}}, \ and\ \bibinfo {author} {\bibfnamefont {Marcello}\
  \bibnamefont {Dalmonte}},\ }\bibfield  {title} {\enquote {\bibinfo {title}
  {Lattice gauge theories and string dynamics in {Rydberg} atom quantum
  simulators},}\ }\href {\doibase 10.1103/PhysRevX.10.021041} {\bibfield
  {journal} {\bibinfo  {journal} {Phys. Rev. X}\ }\textbf {\bibinfo {volume}
  {10}},\ \bibinfo {pages} {021041} (\bibinfo {year} {2020})}\BibitemShut
  {NoStop}%
\bibitem [{\citenamefont {Iadecola}\ and\ \citenamefont
  {Schecter}(2020)}]{Iadecola2020quantum}%
  \BibitemOpen
  \bibfield  {author} {\bibinfo {author} {\bibfnamefont {Thomas}\ \bibnamefont
  {Iadecola}}\ and\ \bibinfo {author} {\bibfnamefont {Michael}\ \bibnamefont
  {Schecter}},\ }\bibfield  {title} {\enquote {\bibinfo {title} {Quantum
  many-body scar states with emergent kinetic constraints and
  finite-entanglement revivals},}\ }\href {\doibase
  10.1103/PhysRevB.101.024306} {\bibfield  {journal} {\bibinfo  {journal}
  {Phys. Rev. B}\ }\textbf {\bibinfo {volume} {101}},\ \bibinfo {pages}
  {024306} (\bibinfo {year} {2020})}\BibitemShut {NoStop}%
\bibitem [{\citenamefont {Banerjee}\ and\ \citenamefont
  {Sen}(2021)}]{Banerjee2021}%
  \BibitemOpen
  \bibfield  {author} {\bibinfo {author} {\bibfnamefont {Debasish}\
  \bibnamefont {Banerjee}}\ and\ \bibinfo {author} {\bibfnamefont {Arnab}\
  \bibnamefont {Sen}},\ }\bibfield  {title} {\enquote {\bibinfo {title}
  {Quantum scars from zero modes in an {Abelian} lattice gauge theory on
  ladders},}\ }\href {\doibase 10.1103/PhysRevLett.126.220601} {\bibfield
  {journal} {\bibinfo  {journal} {Phys. Rev. Lett.}\ }\textbf {\bibinfo
  {volume} {126}},\ \bibinfo {pages} {220601} (\bibinfo {year}
  {2021})}\BibitemShut {NoStop}%
\bibitem [{\citenamefont {Halimeh}\ \emph {et~al.}(2023)\citenamefont
  {Halimeh}, \citenamefont {Barbiero}, \citenamefont {Hauke}, \citenamefont
  {Grusdt},\ and\ \citenamefont {Bohrdt}}]{Halimeh2022robust}%
  \BibitemOpen
  \bibfield  {author} {\bibinfo {author} {\bibfnamefont {Jad~C.}\ \bibnamefont
  {Halimeh}}, \bibinfo {author} {\bibfnamefont {Luca}\ \bibnamefont
  {Barbiero}}, \bibinfo {author} {\bibfnamefont {Philipp}\ \bibnamefont
  {Hauke}}, \bibinfo {author} {\bibfnamefont {Fabian}\ \bibnamefont {Grusdt}},
  \ and\ \bibinfo {author} {\bibfnamefont {Annabelle}\ \bibnamefont {Bohrdt}},\
  }\bibfield  {title} {\enquote {\bibinfo {title} {Robust quantum many-body
  scars in lattice gauge theories},}\ }\href {\doibase
  10.22331/q-2023-05-15-1004} {\bibfield  {journal} {\bibinfo  {journal}
  {{Quantum}}\ }\textbf {\bibinfo {volume} {7}},\ \bibinfo {pages} {1004}
  (\bibinfo {year} {2023})}\BibitemShut {NoStop}%
\bibitem [{\citenamefont {Hudomal}\ \emph {et~al.}(2022)\citenamefont
  {Hudomal}, \citenamefont {Desaules}, \citenamefont {Mukherjee}, \citenamefont
  {Su}, \citenamefont {Halimeh},\ and\ \citenamefont
  {Papi\ifmmode~\acute{c}\else \'{c}\fi{}}}]{Hudomal2022}%
  \BibitemOpen
  \bibfield  {author} {\bibinfo {author} {\bibfnamefont {Ana}\ \bibnamefont
  {Hudomal}}, \bibinfo {author} {\bibfnamefont {Jean-Yves}\ \bibnamefont
  {Desaules}}, \bibinfo {author} {\bibfnamefont {Bhaskar}\ \bibnamefont
  {Mukherjee}}, \bibinfo {author} {\bibfnamefont {Guo-Xian}\ \bibnamefont
  {Su}}, \bibinfo {author} {\bibfnamefont {Jad~C.}\ \bibnamefont {Halimeh}}, \
  and\ \bibinfo {author} {\bibfnamefont {Zlatko}\ \bibnamefont
  {Papi\ifmmode~\acute{c}\else \'{c}\fi{}}},\ }\bibfield  {title} {\enquote
  {\bibinfo {title} {Driving quantum many-body scars in the {PXP} model},}\
  }\href {\doibase 10.1103/PhysRevB.106.104302} {\bibfield  {journal} {\bibinfo
   {journal} {Phys. Rev. B}\ }\textbf {\bibinfo {volume} {106}},\ \bibinfo
  {pages} {104302} (\bibinfo {year} {2022})}\BibitemShut {NoStop}%
\bibitem [{\citenamefont {Desaules}\ \emph
  {et~al.}(2023{\natexlab{a}})\citenamefont {Desaules}, \citenamefont
  {Banerjee}, \citenamefont {Hudomal}, \citenamefont
  {Papi\ifmmode~\acute{c}\else \'{c}\fi{}}, \citenamefont {Sen},\ and\
  \citenamefont {Halimeh}}]{Desaules2022weak}%
  \BibitemOpen
  \bibfield  {author} {\bibinfo {author} {\bibfnamefont {Jean-Yves}\
  \bibnamefont {Desaules}}, \bibinfo {author} {\bibfnamefont {Debasish}\
  \bibnamefont {Banerjee}}, \bibinfo {author} {\bibfnamefont {Ana}\
  \bibnamefont {Hudomal}}, \bibinfo {author} {\bibfnamefont {Zlatko}\
  \bibnamefont {Papi\ifmmode~\acute{c}\else \'{c}\fi{}}}, \bibinfo {author}
  {\bibfnamefont {Arnab}\ \bibnamefont {Sen}}, \ and\ \bibinfo {author}
  {\bibfnamefont {Jad~C.}\ \bibnamefont {Halimeh}},\ }\bibfield  {title}
  {\enquote {\bibinfo {title} {Weak ergodicity breaking in the {Schwinger}
  model},}\ }\href {\doibase 10.1103/PhysRevB.107.L201105} {\bibfield
  {journal} {\bibinfo  {journal} {Phys. Rev. B}\ }\textbf {\bibinfo {volume}
  {107}},\ \bibinfo {pages} {L201105} (\bibinfo {year}
  {2023}{\natexlab{a}})}\BibitemShut {NoStop}%
\bibitem [{\citenamefont {Desaules}\ \emph
  {et~al.}(2023{\natexlab{b}})\citenamefont {Desaules}, \citenamefont
  {Hudomal}, \citenamefont {Banerjee}, \citenamefont {Sen}, \citenamefont
  {Papi\ifmmode~\acute{c}\else \'{c}\fi{}},\ and\ \citenamefont
  {Halimeh}}]{Desaules2022prominent}%
  \BibitemOpen
  \bibfield  {author} {\bibinfo {author} {\bibfnamefont {Jean-Yves}\
  \bibnamefont {Desaules}}, \bibinfo {author} {\bibfnamefont {Ana}\
  \bibnamefont {Hudomal}}, \bibinfo {author} {\bibfnamefont {Debasish}\
  \bibnamefont {Banerjee}}, \bibinfo {author} {\bibfnamefont {Arnab}\
  \bibnamefont {Sen}}, \bibinfo {author} {\bibfnamefont {Zlatko}\ \bibnamefont
  {Papi\ifmmode~\acute{c}\else \'{c}\fi{}}}, \ and\ \bibinfo {author}
  {\bibfnamefont {Jad~C.}\ \bibnamefont {Halimeh}},\ }\bibfield  {title}
  {\enquote {\bibinfo {title} {Prominent quantum many-body scars in a truncated
  {Schwinger} model},}\ }\href {\doibase 10.1103/PhysRevB.107.205112}
  {\bibfield  {journal} {\bibinfo  {journal} {Phys. Rev. B}\ }\textbf {\bibinfo
  {volume} {107}},\ \bibinfo {pages} {205112} (\bibinfo {year}
  {2023}{\natexlab{b}})}\BibitemShut {NoStop}%
\bibitem [{\citenamefont {Aramthottil}\ \emph {et~al.}(2022)\citenamefont
  {Aramthottil}, \citenamefont {Bhattacharya}, \citenamefont
  {Gonz\'alez-Cuadra}, \citenamefont {Lewenstein}, \citenamefont {Barbiero},\
  and\ \citenamefont {Zakrzewski}}]{aramthottil2022scar}%
  \BibitemOpen
  \bibfield  {author} {\bibinfo {author} {\bibfnamefont {Adith~Sai}\
  \bibnamefont {Aramthottil}}, \bibinfo {author} {\bibfnamefont {Utso}\
  \bibnamefont {Bhattacharya}}, \bibinfo {author} {\bibfnamefont {Daniel}\
  \bibnamefont {Gonz\'alez-Cuadra}}, \bibinfo {author} {\bibfnamefont {Maciej}\
  \bibnamefont {Lewenstein}}, \bibinfo {author} {\bibfnamefont {Luca}\
  \bibnamefont {Barbiero}}, \ and\ \bibinfo {author} {\bibfnamefont {Jakub}\
  \bibnamefont {Zakrzewski}},\ }\bibfield  {title} {\enquote {\bibinfo {title}
  {Scar states in deconfined {${\mathbb{Z}}_{2}$} lattice gauge theories},}\
  }\href {\doibase 10.1103/PhysRevB.106.L041101} {\bibfield  {journal}
  {\bibinfo  {journal} {Phys. Rev. B}\ }\textbf {\bibinfo {volume} {106}},\
  \bibinfo {pages} {L041101} (\bibinfo {year} {2022})}\BibitemShut {NoStop}%
\bibitem [{\citenamefont {Biswas}\ \emph {et~al.}(2022)\citenamefont {Biswas},
  \citenamefont {Banerjee},\ and\ \citenamefont {Sen}}]{biswas2022scars}%
  \BibitemOpen
  \bibfield  {author} {\bibinfo {author} {\bibfnamefont {Saptarshi}\
  \bibnamefont {Biswas}}, \bibinfo {author} {\bibfnamefont {Debasish}\
  \bibnamefont {Banerjee}}, \ and\ \bibinfo {author} {\bibfnamefont {Arnab}\
  \bibnamefont {Sen}},\ }\bibfield  {title} {\enquote {\bibinfo {title} {{Scars
  from protected zero modes and beyond in $U(1)$ quantum link and quantum dimer
  models}},}\ }\href {\doibase 10.21468/SciPostPhys.12.5.148} {\bibfield
  {journal} {\bibinfo  {journal} {SciPost Phys.}\ }\textbf {\bibinfo {volume}
  {12}},\ \bibinfo {pages} {148} (\bibinfo {year} {2022})}\BibitemShut
  {NoStop}%
\bibitem [{\citenamefont {Daniel}\ \emph {et~al.}(2023)\citenamefont {Daniel},
  \citenamefont {Hallam}, \citenamefont {Desaules}, \citenamefont {Hudomal},
  \citenamefont {Su}, \citenamefont {Halimeh},\ and\ \citenamefont
  {Papi\ifmmode~\acute{c}\else \'{c}\fi{}}}]{Daniel2023}%
  \BibitemOpen
  \bibfield  {author} {\bibinfo {author} {\bibfnamefont {Aiden}\ \bibnamefont
  {Daniel}}, \bibinfo {author} {\bibfnamefont {Andrew}\ \bibnamefont {Hallam}},
  \bibinfo {author} {\bibfnamefont {Jean-Yves}\ \bibnamefont {Desaules}},
  \bibinfo {author} {\bibfnamefont {Ana}\ \bibnamefont {Hudomal}}, \bibinfo
  {author} {\bibfnamefont {Guo-Xian}\ \bibnamefont {Su}}, \bibinfo {author}
  {\bibfnamefont {Jad~C.}\ \bibnamefont {Halimeh}}, \ and\ \bibinfo {author}
  {\bibfnamefont {Zlatko}\ \bibnamefont {Papi\ifmmode~\acute{c}\else
  \'{c}\fi{}}},\ }\bibfield  {title} {\enquote {\bibinfo {title} {Bridging
  quantum criticality via many-body scarring},}\ }\href {\doibase
  10.1103/PhysRevB.107.235108} {\bibfield  {journal} {\bibinfo  {journal}
  {Phys. Rev. B}\ }\textbf {\bibinfo {volume} {107}},\ \bibinfo {pages}
  {235108} (\bibinfo {year} {2023})}\BibitemShut {NoStop}%
\bibitem [{\citenamefont {Ebner}\ \emph {et~al.}(2024)\citenamefont {Ebner},
  \citenamefont {Sch\"afer}, \citenamefont {Seidl}, \citenamefont {M\"uller},\
  and\ \citenamefont {Yao}}]{ebner2024entanglement}%
  \BibitemOpen
  \bibfield  {author} {\bibinfo {author} {\bibfnamefont {Lukas}\ \bibnamefont
  {Ebner}}, \bibinfo {author} {\bibfnamefont {Andreas}\ \bibnamefont
  {Sch\"afer}}, \bibinfo {author} {\bibfnamefont {Clemens}\ \bibnamefont
  {Seidl}}, \bibinfo {author} {\bibfnamefont {Berndt}\ \bibnamefont
  {M\"uller}}, \ and\ \bibinfo {author} {\bibfnamefont {Xiaojun}\ \bibnamefont
  {Yao}},\ }\bibfield  {title} {\enquote {\bibinfo {title} {Entanglement
  entropy of ($2+1$)-dimensional {SU(2)} lattice gauge theory on plaquette
  chains},}\ }\href {\doibase 10.1103/PhysRevD.110.014505} {\bibfield
  {journal} {\bibinfo  {journal} {Phys. Rev. D}\ }\textbf {\bibinfo {volume}
  {110}},\ \bibinfo {pages} {014505} (\bibinfo {year} {2024})}\BibitemShut
  {NoStop}%
\bibitem [{\citenamefont {Sau}\ \emph {et~al.}(2024)\citenamefont {Sau},
  \citenamefont {Stornati}, \citenamefont {Banerjee},\ and\ \citenamefont
  {Sen}}]{Sau2024}%
  \BibitemOpen
  \bibfield  {author} {\bibinfo {author} {\bibfnamefont {Indrajit}\
  \bibnamefont {Sau}}, \bibinfo {author} {\bibfnamefont {Paolo}\ \bibnamefont
  {Stornati}}, \bibinfo {author} {\bibfnamefont {Debasish}\ \bibnamefont
  {Banerjee}}, \ and\ \bibinfo {author} {\bibfnamefont {Arnab}\ \bibnamefont
  {Sen}},\ }\bibfield  {title} {\enquote {\bibinfo {title} {Sublattice scars
  and beyond in two-dimensional {$U(1)$} quantum link lattice gauge
  theories},}\ }\href {\doibase 10.1103/PhysRevD.109.034519} {\bibfield
  {journal} {\bibinfo  {journal} {Phys. Rev. D}\ }\textbf {\bibinfo {volume}
  {109}},\ \bibinfo {pages} {034519} (\bibinfo {year} {2024})}\BibitemShut
  {NoStop}%
\bibitem [{\citenamefont {Osborne}\ \emph {et~al.}(2024)\citenamefont
  {Osborne}, \citenamefont {McCulloch},\ and\ \citenamefont
  {Halimeh}}]{osborne2024quantum}%
  \BibitemOpen
  \bibfield  {author} {\bibinfo {author} {\bibfnamefont {Jesse}\ \bibnamefont
  {Osborne}}, \bibinfo {author} {\bibfnamefont {Ian~P.}\ \bibnamefont
  {McCulloch}}, \ and\ \bibinfo {author} {\bibfnamefont {Jad~C.}\ \bibnamefont
  {Halimeh}},\ }\bibfield  {title} {\enquote {\bibinfo {title} {Quantum
  many-body scarring in {$2+1$D} gauge theories with dynamical matter},}\
  }\href@noop {} {\bibfield  {journal} {\bibinfo  {journal} {arXiv eprints}\ }
  (\bibinfo {year} {2024})},\ \Eprint {http://arxiv.org/abs/2403.08858}
  {arXiv:2403.08858 [cond-mat.quant-gas]} \BibitemShut {NoStop}%
\bibitem [{\citenamefont {Budde}\ \emph {et~al.}(2024)\citenamefont {Budde},
  \citenamefont {Krstic~Marinkovic},\ and\ \citenamefont
  {Pinto~Barros}}]{budde2024quantum}%
  \BibitemOpen
  \bibfield  {author} {\bibinfo {author} {\bibfnamefont {Thea}\ \bibnamefont
  {Budde}}, \bibinfo {author} {\bibfnamefont {Marina}\ \bibnamefont
  {Krstic~Marinkovic}}, \ and\ \bibinfo {author} {\bibfnamefont {Joao~C.}\
  \bibnamefont {Pinto~Barros}},\ }\bibfield  {title} {\enquote {\bibinfo
  {title} {Quantum many-body scars for arbitrary integer spin in
  {$2+1\mathrm{D}$} {Abelian} gauge theories},}\ }\href {\doibase
  10.1103/PhysRevD.110.094506} {\bibfield  {journal} {\bibinfo  {journal}
  {Phys. Rev. D}\ }\textbf {\bibinfo {volume} {110}},\ \bibinfo {pages}
  {094506} (\bibinfo {year} {2024})}\BibitemShut {NoStop}%
\bibitem [{\citenamefont {Borla}\ \emph {et~al.}(2020)\citenamefont {Borla},
  \citenamefont {Verresen}, \citenamefont {Grusdt},\ and\ \citenamefont
  {Moroz}}]{Borla2019}%
  \BibitemOpen
  \bibfield  {author} {\bibinfo {author} {\bibfnamefont {Umberto}\ \bibnamefont
  {Borla}}, \bibinfo {author} {\bibfnamefont {Ruben}\ \bibnamefont {Verresen}},
  \bibinfo {author} {\bibfnamefont {Fabian}\ \bibnamefont {Grusdt}}, \ and\
  \bibinfo {author} {\bibfnamefont {Sergej}\ \bibnamefont {Moroz}},\ }\bibfield
   {title} {\enquote {\bibinfo {title} {Confined phases of one-dimensional
  spinless fermions coupled to ${Z}_{2}$ gauge theory},}\ }\href {\doibase
  10.1103/PhysRevLett.124.120503} {\bibfield  {journal} {\bibinfo  {journal}
  {Phys. Rev. Lett.}\ }\textbf {\bibinfo {volume} {124}},\ \bibinfo {pages}
  {120503} (\bibinfo {year} {2020})}\BibitemShut {NoStop}%
\bibitem [{\citenamefont {Kebri\ifmmode~\check{c}\else \v{c}\fi{}}\ \emph
  {et~al.}(2021)\citenamefont {Kebri\ifmmode~\check{c}\else \v{c}\fi{}},
  \citenamefont {Barbiero}, \citenamefont {Reinmoser}, \citenamefont
  {Schollw\"ock},\ and\ \citenamefont {Grusdt}}]{kebric2021confinement}%
  \BibitemOpen
  \bibfield  {author} {\bibinfo {author} {\bibfnamefont {Matja\ifmmode
  \check{z}\else~\v{z}\fi{}}\ \bibnamefont {Kebri\ifmmode~\check{c}\else
  \v{c}\fi{}}}, \bibinfo {author} {\bibfnamefont {Luca}\ \bibnamefont
  {Barbiero}}, \bibinfo {author} {\bibfnamefont {Christian}\ \bibnamefont
  {Reinmoser}}, \bibinfo {author} {\bibfnamefont {Ulrich}\ \bibnamefont
  {Schollw\"ock}}, \ and\ \bibinfo {author} {\bibfnamefont {Fabian}\
  \bibnamefont {Grusdt}},\ }\bibfield  {title} {\enquote {\bibinfo {title}
  {Confinement and {Mott} transitions of dynamical charges in one-dimensional
  lattice gauge theories},}\ }\href {\doibase 10.1103/PhysRevLett.127.167203}
  {\bibfield  {journal} {\bibinfo  {journal} {Phys. Rev. Lett.}\ }\textbf
  {\bibinfo {volume} {127}},\ \bibinfo {pages} {167203} (\bibinfo {year}
  {2021})}\BibitemShut {NoStop}%
\bibitem [{\citenamefont {Kebrič}\ \emph {et~al.}(2023)\citenamefont
  {Kebrič}, \citenamefont {Borla}, \citenamefont {Schollwöck}, \citenamefont
  {Moroz}, \citenamefont {Barbiero},\ and\ \citenamefont
  {Grusdt}}]{Kebric2023njp}%
  \BibitemOpen
  \bibfield  {author} {\bibinfo {author} {\bibfnamefont {Matjaž}\ \bibnamefont
  {Kebrič}}, \bibinfo {author} {\bibfnamefont {Umberto}\ \bibnamefont
  {Borla}}, \bibinfo {author} {\bibfnamefont {Ulrich}\ \bibnamefont
  {Schollwöck}}, \bibinfo {author} {\bibfnamefont {Sergej}\ \bibnamefont
  {Moroz}}, \bibinfo {author} {\bibfnamefont {Luca}\ \bibnamefont {Barbiero}},
  \ and\ \bibinfo {author} {\bibfnamefont {Fabian}\ \bibnamefont {Grusdt}},\
  }\bibfield  {title} {\enquote {\bibinfo {title} {Confinement induced
  frustration in a one-dimensional $\mathbb{Z}_2$ lattice gauge theory},}\
  }\href {\doibase 10.1088/1367-2630/acb45c} {\bibfield  {journal} {\bibinfo
  {journal} {New Journal of Physics}\ }\textbf {\bibinfo {volume} {25}},\
  \bibinfo {pages} {013035} (\bibinfo {year} {2023})}\BibitemShut {NoStop}%
\bibitem [{\citenamefont {Homeier}\ \emph {et~al.}(2023)\citenamefont
  {Homeier}, \citenamefont {Bohrdt}, \citenamefont {Linsel}, \citenamefont
  {Demler}, \citenamefont {Halimeh},\ and\ \citenamefont
  {Grusdt}}]{Homeier2023realistic}%
  \BibitemOpen
  \bibfield  {author} {\bibinfo {author} {\bibfnamefont {Lukas}\ \bibnamefont
  {Homeier}}, \bibinfo {author} {\bibfnamefont {Annabelle}\ \bibnamefont
  {Bohrdt}}, \bibinfo {author} {\bibfnamefont {Simon}\ \bibnamefont {Linsel}},
  \bibinfo {author} {\bibfnamefont {Eugene}\ \bibnamefont {Demler}}, \bibinfo
  {author} {\bibfnamefont {Jad~C.}\ \bibnamefont {Halimeh}}, \ and\ \bibinfo
  {author} {\bibfnamefont {Fabian}\ \bibnamefont {Grusdt}},\ }\bibfield
  {title} {\enquote {\bibinfo {title} {Realistic scheme for quantum simulation
  of {$\mathbb{Z}_2$} lattice gauge theories with dynamical matter in {(2 +
  1)D}},}\ }\href {\doibase 10.1038/s42005-023-01237-6} {\bibfield  {journal}
  {\bibinfo  {journal} {Communications Physics}\ }\textbf {\bibinfo {volume}
  {6}},\ \bibinfo {pages} {127} (\bibinfo {year} {2023})}\BibitemShut {NoStop}%
\bibitem [{\citenamefont {Linsel}\ \emph {et~al.}(2024)\citenamefont {Linsel},
  \citenamefont {Bohrdt}, \citenamefont {Homeier}, \citenamefont {Pollet},\
  and\ \citenamefont {Grusdt}}]{linsel2024percolation}%
  \BibitemOpen
  \bibfield  {author} {\bibinfo {author} {\bibfnamefont {Simon~M.}\
  \bibnamefont {Linsel}}, \bibinfo {author} {\bibfnamefont {Annabelle}\
  \bibnamefont {Bohrdt}}, \bibinfo {author} {\bibfnamefont {Lukas}\
  \bibnamefont {Homeier}}, \bibinfo {author} {\bibfnamefont {Lode}\
  \bibnamefont {Pollet}}, \ and\ \bibinfo {author} {\bibfnamefont {Fabian}\
  \bibnamefont {Grusdt}},\ }\bibfield  {title} {\enquote {\bibinfo {title}
  {Percolation as a confinement order parameter in {${\mathbb{Z}}_{2}$} lattice
  gauge theories},}\ }\href {\doibase 10.1103/PhysRevB.110.L241101} {\bibfield
  {journal} {\bibinfo  {journal} {Phys. Rev. B}\ }\textbf {\bibinfo {volume}
  {110}},\ \bibinfo {pages} {L241101} (\bibinfo {year} {2024})}\BibitemShut
  {NoStop}%
\bibitem [{\citenamefont {Kebri\ifmmode~\check{c}\else \v{c}\fi{}}\ \emph
  {et~al.}(2024)\citenamefont {Kebri\ifmmode~\check{c}\else \v{c}\fi{}},
  \citenamefont {Halimeh}, \citenamefont {Schollw\"ock},\ and\ \citenamefont
  {Grusdt}}]{Kebric2023confinement}%
  \BibitemOpen
  \bibfield  {author} {\bibinfo {author} {\bibfnamefont {Matja\ifmmode
  \check{z}\else~\v{z}\fi{}}\ \bibnamefont {Kebri\ifmmode~\check{c}\else
  \v{c}\fi{}}}, \bibinfo {author} {\bibfnamefont {Jad~C.}\ \bibnamefont
  {Halimeh}}, \bibinfo {author} {\bibfnamefont {Ulrich}\ \bibnamefont
  {Schollw\"ock}}, \ and\ \bibinfo {author} {\bibfnamefont {Fabian}\
  \bibnamefont {Grusdt}},\ }\bibfield  {title} {\enquote {\bibinfo {title}
  {Confinement in $(1+1)$-dimensional {${\mathbb{Z}}_{2}$} lattice gauge
  theories at finite temperature},}\ }\href {\doibase
  10.1103/PhysRevB.109.245110} {\bibfield  {journal} {\bibinfo  {journal}
  {Phys. Rev. B}\ }\textbf {\bibinfo {volume} {109}},\ \bibinfo {pages}
  {245110} (\bibinfo {year} {2024})}\BibitemShut {NoStop}%
\bibitem [{\citenamefont {Fromm}\ \emph {et~al.}(2024)\citenamefont {Fromm},
  \citenamefont {Philipsen}, \citenamefont {Spannowsky},\ and\ \citenamefont
  {Winterowd}}]{Fromm2023simulating}%
  \BibitemOpen
  \bibfield  {author} {\bibinfo {author} {\bibfnamefont {Michael}\ \bibnamefont
  {Fromm}}, \bibinfo {author} {\bibfnamefont {Owe}\ \bibnamefont {Philipsen}},
  \bibinfo {author} {\bibfnamefont {Michael}\ \bibnamefont {Spannowsky}}, \
  and\ \bibinfo {author} {\bibfnamefont {Christopher}\ \bibnamefont
  {Winterowd}},\ }\bibfield  {title} {\enquote {\bibinfo {title} {Simulating
  $\mathbb{Z}_2$ lattice gauge theory with the variational quantum
  thermalizer},}\ }\href {\doibase 10.1140/epjqt/s40507-024-00232-2} {\bibfield
   {journal} {\bibinfo  {journal} {EPJ Quantum Technology}\ }\textbf {\bibinfo
  {volume} {11}},\ \bibinfo {pages} {20} (\bibinfo {year} {2024})}\BibitemShut
  {NoStop}%
\bibitem [{\citenamefont {Gustafson}\ \emph {et~al.}(2023)\citenamefont
  {Gustafson}, \citenamefont {Li}, \citenamefont {Khan}, \citenamefont {Kim},
  \citenamefont {Kurkcuoglu}, \citenamefont {Alam}, \citenamefont {Orth},
  \citenamefont {Rahmani},\ and\ \citenamefont {Iadecola}}]{Gustafson2023}%
  \BibitemOpen
  \bibfield  {author} {\bibinfo {author} {\bibfnamefont {Erik~J.}\ \bibnamefont
  {Gustafson}}, \bibinfo {author} {\bibfnamefont {Andy C.~Y.}\ \bibnamefont
  {Li}}, \bibinfo {author} {\bibfnamefont {Abid}\ \bibnamefont {Khan}},
  \bibinfo {author} {\bibfnamefont {Joonho}\ \bibnamefont {Kim}}, \bibinfo
  {author} {\bibfnamefont {Doga~Murat}\ \bibnamefont {Kurkcuoglu}}, \bibinfo
  {author} {\bibfnamefont {M.~Sohaib}\ \bibnamefont {Alam}}, \bibinfo {author}
  {\bibfnamefont {Peter~P.}\ \bibnamefont {Orth}}, \bibinfo {author}
  {\bibfnamefont {Armin}\ \bibnamefont {Rahmani}}, \ and\ \bibinfo {author}
  {\bibfnamefont {Thomas}\ \bibnamefont {Iadecola}},\ }\bibfield  {title}
  {\enquote {\bibinfo {title} {Preparing quantum many-body scar states on
  quantum computers},}\ }\href {\doibase 10.22331/q-2023-11-07-1171} {\bibfield
   {journal} {\bibinfo  {journal} {Quantum}\ }\textbf {\bibinfo {volume} {7}},\
  \bibinfo {pages} {1171} (\bibinfo {year} {2023})}\BibitemShut {NoStop}%
\bibitem [{\citenamefont {Kogut}\ and\ \citenamefont
  {Susskind}(1975)}]{Kogut1975}%
  \BibitemOpen
  \bibfield  {author} {\bibinfo {author} {\bibfnamefont {John}\ \bibnamefont
  {Kogut}}\ and\ \bibinfo {author} {\bibfnamefont {Leonard}\ \bibnamefont
  {Susskind}},\ }\bibfield  {title} {\enquote {\bibinfo {title} {Hamiltonian
  formulation of wilson's lattice gauge theories},}\ }\href {\doibase
  10.1103/PhysRevD.11.395} {\bibfield  {journal} {\bibinfo  {journal} {Phys.
  Rev. D}\ }\textbf {\bibinfo {volume} {11}},\ \bibinfo {pages} {395--408}
  (\bibinfo {year} {1975})}\BibitemShut {NoStop}%
\bibitem [{\citenamefont {Yang}\ \emph
  {et~al.}(2020{\natexlab{b}})\citenamefont {Yang}, \citenamefont {Liu},
  \citenamefont {Gorshkov},\ and\ \citenamefont
  {Iadecola}}]{Yang2020fragmentation}%
  \BibitemOpen
  \bibfield  {author} {\bibinfo {author} {\bibfnamefont {Zhi-Cheng}\
  \bibnamefont {Yang}}, \bibinfo {author} {\bibfnamefont {Fangli}\ \bibnamefont
  {Liu}}, \bibinfo {author} {\bibfnamefont {Alexey~V.}\ \bibnamefont
  {Gorshkov}}, \ and\ \bibinfo {author} {\bibfnamefont {Thomas}\ \bibnamefont
  {Iadecola}},\ }\bibfield  {title} {\enquote {\bibinfo {title} {Hilbert-space
  fragmentation from strict confinement},}\ }\href {\doibase
  10.1103/PhysRevLett.124.207602} {\bibfield  {journal} {\bibinfo  {journal}
  {Phys. Rev. Lett.}\ }\textbf {\bibinfo {volume} {124}},\ \bibinfo {pages}
  {207602} (\bibinfo {year} {2020}{\natexlab{b}})}\BibitemShut {NoStop}%
\bibitem [{\citenamefont {Bravyi}\ \emph {et~al.}(2011)\citenamefont {Bravyi},
  \citenamefont {DiVincenzo},\ and\ \citenamefont {Loss}}]{Bravyi2011}%
  \BibitemOpen
  \bibfield  {author} {\bibinfo {author} {\bibfnamefont {Sergey}\ \bibnamefont
  {Bravyi}}, \bibinfo {author} {\bibfnamefont {David~P.}\ \bibnamefont
  {DiVincenzo}}, \ and\ \bibinfo {author} {\bibfnamefont {Daniel}\ \bibnamefont
  {Loss}},\ }\bibfield  {title} {\enquote {\bibinfo {title}
  {{Schrieffer–Wolff} transformation for quantum many-body systems},}\ }\href
  {\doibase https://doi.org/10.1016/j.aop.2011.06.004} {\bibfield  {journal}
  {\bibinfo  {journal} {Annals of Physics}\ }\textbf {\bibinfo {volume}
  {326}},\ \bibinfo {pages} {2793 -- 2826} (\bibinfo {year}
  {2011})}\BibitemShut {NoStop}%
\bibitem [{\citenamefont {Inokuchi}(1998)}]{Inokuchi156}%
  \BibitemOpen
  \bibfield  {author} {\bibinfo {author} {\bibfnamefont {Shuichi}\ \bibnamefont
  {Inokuchi}},\ }\bibfield  {title} {\enquote {\bibinfo {title} {On behaviors
  of cellular automata with rule 156},}\ }\href {\doibase
  https://doi.org/10.5109/13474} {\bibfield  {journal} {\bibinfo  {journal}
  {Bulletin of Informatics and Cybernetics}\ }\textbf {\bibinfo {volume}
  {30}},\ \bibinfo {pages} {121 -- 131} (\bibinfo {year} {1998})}\BibitemShut
  {NoStop}%
\bibitem [{\citenamefont {Valencia-Tortora}\ \emph {et~al.}(2024)\citenamefont
  {Valencia-Tortora}, \citenamefont {Pancotti}, \citenamefont {Fleischhauer},
  \citenamefont {Bernien},\ and\ \citenamefont {Marino}}]{Valencia2024Rydberg}%
  \BibitemOpen
  \bibfield  {author} {\bibinfo {author} {\bibfnamefont {Riccardo~J.}\
  \bibnamefont {Valencia-Tortora}}, \bibinfo {author} {\bibfnamefont {Nicola}\
  \bibnamefont {Pancotti}}, \bibinfo {author} {\bibfnamefont {Michael}\
  \bibnamefont {Fleischhauer}}, \bibinfo {author} {\bibfnamefont {Hannes}\
  \bibnamefont {Bernien}}, \ and\ \bibinfo {author} {\bibfnamefont {Jamir}\
  \bibnamefont {Marino}},\ }\bibfield  {title} {\enquote {\bibinfo {title}
  {Rydberg platform for nonergodic chiral quantum dynamics},}\ }\href {\doibase
  10.1103/PhysRevLett.132.223201} {\bibfield  {journal} {\bibinfo  {journal}
  {Phys. Rev. Lett.}\ }\textbf {\bibinfo {volume} {132}},\ \bibinfo {pages}
  {223201} (\bibinfo {year} {2024})}\BibitemShut {NoStop}%
\bibitem [{\citenamefont {Maity}\ and\ \citenamefont
  {Hamazaki}(2024)}]{Maity2024kinetically}%
  \BibitemOpen
  \bibfield  {author} {\bibinfo {author} {\bibfnamefont {Somnath}\ \bibnamefont
  {Maity}}\ and\ \bibinfo {author} {\bibfnamefont {Ryusuke}\ \bibnamefont
  {Hamazaki}},\ }\bibfield  {title} {\enquote {\bibinfo {title} {Kinetically
  constrained models constructed from dissipative quantum dynamics},}\ }\href
  {\doibase 10.1103/PhysRevB.110.014301} {\bibfield  {journal} {\bibinfo
  {journal} {Phys. Rev. B}\ }\textbf {\bibinfo {volume} {110}},\ \bibinfo
  {pages} {014301} (\bibinfo {year} {2024})}\BibitemShut {NoStop}%
\bibitem [{\citenamefont {Sala}\ \emph {et~al.}(2020)\citenamefont {Sala},
  \citenamefont {Rakovszky}, \citenamefont {Verresen}, \citenamefont {Knap},\
  and\ \citenamefont {Pollmann}}]{Sala20}%
  \BibitemOpen
  \bibfield  {author} {\bibinfo {author} {\bibfnamefont {Pablo}\ \bibnamefont
  {Sala}}, \bibinfo {author} {\bibfnamefont {Tibor}\ \bibnamefont {Rakovszky}},
  \bibinfo {author} {\bibfnamefont {Ruben}\ \bibnamefont {Verresen}}, \bibinfo
  {author} {\bibfnamefont {Michael}\ \bibnamefont {Knap}}, \ and\ \bibinfo
  {author} {\bibfnamefont {Frank}\ \bibnamefont {Pollmann}},\ }\bibfield
  {title} {\enquote {\bibinfo {title} {Ergodicity breaking arising from hilbert
  space fragmentation in dipole-conserving {Hamiltonians}},}\ }\href {\doibase
  10.1103/PhysRevX.10.011047} {\bibfield  {journal} {\bibinfo  {journal} {Phys.
  Rev. X}\ }\textbf {\bibinfo {volume} {10}},\ \bibinfo {pages} {011047}
  (\bibinfo {year} {2020})}\BibitemShut {NoStop}%
\bibitem [{\citenamefont {Khemani}\ \emph {et~al.}(2020)\citenamefont
  {Khemani}, \citenamefont {Hermele},\ and\ \citenamefont
  {Nandkishore}}]{Khemani20}%
  \BibitemOpen
  \bibfield  {author} {\bibinfo {author} {\bibfnamefont {Vedika}\ \bibnamefont
  {Khemani}}, \bibinfo {author} {\bibfnamefont {Michael}\ \bibnamefont
  {Hermele}}, \ and\ \bibinfo {author} {\bibfnamefont {Rahul}\ \bibnamefont
  {Nandkishore}},\ }\bibfield  {title} {\enquote {\bibinfo {title}
  {Localization from {Hilbert} space shattering: From theory to physical
  realizations},}\ }\href {\doibase 10.1103/PhysRevB.101.174204} {\bibfield
  {journal} {\bibinfo  {journal} {Phys. Rev. B}\ }\textbf {\bibinfo {volume}
  {101}},\ \bibinfo {pages} {174204} (\bibinfo {year} {2020})}\BibitemShut
  {NoStop}%
\bibitem [{\citenamefont {Lesanovsky}\ and\ \citenamefont
  {Katsura}(2012)}]{Lesanovsky2012}%
  \BibitemOpen
  \bibfield  {author} {\bibinfo {author} {\bibfnamefont {Igor}\ \bibnamefont
  {Lesanovsky}}\ and\ \bibinfo {author} {\bibfnamefont {Hosho}\ \bibnamefont
  {Katsura}},\ }\bibfield  {title} {\enquote {\bibinfo {title} {{Interacting
  Fibonacci anyons in a Rydberg gas}},}\ }\href {\doibase
  10.1103/PhysRevA.86.041601} {\bibfield  {journal} {\bibinfo  {journal} {Phys.
  Rev. A}\ }\textbf {\bibinfo {volume} {86}},\ \bibinfo {pages} {041601}
  (\bibinfo {year} {2012})}\BibitemShut {NoStop}%
\bibitem [{\citenamefont {Chandrasekharan}\ and\ \citenamefont
  {Wiese}(1997)}]{Chandrasekharan1997}%
  \BibitemOpen
  \bibfield  {author} {\bibinfo {author} {\bibfnamefont {S}~\bibnamefont
  {Chandrasekharan}}\ and\ \bibinfo {author} {\bibfnamefont {U.-J}\
  \bibnamefont {Wiese}},\ }\bibfield  {title} {\enquote {\bibinfo {title}
  {Quantum link models: {A} discrete approach to gauge theories},}\ }\href
  {\doibase https://doi.org/10.1016/S0550-3213(97)80041-7} {\bibfield
  {journal} {\bibinfo  {journal} {Nuclear Physics B}\ }\textbf {\bibinfo
  {volume} {492}},\ \bibinfo {pages} {455 -- 471} (\bibinfo {year}
  {1997})}\BibitemShut {NoStop}%
\bibitem [{\citenamefont {Wiese}(2013)}]{Wiese_review}%
  \BibitemOpen
  \bibfield  {author} {\bibinfo {author} {\bibfnamefont {U.-J.}\ \bibnamefont
  {Wiese}},\ }\bibfield  {title} {\enquote {\bibinfo {title} {Ultracold quantum
  gases and lattice systems: quantum simulation of lattice gauge theories},}\
  }\href {\doibase 10.1002/andp.201300104} {\bibfield  {journal} {\bibinfo
  {journal} {Annalen der Physik}\ }\textbf {\bibinfo {volume} {525}},\ \bibinfo
  {pages} {777--796} (\bibinfo {year} {2013})}\BibitemShut {NoStop}%
\bibitem [{\citenamefont {Desaules}\ \emph {et~al.}(2024)\citenamefont
  {Desaules}, \citenamefont {Su}, \citenamefont {McCulloch}, \citenamefont
  {Yang}, \citenamefont {Papi{\'{c}}},\ and\ \citenamefont
  {Halimeh}}]{Desaules2024ergodicitybreaking}%
  \BibitemOpen
  \bibfield  {author} {\bibinfo {author} {\bibfnamefont {Jean-Yves}\
  \bibnamefont {Desaules}}, \bibinfo {author} {\bibfnamefont {Guo-Xian}\
  \bibnamefont {Su}}, \bibinfo {author} {\bibfnamefont {Ian~P.}\ \bibnamefont
  {McCulloch}}, \bibinfo {author} {\bibfnamefont {Bing}\ \bibnamefont {Yang}},
  \bibinfo {author} {\bibfnamefont {Zlatko}\ \bibnamefont {Papi{\'{c}}}}, \
  and\ \bibinfo {author} {\bibfnamefont {Jad~C.}\ \bibnamefont {Halimeh}},\
  }\bibfield  {title} {\enquote {\bibinfo {title} {Ergodicity breaking under
  confinement in cold-atom quantum simulators},}\ }\href {\doibase
  10.22331/q-2024-02-29-1274} {\bibfield  {journal} {\bibinfo  {journal}
  {{Quantum}}\ }\textbf {\bibinfo {volume} {8}},\ \bibinfo {pages} {1274}
  (\bibinfo {year} {2024})}\BibitemShut {NoStop}%
\bibitem [{\citenamefont {Kay}(2010)}]{Kay2010perfect}%
  \BibitemOpen
  \bibfield  {author} {\bibinfo {author} {\bibfnamefont {Alastair}\
  \bibnamefont {Kay}},\ }\bibfield  {title} {\enquote {\bibinfo {title}
  {Perfect, efficient, state transfer and its application as a constructive
  tool},}\ }\href {\doibase 10.1142/S0219749910006514} {\bibfield  {journal}
  {\bibinfo  {journal} {International Journal of Quantum Information}\ }\textbf
  {\bibinfo {volume} {08}},\ \bibinfo {pages} {641--676} (\bibinfo {year}
  {2010})}\BibitemShut {NoStop}%
\bibitem [{\citenamefont {Oganesyan}\ and\ \citenamefont
  {Huse}(2007)}]{Oganesyan07}%
  \BibitemOpen
  \bibfield  {author} {\bibinfo {author} {\bibfnamefont {Vadim}\ \bibnamefont
  {Oganesyan}}\ and\ \bibinfo {author} {\bibfnamefont {David~A.}\ \bibnamefont
  {Huse}},\ }\bibfield  {title} {\enquote {\bibinfo {title} {Localization of
  interacting fermions at high temperature},}\ }\href {\doibase
  10.1103/PhysRevB.75.155111} {\bibfield  {journal} {\bibinfo  {journal} {Phys.
  Rev. B}\ }\textbf {\bibinfo {volume} {75}},\ \bibinfo {pages} {155111}
  (\bibinfo {year} {2007})}\BibitemShut {NoStop}%
\bibitem [{\citenamefont {Atas}\ \emph {et~al.}(2013)\citenamefont {Atas},
  \citenamefont {Bogomolny}, \citenamefont {Giraud},\ and\ \citenamefont
  {Roux}}]{Atas13}%
  \BibitemOpen
  \bibfield  {author} {\bibinfo {author} {\bibfnamefont {Y.~Y.}\ \bibnamefont
  {Atas}}, \bibinfo {author} {\bibfnamefont {E.}~\bibnamefont {Bogomolny}},
  \bibinfo {author} {\bibfnamefont {O.}~\bibnamefont {Giraud}}, \ and\ \bibinfo
  {author} {\bibfnamefont {G.}~\bibnamefont {Roux}},\ }\bibfield  {title}
  {\enquote {\bibinfo {title} {Distribution of the ratio of consecutive level
  spacings in random matrix ensembles},}\ }\href {\doibase
  10.1103/PhysRevLett.110.084101} {\bibfield  {journal} {\bibinfo  {journal}
  {Phys. Rev. Lett.}\ }\textbf {\bibinfo {volume} {110}},\ \bibinfo {pages}
  {084101} (\bibinfo {year} {2013})}\BibitemShut {NoStop}%
\bibitem [{Note1()}]{Note1}%
  \BibitemOpen
  \bibinfo {note} {This is because unfrozen spins on odd sites are always in
  the opposite state as unfrozen spins on even sites. When an even site $j$ is
  $\downarrow $, this allows the term $\protect \hat {P}_j\protect \hat
  {Z}_{j+1}\protect \hat {Q}_{j+2}$ to be $+1$. But at the same time, the odd
  site $k=j+3$ is $\uparrow $, allowing $\protect \hat {Q}_k\protect \hat
  {Z}_{k+1}\protect \hat {P}_{k+2}$ to be $-1$. The same can be seen when $j$
  is $\uparrow $ and $k$ is $\downarrow $, but with the terms $\protect \hat
  {P}_{j-1}\protect \hat {Z}_{j-1}\protect \hat {Q}_{j}$ and $\protect \hat
  {Q}_{k-2}\protect \hat {Z}_{k-1}\protect \hat {P}_{k}$.}\BibitemShut {Stop}%
\bibitem [{\citenamefont {Desaules}\ \emph {et~al.}(2021)\citenamefont
  {Desaules}, \citenamefont {Hudomal}, \citenamefont {Turner},\ and\
  \citenamefont {Papi\'{c}}}]{Desaules2021TFH}%
  \BibitemOpen
  \bibfield  {author} {\bibinfo {author} {\bibfnamefont {Jean-Yves}\
  \bibnamefont {Desaules}}, \bibinfo {author} {\bibfnamefont {Ana}\
  \bibnamefont {Hudomal}}, \bibinfo {author} {\bibfnamefont {Christopher~J.}\
  \bibnamefont {Turner}}, \ and\ \bibinfo {author} {\bibfnamefont {Zlatko}\
  \bibnamefont {Papi\'{c}}},\ }\bibfield  {title} {\enquote {\bibinfo {title}
  {Proposal for realizing quantum scars in the tilted {1D} {Fermi-Hubbard}
  model},}\ }\href {\doibase 10.1103/PhysRevLett.126.210601} {\bibfield
  {journal} {\bibinfo  {journal} {Phys. Rev. Lett.}\ }\textbf {\bibinfo
  {volume} {126}},\ \bibinfo {pages} {210601} (\bibinfo {year}
  {2021})}\BibitemShut {NoStop}%
\bibitem [{\citenamefont {Bastianello}\ \emph {et~al.}(2022)\citenamefont
  {Bastianello}, \citenamefont {Borla},\ and\ \citenamefont
  {Moroz}}]{Bastianello22}%
  \BibitemOpen
  \bibfield  {author} {\bibinfo {author} {\bibfnamefont {Alvise}\ \bibnamefont
  {Bastianello}}, \bibinfo {author} {\bibfnamefont {Umberto}\ \bibnamefont
  {Borla}}, \ and\ \bibinfo {author} {\bibfnamefont {Sergej}\ \bibnamefont
  {Moroz}},\ }\bibfield  {title} {\enquote {\bibinfo {title} {Fragmentation and
  emergent integrable transport in the weakly tilted {Ising} chain},}\ }\href
  {\doibase 10.1103/PhysRevLett.128.196601} {\bibfield  {journal} {\bibinfo
  {journal} {Phys. Rev. Lett.}\ }\textbf {\bibinfo {volume} {128}},\ \bibinfo
  {pages} {196601} (\bibinfo {year} {2022})}\BibitemShut {NoStop}%
\bibitem [{RDa(2025)}]{RData}%
  \BibitemOpen
  \bibfield  {title} {\enquote {\bibinfo {title} {Research data for
  \textit{{Mass}-{Assisted} {Local} {Deconfinement} in a {Confined}
  {$\mathbb{Z}_2$} {Lattice} {Gauge} {Theory}}},}\ }\href {\doibase
  10.15479/AT:ISTA:19791} {\bibfield  {journal} {\bibinfo  {journal} {Institute
  of Science and Technology Austria}\ } (\bibinfo {year} {2025}),\
  10.15479/AT:ISTA:19791}\BibitemShut {NoStop}%
\bibitem [{\citenamefont {Khor}\ \emph {et~al.}(2024)\citenamefont {Khor},
  \citenamefont {K{\"u}rkc{\"u}oglu}, \citenamefont {Hobbs}, \citenamefont
  {Perdue},\ and\ \citenamefont {Klich}}]{Khor23}%
  \BibitemOpen
  \bibfield  {author} {\bibinfo {author} {\bibfnamefont {Brian J.~J.}\
  \bibnamefont {Khor}}, \bibinfo {author} {\bibfnamefont {D.~M.}\ \bibnamefont
  {K{\"u}rkc{\"u}oglu}}, \bibinfo {author} {\bibfnamefont {T.~J.}\ \bibnamefont
  {Hobbs}}, \bibinfo {author} {\bibfnamefont {G.~N.}\ \bibnamefont {Perdue}}, \
  and\ \bibinfo {author} {\bibfnamefont {Israel}\ \bibnamefont {Klich}},\
  }\bibfield  {title} {\enquote {\bibinfo {title} {Confinement and kink
  entanglement asymmetry on a quantum {Ising} chain},}\ }\href {\doibase
  10.22331/q-2024-09-06-1462} {\bibfield  {journal} {\bibinfo  {journal}
  {Quantum}\ }\textbf {\bibinfo {volume} {8}},\ \bibinfo {pages} {1462}
  (\bibinfo {year} {2024})}\BibitemShut {NoStop}%
\end{thebibliography}%

\end{document}